\documentclass[nonacm]{acmart}
\AtBeginDocument{%
  \providecommand\BibTeX{{%
    \normalfont B\kern-0.5em{\scshape i\kern-0.25em b}\kern-0.8em\TeX}}}

\usepackage{subfigure}
\usepackage{multirow}
\usepackage{array}
\usepackage{makecell}
\usepackage{graphicx}
\usepackage{colortbl}

\begin{document}

\title{Information Cascade Prediction under Public Emergencies: A Survey}

\author{Qi Zhang}
\email{230228520@seu.edu.cn}
\orcid{0000-0002-3695-1161}
\affiliation{%
  \institution{southeast university}
  \city{NanJing}
  \state{JiangSU}
  \country{China}
  \postcode{211189}
}
\author{Guang Wang}
\email{guang@cs.fsu.edu}
\affiliation{%
  \institution{Florida State University}
  \streetaddress{}
  \city{}
  \state{}
  \country{USA}
  \postcode{}
}
\author{Li Lin}
\email{linli321@seu.edu.cn}
\affiliation{%
  \institution{southeast university}
  \city{NanJing}
  \state{JiangSU}
  \country{China}
  \postcode{211189}
}

\author{Kaiwen Xia}
\email{230228518@seu.edu.cn}
\affiliation{%
  \institution{southeast university}
  \city{NanJing}
  \state{JiangSU}
  \country{China}
  \postcode{211189}
}

\author{Shuai Wang}
\authornotemark[1]
\email{shuaiwang@seu.edu.cn}
\affiliation{%
  \institution{southeast university}
  \city{NanJing}
  \state{JiangSU}
  \country{China}
  \postcode{211189}
}

\renewcommand{\shortauthors}{, et al.}

\begin{abstract}
With the advent of the era of big data, massive information, expert experience, and high-accuracy models bring great opportunities to the information cascade prediction of public emergencies. However, the involvement of specialist knowledge from various disciplines has resulted in a primarily application-specific focus (e.g., earthquakes, floods, infectious diseases) for information cascade prediction of public emergencies. The lack of a unified prediction framework poses a challenge for classifying intersectional prediction methods across different application fields. This survey paper offers a systematic classification and summary of information cascade modeling, prediction, and application. We aim to help researchers identify cutting-edge research and comprehend models and methods of information cascade prediction under public emergencies. By summarizing open issues and outlining future directions in this field, this paper has the potential to be a valuable resource for researchers conducting further studies on predicting information cascades.
\end{abstract}



\maketitle

\section{Introduction}
\subsection{Background}
\begin{figure}[htbp]
\centering  

\setlength{\abovecaptionskip}{0.2cm}   
\setlength{\belowcaptionskip}{-0.8cm}   
\subfigure[Global Distribution from Natural Disasters, 2000 to 2023]{   
\centering    
\includegraphics[width=0.44\linewidth]{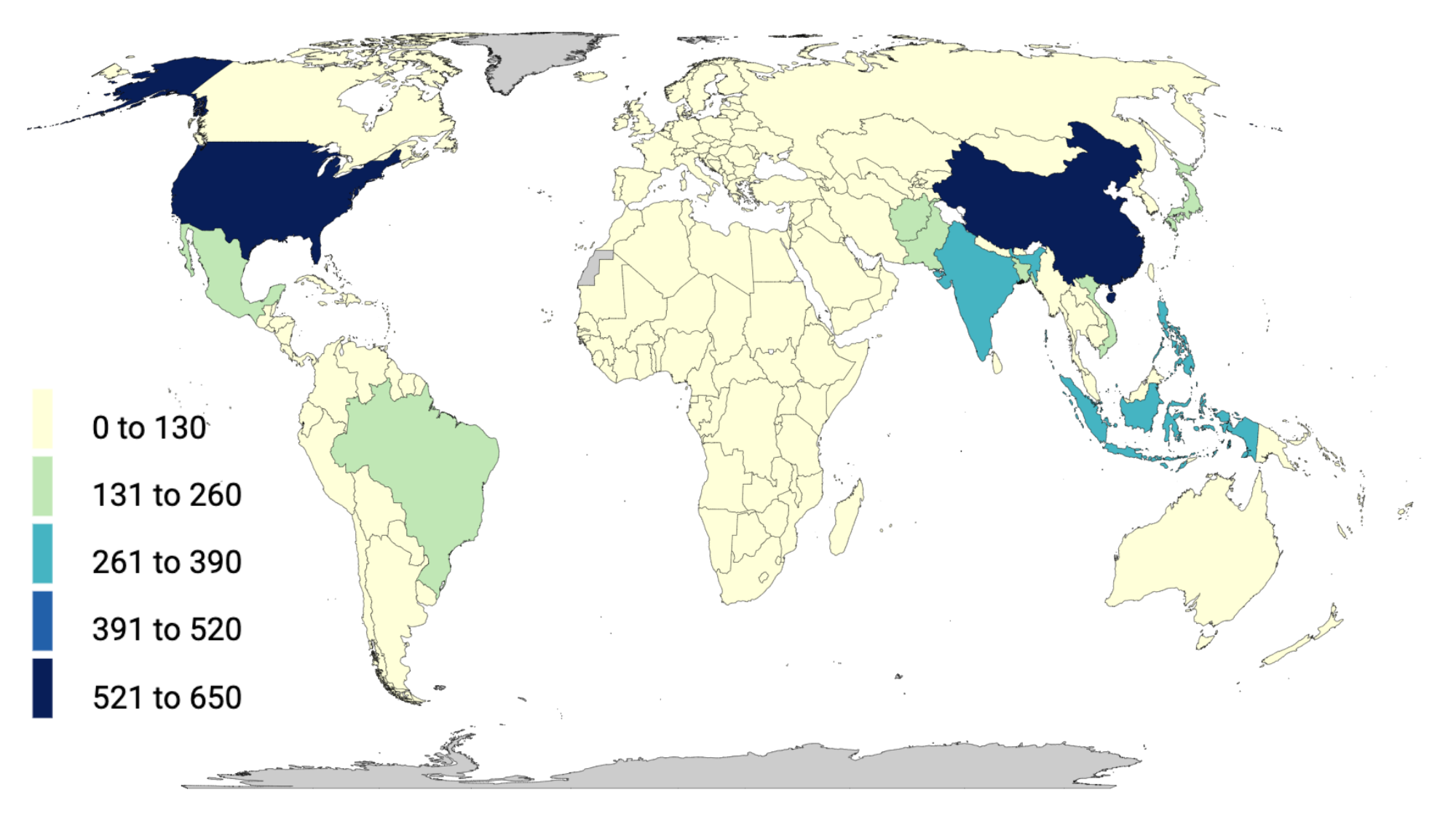} 
}
\centering  
\subfigure[Global Distribution from Technological Disasters, 2000 to 2023]{ 
\centering    
\includegraphics[width=0.44\linewidth]{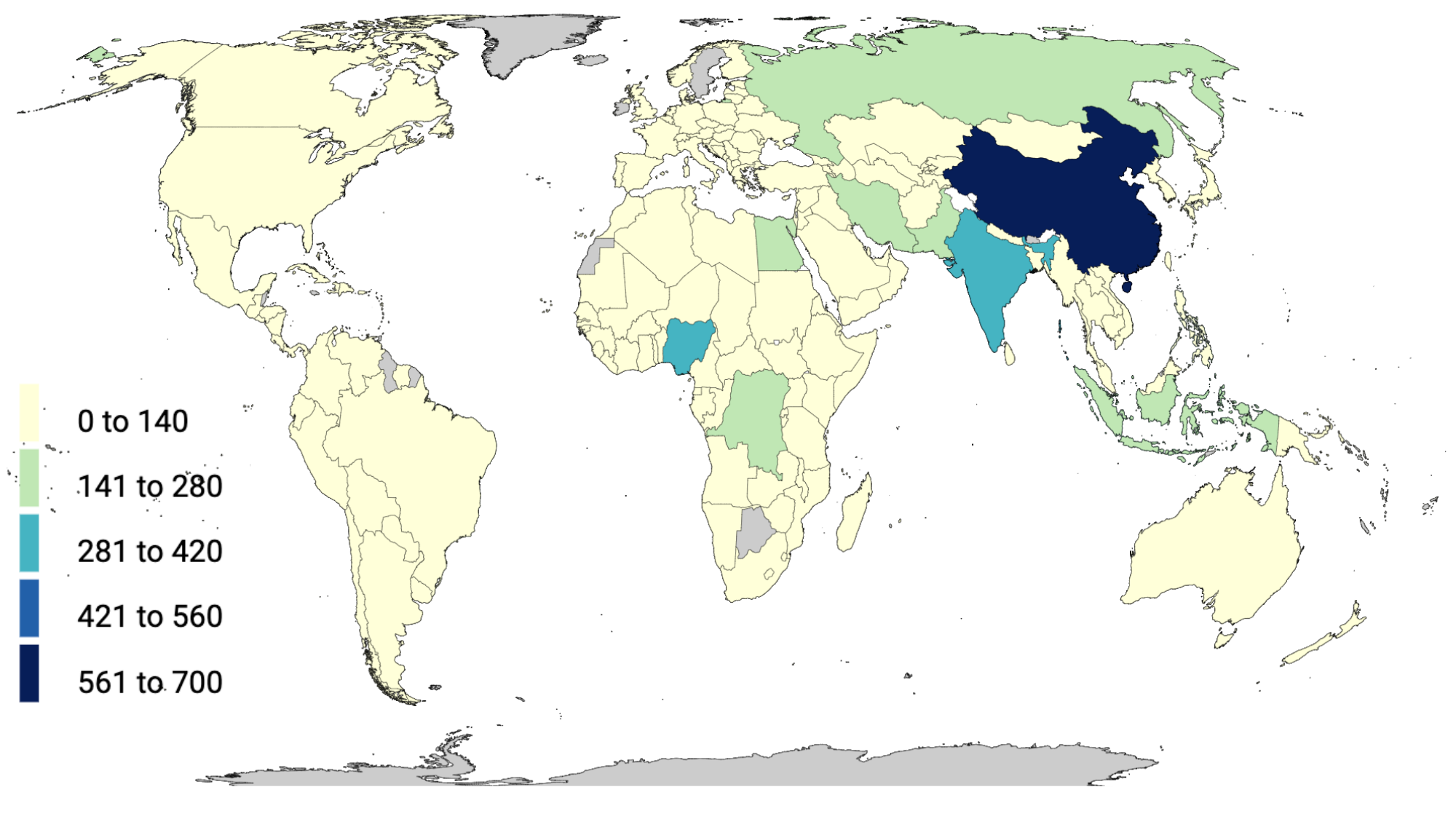}
}
\caption{Global Distribution of Public Emergencies}    
\label{fig:1}    
\end{figure}

Information Cascade Prediction is a critical research field that has implications in a variety of domains, such as healthcare, business, cyber domain, politics, and entertainment, ultimately impacting nearly every aspect of our lives \cite{disaster}. These emergencies are unexpected events that occur suddenly and result in or have the potential to result in significant casualties, property damage, ecological harm, and serious social consequences \cite{disaster2}. Throughout history, natural disasters (such as earthquakes, tsunamis, volcanic eruptions, storms, floods, avalanches, droughts, and wildfires) and accident disasters (including environmental disasters, traffic accidents, explosions, and gas leaks) have caused numerous fatalities, infrastructure damage, and extensive economic loss. According to the Emergencies Database (EM-DAT), between 2000 and 2023, 5,922 public emergencies occurred, leading to 480,000 casualties and 3.5 trillion in economic losses, as shown in \textit{Figure \ref{fig:1}} \cite{emdat}. Therefore, it is increasingly vital to use data, information, and various models to predict potential public emergencies that jeopardize public safety and well-being. Predicting the cascade of information in the event deduction process under public emergencies assists governments, organizations, and individuals in taking proactive measures to mitigate the impact of emergencies and minimize damage.

Public emergencies are classified into different categories. The most common categories of public emergencies include \textbf{(1) Natural disasters}, \textbf{(2) Accident disasters}. 
\subsubsection{Natural Disaster}
Natural disasters pose a significant and recurring threat to human populations and infrastructure worldwide. These disasters range from sudden, catastrophic events such as volcanic eruptions, earthquakes, floods, hurricanes, and storms to slower, gradual processes like land desertification, soil erosion, environmental degradation, and major infectious disease outbreaks, mass unexplained diseases. For instance, the COVID-19 pandemic, which began in December 2019, has had a devastating impact globally, with over 6.5 million deaths reported to the World Health Organization as of 21 October 2022, and more than 623 million confirmed cases. In addition to the immediate impact of natural disasters, derivative disasters often follow, which exacerbate the initial damage caused by the event. Therefore, it is crucial to recognize the complex and interconnected nature of natural disasters, as one disaster trigger or contribute to several others.

\subsubsection{Accident Disaster}
Accident Disaster is a term that refers to unexpected events occurring during people's production or daily life, typically caused by human activities, and resulting in a substantial number of casualties, economic losses, or environmental pollution. Examples of Accident Disasters include fires, explosions, and toxic leaks during industrial production; collapses, gas and coal dust explosions, and flooding during mining; and shipwrecks, major traffic accidents, and aircraft accidents during river-sea transportation. Accident Disasters are a type of public emergency with universality, randomness, inevitability, causal correlation, mutation, latency, and harmfulness. Disasters significantly impact people's lives and production, resulting in severe social consequences.

\subsection{Information Cascade Prediction under Public Emergencies}
Public emergencies often create uncertainty and panic, leading individuals to be influenced by the behavior of those around them. The phenomenon where people rely on the actions of others rather than their own judgment or the development of events to form impressions, often due to external factors, is referred to as information cascades. The cascading information predicts events, anticipate their progress and human behavior, and predicts their impact on other events or sub-events. Information Cascade Prediction involves analyzing and predicting how information spreads and affects decision-making during crises or emergencies. Researchers and practitioners utilize diverse algorithms to predict and manage information cascades during public emergencies.

Decision Trees are a type of supervised learning algorithm that works by recursively partitioning the data based on various features or parameters until a stopping criterion is met. Bayesian Networks are probabilistic graphical models that represent and reason about uncertain relationships between various factors. They are used to predict the likelihood of cascading events or hazards by incorporating various factors and their probabilistic relationships.

Matrix Factorization algorithms, such as Singular Value Decomposition and Non-negative Matrix Factorization, are effective tools for analyzing and predicting the spread of a pandemic or other public health crises. Support Vector Machines are commonly used to predict the path and intensity of natural disasters like hurricanes or typhoons, taking factors like wind speed and atmospheric pressure into account. Similarly, Random Forests predict the likelihood of cascading failures during a cyber attack on critical infrastructure by analyzing various factors related to the attack and the resilience of the infrastructure.

Neural Networks, including Long Short-Term Memory Networks (LSTMs) and Graph Neural Networks (GNNs), are capable of learning complex patterns and relationships in data, which leads to improved prediction accuracy. LSTMs predict future actions based on past behavior patterns, while GNNs incorporate social network data to predict the spread of behavior patterns. Other algorithms like Deep Learning models, Generative Adversarial Networks (GANs), Autoencoders, Attention mechanisms, and Transformer models are also valuable for learning complex patterns and relationships in data and improving prediction accuracy. Each of these algorithms is used to predict various events and hazards using different types of data and relationships between them.

In summary, the choice of algorithm for information cascade prediction in public emergencies depends on the specific task and the level of complexity required. Each algorithm has its strengths and limitations, and the appropriate choice depends on the nature of the problem and the available data.

\subsection{Summary of Existing Surveys}

It is well-known that data-driven approaches are commonly applied at 4 stages of public emergencies, including mitigation, preparedness, response, and recovery. However, predicting public emergencies is challenging due to the uncertain characteristics of events. In recent years, the rapid development of machine learning methods has led to significant progress in the development and application of information cascade prediction technology. The various data-driven methods are applied at each stage of a public emergency as shown in \textit{Table \ref{tab1}}.

\begin{table}[htbp]
\small 
\vspace{-0.2cm}  
\setlength{\abovecaptionskip}{0.2cm}   
\caption{Compare With Existing Survey}
\label{tab1}
\centering
\begin{tabular}{c|c|c|ccc|cccc}
\hline
Surveys& Regular Events&\multicolumn{7}{c}{\textbf{Public Emergency}}\\
\hline

&& Model &\multicolumn{3}{c|}{Prediction}&\multicolumn{3}{c}{Multi—stage Application} \\
\hline

&&&Direct Risk&\textbf{Sequence Risk}& \textbf{Cascading Risk}& Before&During& After\\
\hline

\rowcolor{gray!20} \cite{63}  &\checkmark&  & &&&&&\\

\cite{1-70} &\checkmark& & & &&&&\\

\rowcolor{gray!20} \cite{144} &\checkmark& &&&&&&\\

\cite{185}&\checkmark& & &&&&&\\
 
\rowcolor{gray!20} \cite{181}&&\checkmark &\checkmark & &&&&\\

\cite{222}& &\checkmark &\checkmark &&&&\checkmark&\\

\rowcolor{gray!20}  \cite{1-47} &&  & \checkmark&&&&\checkmark& \\

\cite{1-29} &&  & &\checkmark&&&&\\

\rowcolor{gray!20} \textbf{Our Work} &&\checkmark &\checkmark &\checkmark&\checkmark&\checkmark&\checkmark&\checkmark\\
\hline

\end{tabular}
\vspace{-0.5cm}  
\end{table}

In recent years, several studies\cite{1-29, 1-47, 1-70, 63, 144, 181, 185, 222} have reviewed prediction algorithms for public emergencies. However, none of these studies have presented a comprehensive overview of methods that effectively incorporate information cascade prediction modeling, prediction, and application. Gao and Duan et al. \cite{1-70, 63} have emphasized event modeling and simple risk forecasting using 1D structured data. Nuno, Alexandru, and Gabor et al. \cite{144, 181, 185, 222} have focused on summarizing information cascade prediction methods under public emergencies and analyzing the application of various feature engineering methods. Ali and Robert et al. \cite{1-29, 1-47} have considered information cascade prediction for emergency public events and summarized application methods for different stages of emergency events. Although these reviews have approached public emergency prediction methods from different perspectives, they fail to consider the influence of each stage of a public emergency and the event evolution caused by the information cascade, as well as time, place, semantics, deduction, and other features.

To address this gap, our work focuses on developing systematic and standardized summary methods to evaluate various information cascade prediction methods. These efforts help identify bottlenecks, pitfalls, open problems, and potentially fruitful future research directions.

\begin{figure}
\setlength{\belowcaptionskip}{-0.5cm}   
\centering
\includegraphics[width=0.9\linewidth]{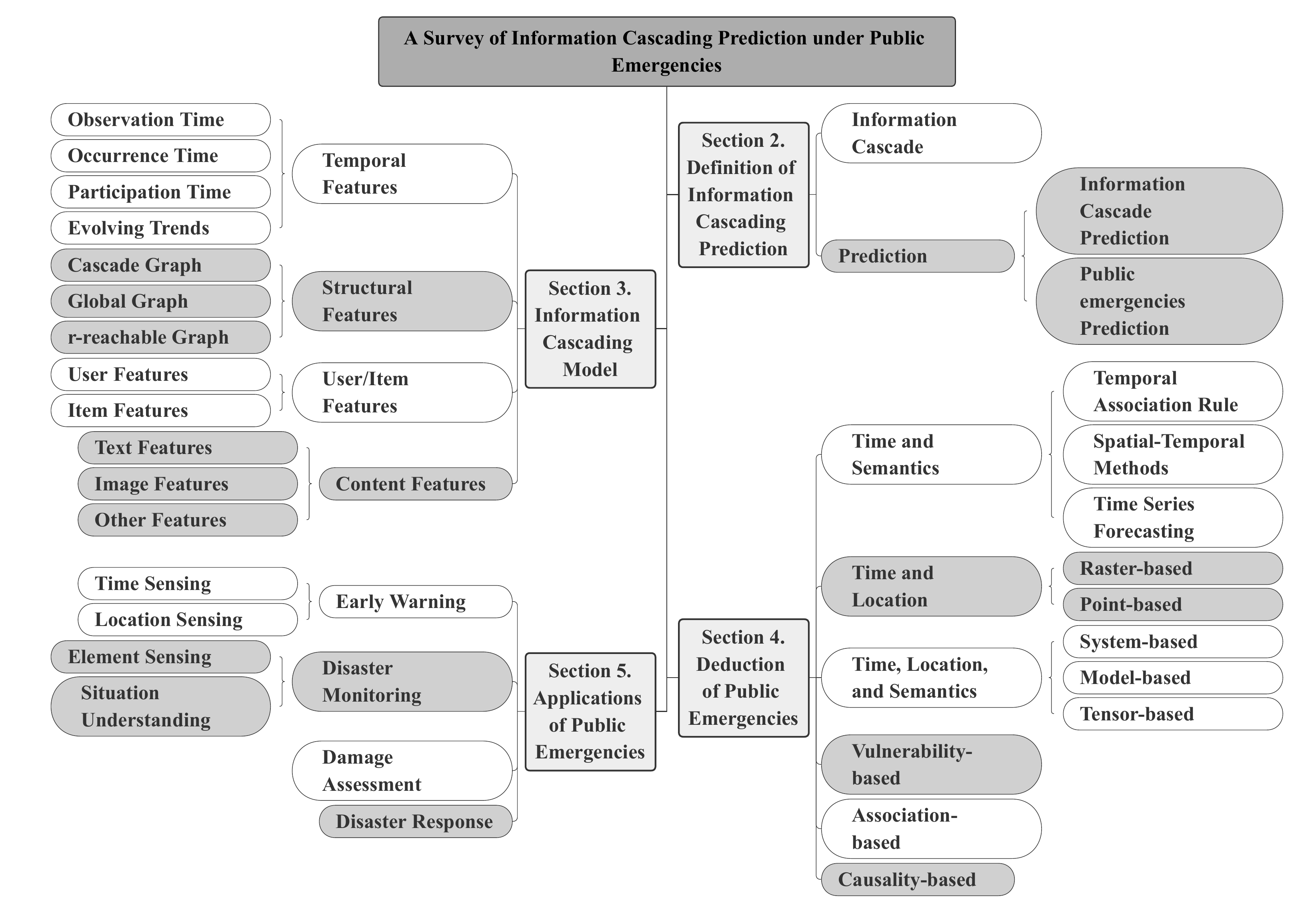}
\caption{The Overall Structure Of the Survey} \label{fig1}
\vspace{-0.4cm}  
\end{figure}
\subsection{Contributions of This Paper}

The motivation behind conducting a comprehensive review of the field of information cascade prediction is to gain a deeper understanding of the methods and techniques used to predict information cascades and analyze the diffusion of individuals. By examining current research in these areas, we identify gaps in knowledge, and potential areas for improvement, and determine the most effective approaches and people flow at different levels of interaction. Different from previous surveys, our work focuses on information cascade prediction methods under emergencies. This review aims to provide a comprehensive overview of the field, highlighting key findings and advancements in information cascade prediction and people flow analysis. Compared to previous investigations, this paper makes several unique contributions.

\begin{itemize}
\item[$\bullet$] Firstly, it systematically classifies and summarizes existing technologies, categorizing them according to event aspects, problem formulation, and corresponding techniques. This taxonomy helps domain experts find the most useful techniques for their target problem settings. 
\item[$\bullet$] Secondly, the paper conducts a finer-grained analysis of existing information cascade prediction methods for emergencies. It also considers the information cascades and prevalence prediction of emergencies, explaining and analyzing their characteristics and methods.
\item[$\bullet$] Thirdly, the paper provides a comprehensive and up-to-date literature review of information cascade prediction methods, covering both the macro-model of collective behavior and the micro-model of individual user responses. These approaches incorporate recent advances in modeling and predicting information popularity.
\item[$\bullet$] Lastly, the paper discusses the research status and future trends in the field, outlining the overall situation and shape of the current research front. It concludes with new insights into bottlenecks, pitfalls, and open issues, and discusses possible future directions.
\end{itemize} 

\subsection{Roadmap}
The remainder of this paper is organized as shown in Figure \ref{fig1}. Section 2 presents Problem Formulation and Performance Evaluation. Section 3 introduces the Information Cascade Model under Public Emergencies. Section 4 introduces the Deduction of Public Emergencies by joint information cascade prediction, including joint time and semantics (Section 4.1), joint time and location (Section 4.2), joint time, location, and semantics (Section 4.3), Vulnerability Prediction (Section 4.4), Association-based Impact Prediction (Section 4.5), and Causality-based Prediction (Section 4.6). Section 5 introduces the Application to Public Emergencies and points out directions for future work, and the survey concludes in Section 5.
\section{Definition of Information Cascade Prediction}
Due to the particularity of public emergencies, there are usually information cascade items between public emergencies. This information is usually heterogeneous, sparse, and biased data due to different acquisition channels and methods. This section introduces the formulas and classifications used in the information cascade prediction process of public emergencies: \textbf{(1) Information Cascade}; \textbf{(2) Prediction}.

\subsection{Information Cascade}
Information-level correlation in public emergencies often refers to the information correlation between sub-events or between events and events in public emergencies. The information cascade prediction of public emergencies is essentially a regression problem. The development trend of public emergencies is represented by information diffusion.

\textbf{Definition 2.1 (Information Cascade)} Assume that there are $M$ information items ${I_1, I_2,..., I_M}$ for event $y$. $I_i$ represents the cascading degree of events at time $t$,  that is, the probability that event $y$ may affect subsequent events. and information cascade prediction aims to predict the occurrence probability $P_i(t)$ at time $t$ in the future.

Information Cascade Prediction aims to predict whether a cascading event occurs given a predefined absolute/relative threshold. For example, whether the event occurs \cite{149, 155}, and whether the public emergency causes other cascading events in the future \cite{39, 46}, in addition, the classification task is also used to predict the probability of the occurrence of the emergency. It is possible to fall into which interval \cite{61, 81, 135, 134}, which is the classic multi-class classification task. Since the classification task is similar to the low-level version of the regression task, it has lower dimensions in the data parameters, making it easier to achieve good results. Developing regression strategies relies on expert knowledge and requires a fine-grained scope to analyze which factors affect predicting cascading events in public emergencies. Due to the complexity of cascades among public emergencies, accurate regression predictions usually require more information about events as well as cascades \cite{60}. Furthermore, it suffers from undesirable problems such as overfitting, inductive bias, and accumulation of prediction errors \cite{214}.

\subsection{Prediction} 
Public emergencies contain three important information items: time, location, and semantic \cite{1-187}. Let sudden public emergencies be represented by $y = (t,l,s)$, $t \in T$ represents the time factor, $l \in L$ represents the positioning element, and $s \in S$ represents the semantic element. $T$, $L$, and S represent the time, location, and semantic domains. Among them, the semantic domain contains any type of semantic feature useful in detailing the semantics of various aspects of an event, including its actors, objects, actions, size, textual description, and other analytical information. 

In the context of predicting information cascades during public emergencies, it is possible to independently predict the time, location, and semantics of disasters while predicting other factors of the emergency through different elements of the three. Alternatively, the cascading nature of public emergencies is used to predict the occurrence of public emergencies, with precursor events and subsequent events being jointly predicted. The emergency information is represented by $X \subseteq T*L*F$, and F represents the set of semantic information. The current time is expressed as $t$, past time and future time are expressed as $T^-\equiv \{ t|t_{now} \le t,t \in T\}$ and $T^+ \equiv \{ t |t > t_{now},t \in T \}$, the public emergencies formula is as follows:

\textbf{Definition 2.2 (Public Emergency Prediction)} Let the set of precursory events be expressed as $X\subseteq T^-*L*S$, and the data of subsequent precursory events be expressed as $Y_0\subseteq T^-*L*S$. Based on the information of precursor events and subsequent events, the information of public emergencies is expressed as $\hat {Y} \subseteq T^-*L*S$. So that each future emergency is expressed as $\hat{y} = (t,l,s)\in \hat{Y} $ where $t>t_{now }$.
\section{Information Cascade Model}
The Information Cascade Model is a predictive model that utilizes feature extraction techniques to predict the spread and severity of public emergencies. This model is designed to identify relevant factors that influence the occurrence and spread of emergencies and to incorporate these factors into a machine learning framework for predictive modeling. The Information Cascade Model incorporates various types of features, including \textit{Temporal Features}, \textit{Structural Features}, \textit{User/Ttem Features}, and \textit{Content Features}, to capture different aspects of public emergency occurrences. Temporal features capture changes in emergency occurrences over time, while structural features capture relationships between emergency response networks. User/item features capture relevant demographic and socioeconomic factors, while content features capture emergency data.

\subsection{Temporal Features}
Temporal features play a crucial role in predicting the spread of information items or cascades during public emergencies. In this section, we discuss the importance of different types of temporal features in feature engineering for prediction. Specifically, we focus on four subtopics: \textbf{Observation Time, Occurrence Time, Participation Time, and Evolving Trends} as shown in \textit{Table \ref{tab:temporal-features}}.
\begin{table}[htbp]
\small
\vspace{-0.3cm}
\setlength{\abovecaptionskip}{0.2cm}   
\caption{Temporal and Structural Features of Public Emergencies}
\label{tab:temporal-features}
\centering
\begin{tabular}{m{0.15\linewidth}<{\centering}|m{0.35\linewidth}|m{0.22\linewidth}}
\hline
\textbf{Feature} & \textbf{Definition} & \textbf{References} \\
\hline
Observation Time & The observation time of early features and precursor events& \cite{224,39,181} \\
\hline
Occurrence Time & The time at which a public emergency begins& \cite{109,137,181,186,61,210} \\
\hline
Participation Time & The time at which an individual participates in a public emergency& \cite{17, 190,155} \\
\hline
Evolving Trends & The Developments in public opinion, behavior, or Events over Time & \cite{74,116,224} \\
\hline
Cascade Graph & A directed graph representing the spread of information from an initial node& \cite{1-241, 1-14, 1-58, 1-240} \\
\hline
Global Graph &A fundamental representation of the relationships between nodes & \cite{1-39, 1-60,1-61, 1-66,1-49, 1-222,1-83,1-84,1-121,1-147, 1-164,1-165,1-203}\\
\hline
r-reachable Graph & A sub-graph extracted from the global graph based on cascade nodes& \cite{1-61, 1-135} \\
\hline
\end{tabular}
\vspace{-0.6cm}  
\end{table}
\subsubsection{Observation Time} 
Temporal features play a critical role in predicting the popularity of public emergencies. These features are typically extracted based on the peeking strategy, which involves observing a small number of early features and their action time to obtain a sequence of timestamps that are utilized for feature selection. However, the length of the time series is highly irregular, and directly utilizing timestamps as a feature is often ineffective in practice. To address this, transformations are often applied in advance \cite{224}. For example, the period may be divided into evenly distributed intervals, and the cumulative or incremental popularity may be calculated, or a fixed number of early features may be observed \cite{39}. To predict the popularity $P_i(t_p)$ at prediction time $t_p$ based on the information observed at time $t_o$, previous studies have analyzed the relationships between the log-transformed popularity $P_i(t_p)$ and $P_i(t_o)$. One such study \cite{181} found a high correlation between early-stage and future popularity and used a simple linear prediction model that takes the early observed popularity as input to predict future popularity.

\subsubsection{Occurrence Time}
Temporal features are crucial in predicting the popularity of public emergencies, with occurrence time ($t_0$) being a particularly important factor. Previous studies \cite{109,137,181,186,190} have highlighted the strong relationship between the popularity of public emergencies and the time they occurred, with items posted during the day generally more popular than those posted at midnight. To overcome the effect of user activity periodicity, researchers have proposed various solutions. For instance,  local models were designed with each model trained on samples published at a specific hour during the day \cite{155}. Tweet time was utilized to eliminate the imbalanced diurnal effect of user activities \cite{61}. Other temporal factors, such as source time \cite{190} and activeness variability \cite{210}, have also been employed to improve the robustness of prediction models. However, to improve the accuracy of prediction models, other temporal features, such as the age of the tweet and the time since the event started, should also be considered. For example, a study \cite{211} showed that the age of a tweet has a negative correlation with its popularity, indicating that tweets posted early during an event are more likely to be popular than those posted later. Therefore, future research should explore the use of various temporal features to improve the accuracy of prediction models for public emergencies.

\subsubsection{Participation Time}
The first arrival time $t_1$ is another crucial temporal feature to consider in forecasting public emergencies. Moreover, various more sophisticated temporal features have been proposed in the literature, such as mean arrival time $1/M {\textstyle \sum_{M}^{j=1}} t_j$, mean reaction time $1/M {\textstyle \sum_{M}^{j=1}} (t_j-t_{j-1})$, change rate, dormant period, and peek fraction. For instance, research has shown that human reaction time typically follows a log-normal distribution, such as people's reactions to calls, emails, and social networks \cite{17}. These temporal features provide useful insights for predicting the spread and impact of public emergencies.

\subsubsection{Evolving Trends}
Analyzing the changing trends of public emergencies has been shown to provide valuable insights for forecasting their severity and impact. Previous studies \cite{74, 116, 224} have demonstrated that identifying the temporal patterns of these events helps predict their popularity and impact on society. These patterns are categorized into various types, such as steadily increasing or rapidly fluctuating, depending on the clustering algorithm.

\subsection{Structural Features}
The spread of public emergencies, such as pandemics, has garnered significant attention from scholars across various fields. In particular, the structure of cascades, also known as information diffusion, has been the subject of extensive research. It explains how information and contagions spread through populations \cite{14, 39, 58, 60, 238, 241}. Scholars have approached the modeling of cascades in different ways, which are categorized into three types. The first type is \textit{Cascade Graph} \cite{1-241, 1-14, 1-58, 1-240}, which involves the analysis of cascade graphs to understand how information spreads among participants in a network. The second type is \textit{Global Graph} \cite{1-39, 1-60,1-61, 1-66,1-49, 1-222,1-83,1-84,1-121,1-147, 1-164,1-165,1-203}, which considers both participants and non-participants in the network, considering the external factors that influence the spread of the emergency. Finally, the third type is \textit{r-reachable modeling} \cite{1-61, 1-135}, which seeks to strike a balance between the first two types, by expanding the cascade graph to include non-participants within the global graph as shown in \textit{Table \ref{tab:temporal-features}}. 

\subsubsection{Cascade Graph}
A cascade graph is constructed based on its participants and their interactions: 

\textbf{Definition3.1 (Cascade Graph)} Given an information item $I_i$ and the corresponding cascade $C_i$, a cascade graph is defined as $G_c$ $=\{V_c, E_c\}$, where nodes $Vc$ $=\{u_0,u_1,...,u_N\}$are participants of cascade $C_i$, and matrix $E_c$ $\subset V_c * V_c$ contains a set of edges representing immediate relationships between $V_c$ in a cascade.

During public emergencies, the spread of information is crucial to inform and mobilize individuals to take necessary actions. The study of information diffusion through cascades has been widely researched, with a cascade graph being a fundamental concept in the field \cite{1-58, 1-240}. A cascade graph, denoted as $G_c ={V_c, E_c}$, consists of a set of nodes, $V_c$, representing participants of the cascade, and a set of edges, $E_c$, representing the immediate relationships between the participants as shown in \textit{Table \ref{tab:global-graphs}}. Cascade graphs play a vital role in characterizing the process of information diffusion of a particular item, such as the spreading directions and graph topology. For instance, one study examined the correlation between cascade popularity and two structural features, namely edge density, and depth, among early participants in microblogging networks \cite{1-14}. However, it is essential to note that the topological structure of cascade graphs varies significantly, even when the number of nodes is in the same \cite{1-240}. The depth, structural virality, and other structural measurements, such as node degree or PageRank, may not accurately predict whether an information item would be popular. As cascades grow over time, the initial structural features may become less important in determining their popularity \cite{1-241}. Therefore, further research is needed to explore the complex mechanisms of information diffusion during public emergencies and develop effective information dissemination strategies.

\subsubsection{Global Graph}
\begin{table}[htbp]
\vspace{-0.4cm}  
\setlength{\abovecaptionskip}{0.2cm}   
\small
\centering
\caption{Structural Features  of Global Graphs}
\label{tab:global-graphs}
\begin{tabular}{m{0.21\linewidth}|m{0.3\linewidth}|m{0.15\linewidth}|m{0.15\linewidth}}
\hline
\textbf{Structural features} & \textbf{Definition} & \textbf{Examples} & \textbf{References} \\ \hline
Follower/Followee Graph & A graph where nodes represent users and edges represent the relationships& Twitter, Instagram, Facebook & \cite{1-83,1-203}\\ \hline
Hidden Friend Graph &A graph representing less obvious or non-publicly visible relationships & Twitter, Facebook & \cite{1-39, 1-60, 1-66, 1-147, 1-164}  \\ \hline
Co-Participation Graph & A graph that captures the relationships between users based on their participation& Digg, Reddit & \cite{1-84,1-85,1-89} \\ \hline
Interaction Global Graph & A graph that captures interactions between users & Twitter, YouTube, Instagram & \cite{1-60, 1-61, 1-222, 1-121, 1-165} \\ \hline
\end{tabular}
\vspace{-0.5cm}  
\end{table}

Understanding the spread of information across different types of networks is crucial for effective crisis communication as shown in \textit{Table \ref{tab:global-graphs}}. While cascade graphs provide insight into the local spread of information, exploring global graphs is also important \cite{1-39, 1-60, 1-66, 1-147, 1-164}. These interaction global graphs are useful in various prediction tasks, particularly when the explicit social graph is not available, and historical behaviors serve as a suitable representation of the actual diffusion of information \cite{1-203}. Therefore, understanding the characteristics of different types of global graphs can aid in designing effective crisis communication strategies during public emergencies.

\textbf{Definition 3.2 (Global Graph)} A global graph $G_g=(V_g, E_g)$ is a fundamental representation of the relationships between nodes, where $V_g$ is a set of nodes, and $E_g \subset V_g * V_g$ is a set of edges that connect nodes based on their relationships. The global graph is further defined based on additional characteristics such as edge direction, edge weight, node/edge attributes, and node/edge features.

By providing a macro perspective of the relationships between nodes, a global graph offers a valuable tool to analyze the spread of information to individuals and communities. In contrast to the cascade graph, which shows the local spread patterns for information cascade, the global graph describes the relationships between users and potential routes for diffusion \cite{iot_9}. In social networking platforms, the discovery and dissemination of information items primarily occur through the users’ social networks. This highlights the importance of understanding the relationships between users in the global graph to effectively analyze and mitigate public emergencies such as the spread of misinformation, epidemics, and natural disasters. \cite{1-181}

\subsubsection{r-reachable Graph}
Drawing upon the concepts of cascade and global graphs, a sub-graph extracted from the global graph, termed an r-reachable graph, is defined as follows:

\textbf{Definition 3.3 (r-reachable Graph)} Consider a global graph $G_g$ and its cascade sub-graph $G_c$. An r-reachable graph of $G_c$, denoted by $G_c^r ={V_c^r, E_c^r}$, is a graph comprising the nodes in $V_c$ and those in $V_g$ that are within $r$-hops of nodes in $V_c$. For instance, if $r$ = 1, then $V_c^r$ contains all nodes in $V_c$ and their immediate neighbors.

The r-reachable graph has been employed in several studies to aid in predicting the popularity of cascades in social media \cite{1-135}. A global graph $G_g$ and several 1-reachable graphs $G_c^1$ were constructed from retweet cascades in Weibo, and structural features were extracted from $G_g$ and $G_c^1$ to forecast tweet popularity \cite{1-61}. A comparative analysis of several content and structural features revealed that the size of the cascade graph and the 1-reachable graph are the two most predictive of 53 features for predicting the popularity of Twitter hashtags \cite{1-135}. Nevertheless, constructing and computing an r-reachable graph are computationally intensive; for instance, the 2-reachable graph of a cascade graph with dozens of nodes may contain tens of thousands of nodes \cite{iot_10}.

\subsubsection{Summary}
Understanding the early structure and patterns of information diffusion is crucial for effective crisis management and response \cite{1-39}. As the spread of misinformation and rumors exacerbate the impact of a crisis, identifying influential users and predicting the popularity of information items are essential tasks \cite{1-103, 1-115, 1-118}. Previous studies have shown that the structure of cascade graphs is not a reliable predictor of final item popularity, and alternative approaches such as identifying influential users must be considered. Furthermore, the types of items and their contents also affect information diffusion behavior during crises \cite{1-67}. Therefore, analyzing the structural patterns of information diffusion and identifying influential users in the context of public emergencies provide valuable insights for crisis management and response \cite{1-121, 1-241}.

\subsection{User/Item Features}
The unprecedented occurrence of public emergencies has heightened the need for accurate and timely predictions of their impact on society. However, obtaining early observations and monitoring temporal and structural features is impractical in the context of rapidly unfolding events. As such, some works try to examine the features inherent to users and information items, which possess distinct characteristics and inherent appeal that render them valuable in forecasting popularity before dissemination.

\subsubsection{User Features}
During public emergencies, such as natural disasters or pandemics, information dissemination becomes crucial in managing and mitigating the situation. User behaviors play a critical role in this process, as they determine the speed and reach of information spread. However, it is important to note that large cascades of information are not produced by influential users such as celebrities and news organizations but also originate from normal users. Therefore, it is essential to study and analyze large cascades produced by many types of users during public emergencies. Various other features have been extensively explored and studied for analyzing and predicting the \textit{popularity} of information items during public emergencies. These features include \textit{user profiles, historical behaviors, user interests, collectivity, similarity, activity/passivity, discoveries, affinities, and responsiveness }\cite{1-81, 1-140, 1-130,  1-175, 1-225, 1-227, 1-236}. Understanding and utilizing these features predict and manage the spread of information during public emergencies, ultimately leading to more effective communication and response.

\subsubsection{Item Features}

\begin{figure}
\vspace{-0.2cm}  
\setlength{\abovecaptionskip}{-0.2cm}   
\setlength{\belowcaptionskip}{-0.5cm}   
\centering
\includegraphics[width=0.75\linewidth]{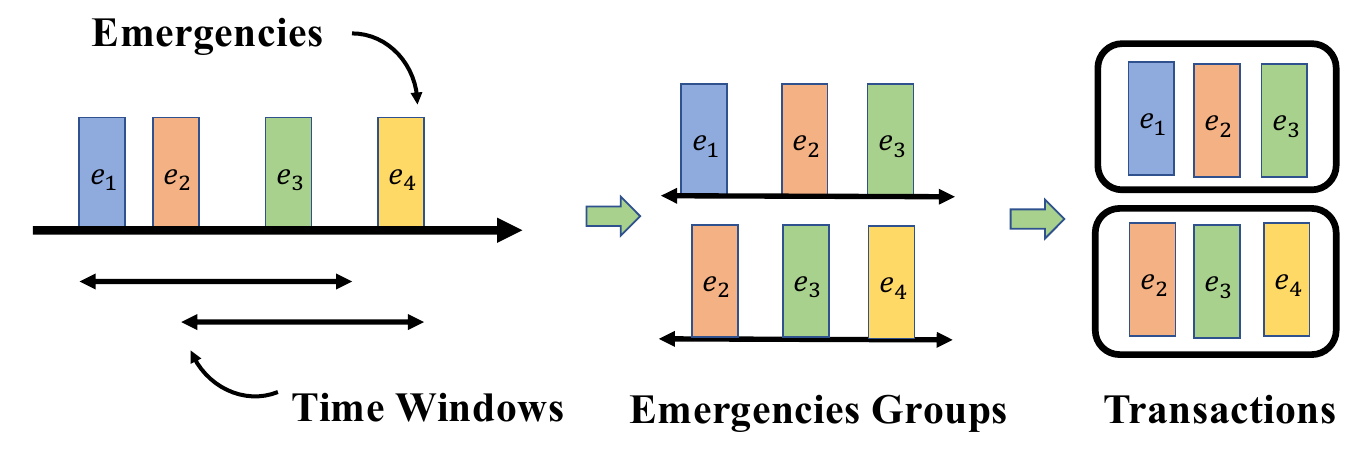}
\caption{From Emergency Time-line to Transactions} \label{item}
\end{figure}

Understanding the impact of project characteristics on information dissemination has a significant impact on the dissemination of key messages as shown in \textit{Figure \ref{item}}. Previous literature explored various item characteristics and their impact on information dissemination. For example, one study analyzed how user interfaces on social media platforms affect item visibility \cite{1-114}. Furthermore, using entropy computed across information categories and topics provides insight into the diversity of information and its impact on dissemination \cite{1-84}. Using popularity variability to analyze how the popularity of different types of items changes over time helps identify patterns and trends in information dissemination during public emergencies \cite{1-210}.

\subsubsection{Summary}
During public emergencies, understanding the factors that contribute to information diffusion is critical. User and item characteristics play a significant role, with some features being more easily understood than others. However, factors like user influence, preferences, and similarities require more sophisticated algorithms and calculations. Advanced techniques enable better predictions and are particularly useful in such situations where timely and accurate information dissemination is crucial \cite{1-107, 1-19, 1-38}.

\subsection{Content Features}
From the perspective of predicting public emergencies, content is widely recognized as an essential driving force and one of the key factors that contribute to the success of any emergency response effort. For instance, breaking news, rumors, and fake news, as well as hot spots and controversial or peculiar topics, disinformation, and misinformation, tend to attract considerably more attention than regular content.

\subsubsection{Text Content}
\begin{figure}[htbp]
\vspace{-0.4cm}  
\setlength{\abovecaptionskip}{-0.2cm}   
\setlength{\belowcaptionskip}{-0.6cm}   
\subfigure[Text Content]{   
\includegraphics[width=0.49\textwidth]{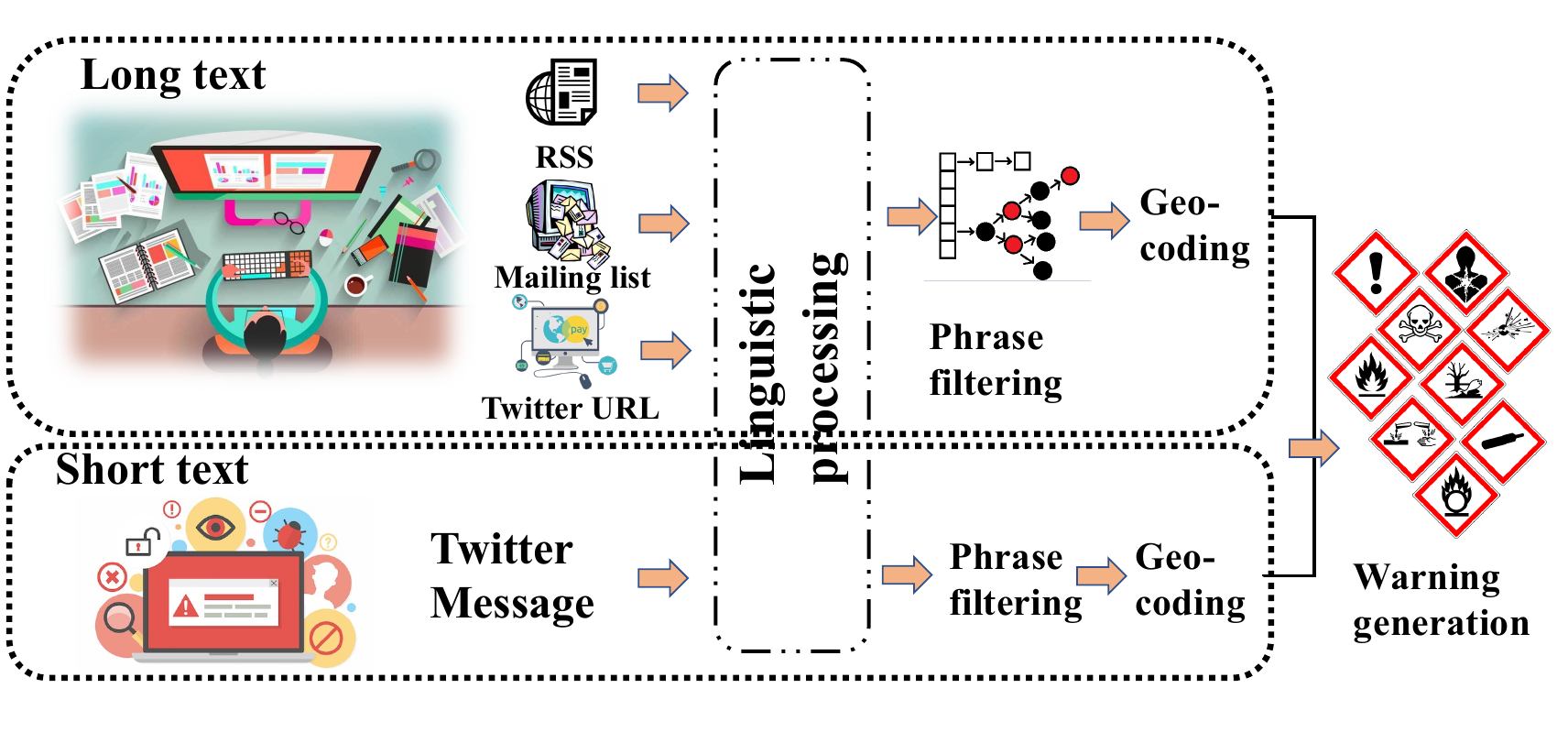}
\label{text}
}
\subfigure[Image Content]{ 
\includegraphics[width=0.47\textwidth]{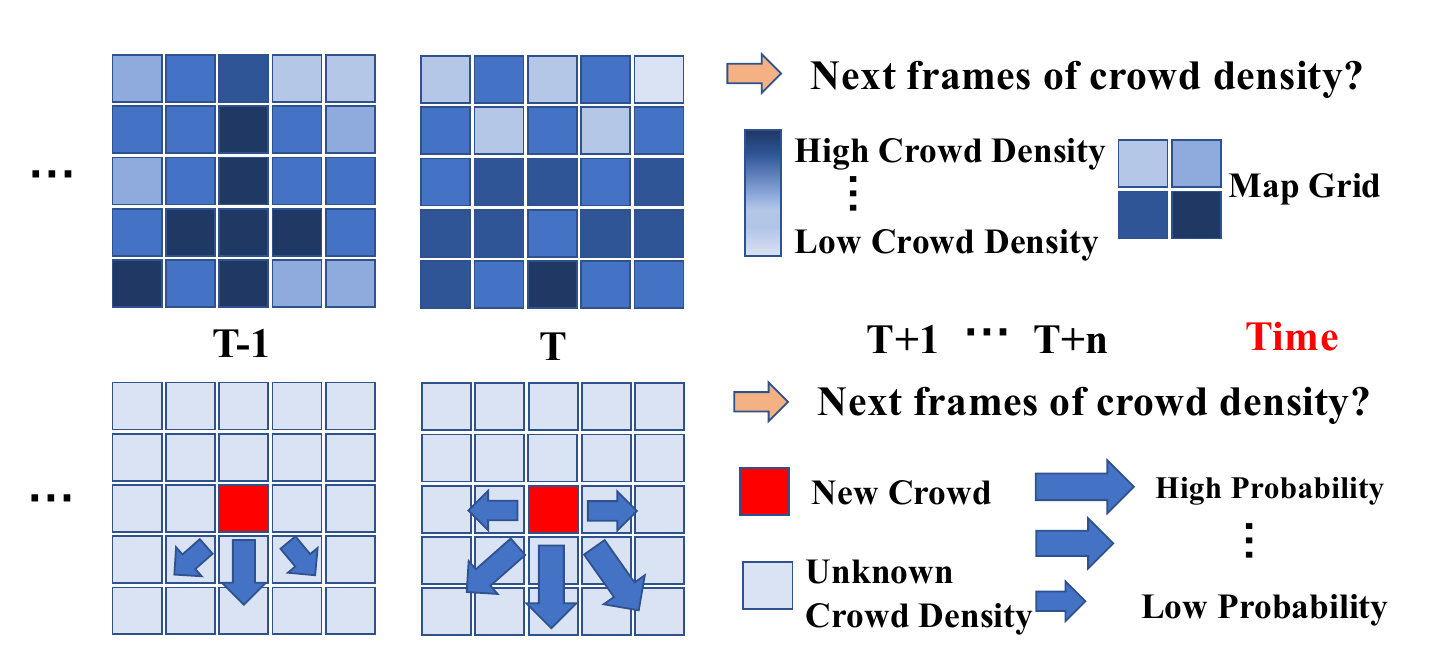}
\label{image}
}
\caption{Content Features}    
\label{Content Features}    
\end{figure}

From the perspective of predicting public emergencies, text feature is a crucial component of existing prediction models. Textual information is pervasive in articles, microblogs, image/audio/video captions/descriptions, and even retrieved from multimedia sources. Researchers have employed various language models such as Term \textbf{Frequency-Inverse Document Frequency (TF-IDF)} and \textbf{Latent Dirichlet Allocation (LDA)} \cite{1-21}, along with typical Machine Learning models like naive Bayes, SVM, and linear regression to predict item popularity. TF-IDF and LDA are used to learn the topic distributions of tweets \cite{1-81}. TF-IDF is utilized to estimate the importance of keywords in user tweets and to calculate the mutual correlation between a user's historical content and a specific item to measure their likelihood of adopting that item as shown in \textit{Figure \ref{text}} \cite{1-227}. Authors analyze various semantic and statistical content features of Digg comments to identify the characteristics of content that people prefer to retweet \cite{1-90}.

\subsubsection{Image Features}

The retrieval and analysis of image features require techniques from computer vision learning, distinct from text-based methods. Basic features of an image, such as \textit{size, date, orientation, dominant color, resolution, location, caption, and tag}. For example, during the COVID-19 pandemic, computer vision techniques were utilized to analyze images from social media platforms and identify violations of social distancing protocols, such as large gatherings or people not wearing masks \cite{1-132}. Lu et al. \cite{1-138} provide an analysis of these attributes, while Michael et al. \cite{1-91} study the correlations between Flickr image features and their normalized popularity as shown in \textit{Figure \ref{image}}. Content features of images, categorized as \textit{simple human-interpretable, low-, and high-level image features}, contribute significantly to improving prediction performance. During natural disasters like hurricanes or earthquakes, computer vision techniques are utilized to analyze satellite images and identify areas that have been affected and need assistance. Li et al. \cite{1-132} extract texture and color features and use VGG19 \cite{1-176} to extract deep features. During a pandemic, computer vision techniques are utilized to analyze chest X-rays and identify patterns that could indicate COVID-19 infection.

\subsubsection{Summary}

The challenge of predicting item features based solely on content features is still prevalent, despite previous studies exploring the linguistic and visual characteristics of items. This difficulty arises because even items with the same content have varying features, making it challenging to distinguish the effects of descriptive factors from intrinsic content. This unpredictability is particularly relevant in predicting the spread of misinformation during a public health crisis.

\section{Deduction of Public Emergencies}

Predicting the joint information of public emergencies is a crucial task in disaster management. Research in this area are categorized into six types: \textbf{(1)} Joint Time and Semantic Prediction; \textbf{(2)} Joint Time and Location Prediction; \textbf{(3)} Joint Time, Location, and Semantic Prediction; \textbf{(4)} Vulnerability Prediction; \textbf{(5)} Association-based Impact Prediction; \textbf{(6)} Causality-based Prediction.

\subsection{Time and Semantics}
In the prediction of public emergencies, various methods have been developed to jointly predict the joint time and semantics. Existing work in this area is broadly categorized into the following three types: \textit{Temporal Association Rule; Spatial-Temporal Methods; Time Series Forecasting-based Methods.}

\subsubsection{Temporal Association Rule}

Temporal association rules are used to embed additional temporal information into the occurrence of events and to redefine the meaning of co-occurrence and association with temporal constraints as shown in Table \ref{table:references114}. In that case, they define the \textit{Left-Hand Side (LHS)} of the association rule as a tuple $(E_L,\tau)$, where $\tau$ represents the time window before the target flood emergency occurrence in the \textit{Right-Hand Side (RHS)} predefined by the user. The events that occur within this time window before the flood emergency satisfy the LHS. However, defining the time window beforehand is challenging and may not suit different target events. To overcome this, researchers have proposed a way to automatically identify information on a continuous time interval from the data \cite{1-225}. Here, the transaction comprises items and continuous time duration information as shown in \textit{figure \ref{fig:5}}. LHS is a set of items (e.g., previous emergencies), and RHS is a tuple $(E_R,[t_1,t_2])$ consisting of a future emergency's semantic representation and its time interval of occurrence. To learn the time interval in RHS, two different methods have been proposed: confidence-interval-based and minimal temporal region selection \cite{1-225}.

\begin{table}[htbp]
\setlength{\abovecaptionskip}{0.2cm}   
\small
\centering
\caption{Deduction of Public Emergencies}
\label{table:references114}
\begin{tabular}{m{0.15\linewidth}|m{0.2\linewidth}|m{0.3\linewidth}|m{0.2\linewidth}}
\hline
\textbf{Category}&\textbf{Direction}& \textbf{Method}& \textbf{References}\\\hline
\multirow{7}{*}{Risk Occurrence}&\multirow{5}{*}{Binary Classification} & Simple Threshold-based & \cite{1-212,1-217,1-214,1-215} \\ \cline{3-4} 
&& Logistic regression &  \cite{1-209,1-208,1-211}\\ \cline{3-4}
&& Support Vector Machines & \cite{1-85,1-86,1-84}\\ \cline{3-4}
&& Neural Networks& \cite{1-117,1-116,1-118} \\ \cline{3-4}
&& Decision trees & \cite{1-47,1-48,1-49}\\ \cline{2-4}
&Anomaly detection& One-classification, hypotheses testing & \cite{1-83, 1-160}\\ \cline{2-4}
&Regression & Auto-regression, linear regression,   ordinal regression & \cite{1-60, 1-68, 1-194}\\ \hline

\multirow{5}{*}{Discrete-time}&Time windows & Auto-regression, ARIMA, ordinal regression or classification & \cite{1-126,1-131,1-162} \\ \cline{2-4}
&Time scales & Regression or classification & \cite{1-172,1-173,1-175} \\ \cline{2-4}
&\multirow{2}{*}{Time series} & Autoregressive, burstiness detection, change detection & \cite{1-20,1-24,1-92} \\ \cline{3-4} 
&& Learning event characterization& \cite{1-107,1-106,1-108,1-109} \\ \hline

Rule-based & Associative learning & Frequent set mining& \cite{1-92,1-124,1-206,1-252} \\
\hline
Cross-chain& Generalizing event graphs & Graph representation& \cite{1-11,1-195} \\
\hline
\multirow{3}{*}{Semantic causation} &  Event representation& RDF-based &\cite{1-57,1-39, 1-111,1-112, 1-249}\\
\cline{2-4}
&  Event inference & probability estimate &  \cite{1-2, 1-155,1-168}\\
\cline{2-4}
& Future event inference& Circular binary search tree&\cite{1-45,1-123, 1-168,1-249}\\
\hline
\multirow{6}{*}{Semantic model} & \multirow{3}{*}{Feature-based} & Aggregation-based & \cite{1-198,1-199,1-201,1-203} \\ \cline{3-4}
& & Compositional-based & \cite{1-79,1-89} \\ \cline{3-4}
&   & Markov models & \cite{1-7,1-8,1-121,1-241,1-224} \\ \cline{2-4}
& Prototype-based & Clustering-based & \cite{1-3,1-1,1-5} \\ \cline{2-4}
& Attribute-based & Feature extraction-based &\cite{ 1-33,1-88, 1-134,1-131} \\ \cline{2-4}
& Descriptive-based & Text-based & \cite{1-96, 1-97, 1-139, 1-195, 1-221,1-231}\\ \hline
\end{tabular}
\vspace{-0.4cm}  
\end{table}

\subsubsection{Spatial-Temporal Methods}
\begin{figure}[htbp]
\setlength{\abovecaptionskip}{-0.1cm}   
\setlength{\belowcaptionskip}{-0.6cm}   
\begin{minipage}[htbp]{0.48\linewidth}
\includegraphics[width=\linewidth]{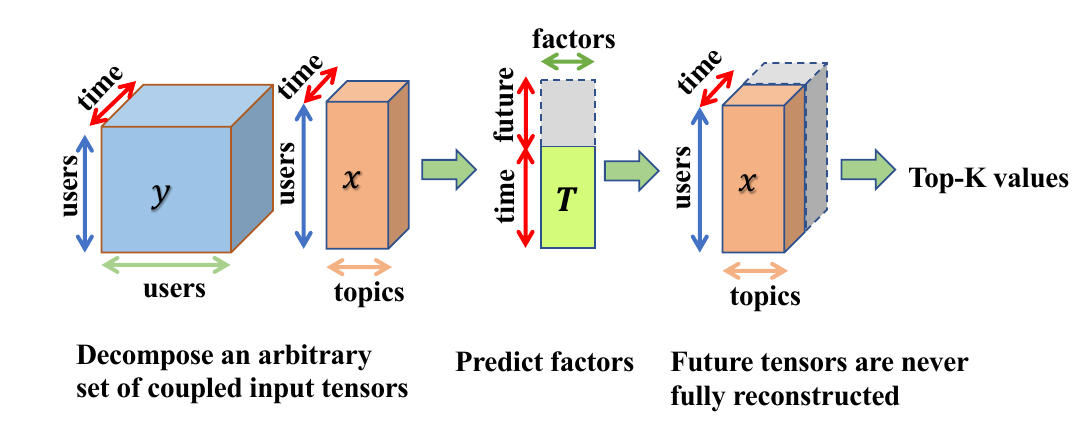}
\caption{Tensor Decomposition and Forecasting for Complex Time-stamped events}
\label{fig:5}
\end{minipage}
   \hfill
\begin{minipage}[htbp]{0.48\linewidth}
\includegraphics[width=\linewidth]{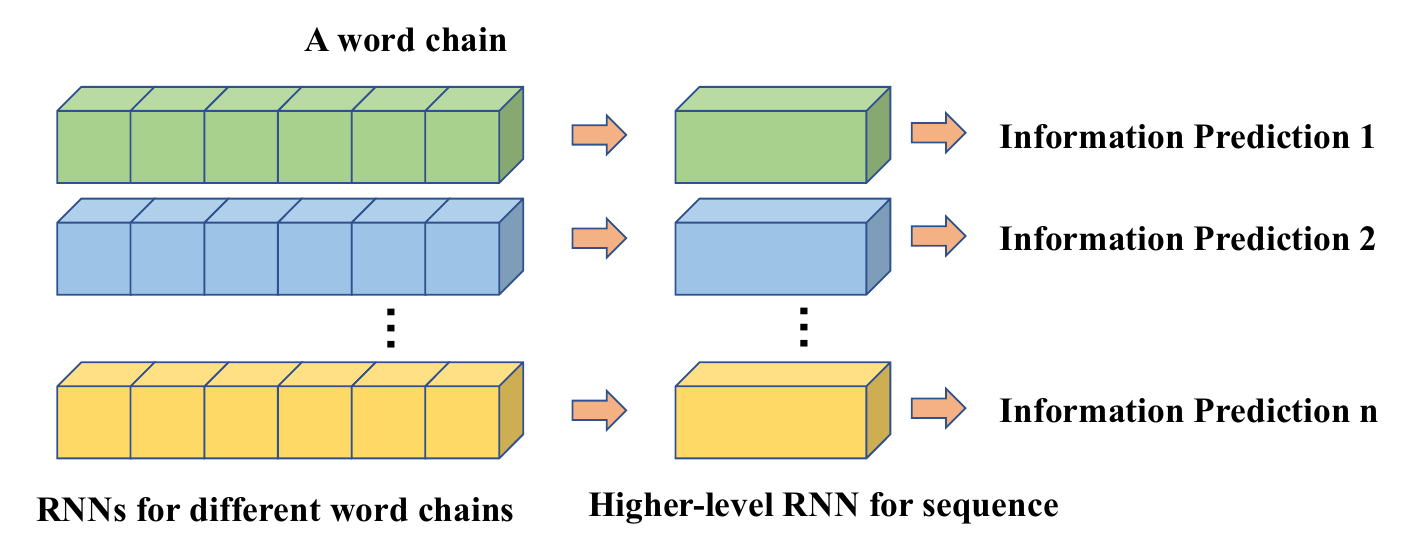}
\caption{Generic framework for hierarchical RNN-based information cascade prediction under public emergencies}
\label{fig:my_label}
\end{minipage}
\end{figure}

Spatial-temporal methods are used to predict public emergencies by incorporating spatial and temporal information about emergencies. For instance, a spatio-temporal model is proposed to predict emergency incidence \cite{1-32}. It considers the correlation between the spatial and temporal dimensions of emergencies and incorporates spatial factors such as population density and road network density. A multi-view learning-based spatio-temporal model is proposed, which learns the features of spatial and temporal domains separately and then fuses them for prediction \cite{1-183}. Another approach is to leverage the deep learning model for prediction \cite{1-194}. A spatiotemporal Graph Convolutional Network is proposed that learns the spatial and temporal correlations of emergencies using graph convolutional networks as shown in \textit{Figure \ref{fig:my_label}}.

\subsubsection{Time Series Forecasting-based Methods}

Methods for forecasting public emergencies based on time series are divided into two categories: 

\textbf{Direct methods} typically approach the problem of predicting emergency types as a multivariate time series forecasting problem, where each variable corresponds to an emergency type $E_i(i=1, ...)$ and the predicted emergency type at future time $t$ is calculated as $\hat{s}_{\hat{t}} = argmax{E_i}$ $f(s_{\hat{t}} = E_i|X)$. For instance, in \cite{1-128}, a longitudinal support vector regressor is utilized to forecast multi-dimensional emergencies. They build $n$ Support Vector Regressors, each of which corresponds to a specific attribute to achieve the goal of predicting the next time point's attribute value. To predict multiple emergency types, Weiss and Page \cite{1-219} use multiple-point process models. To improve the accuracy of their predictions, Biloš et al. \cite{1-25} employ RNN to learn the historical representation of emergencies, and then input the results into a Gaussian process model to predict future emergency types. To better capture the dynamics across multiple variables in the time series, Brandt et al. \cite{1-30} extend this to Bayesian vector autoregression. 

\textbf{Indirect methods} focus on learning a mapping from observed emergency types to low-dimensional latent-topic space using tensor decomposition-based techniques. Regarding information cascade prediction under public emergencies, Matsubara et al. \cite{1-142} propose a 3-way analysis of the original observed emergency tensor $Y_0 \in R^{D_oD_aD_e}$, which consists of three factors: actors, objects, and time. They decompose this tensor into latent variables via three corresponding low-rank matrices $P_o\in R^{D_kD_o}$, $P_a \in R_{D_kD_a}$, and $P_e \in R_{D_k*D_e}$, where $D_k$ is the number of latent topics. Through multivariate time series forecasting, the time matrices $P_e$ are predicted into the future to estimate future emergency tensors by recovering a "future emergency tensor" $\hat{Y}$ through the multiplication of the predicted time matrix $P_e$ with the known actor matrix $P_a$ and object matrix $P_o$. This approach provides insights into the different types of emergencies based on latent topics.

\subsection{Time and Location}
\begin{figure}
\setlength{\abovecaptionskip}{-0.1cm}   
\centering
\includegraphics[width=0.85\linewidth]{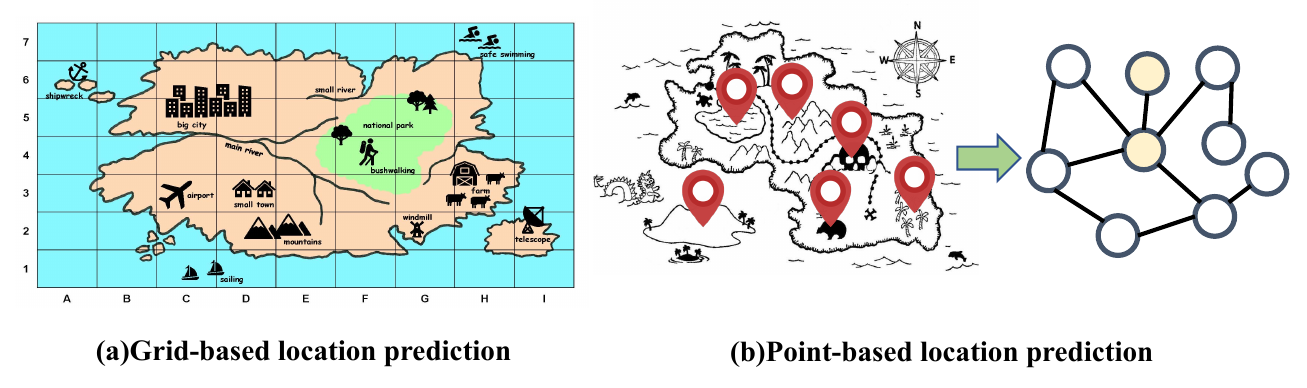}
\caption{The Type Of Location-based Prediction} \label{fig12}
\vspace{-0.5cm}  
\end{figure}

The prediction of the location and time of future public emergencies is a critical area of research for emergency management and disaster response. One relevant category of methods for such predictions is based on the joint consideration of time and location. These methods are classified into two subtypes based on their approach as shown in \textit{Figure \ref{fig12}}.

\subsubsection{Grid-based}
Grid-based methods are useful in predicting the location and time of future public emergencies. In recent years, several techniques have been proposed to capture spatial and temporal information for information cascade prediction under public emergencies as shown in \textit{Table \ref{tab:sample}}.
\begin{table}[htbp]
\vspace{-0.1cm}  
\setlength{\abovecaptionskip}{0.1cm}   
\setlength{\belowcaptionskip}{-0.2cm}   
\small
\centering
\caption{Grid-based Location Prediction}
\label{tab:sample}
\begin{tabular}{m{0.17\linewidth}|m{0.45\linewidth}|m{0.2\linewidth}}
\hline
\textbf{Category} & \textbf{Description} & \textbf{References} \\ \hline
Spatial Clustering & Grouping contiguous regions that collectively exhibit significant patterns. & \cite{1-105, 1-211, 1-220} \\ \hline
Spatial Interpolation & Estimating the probability of event occurrence at locations without historical data, resulting in spatial smoothness. & \cite{1-5, 1-93, 1-105,1-114, 1-176} \\ \hline
Spatial Convolution & Learning and representing spatial patterns using Convolutional Neural Networks (CNNs).& \cite{1-46, 1-88, 1-19, 1-150, 1-164, 1-209}\\ \hline
Trajectory Destination& Predicting population-based events by interpreting the collective behaviors of individuals. & \cite{1-214, 1-203, 1-116, 1-220} \\ \hline
\end{tabular}
\vspace{-0.4cm}  
\end{table}
A straightforward approach to incorporating spatial information is to include location data as an input feature and use it in predictive models, such as \textit{Linear Regression} \cite{1-250}, \textit{LSTM} \cite{1-174}, and \textit{Gaussian Processes} \cite{1-118}. Zhao et al. \cite{1-250} utilized the spatiotemporal dependency to regularize their model parameters during training. However, most of the methods in this domain aim to jointly consider the spatial and temporal dependencies for predictions \cite{1-64}. To achieve this, the multi-attributed spatial information for each time point is organized as a series of multi-channel images that are encoded using convolution-based operations\cite{iot_8}.


The multi-attributed spatial information is represented as a spatiotemporal grid, where each cell in the grid represents a specific location and time point. Let $X_{i,j,t}$ denote the value of the $i$-th attribute at the $(i,j)$-th location and $t$-th time point. The spatiotemporal grid is represented as:
\begin{equation}
    X = [X_{i,j,t}]_{i=1,\dots,n; j=1,\dots,m; t=1,\dots,T}
\end{equation}
where $n$ and $m$ represent the number of locations in the grid, and $T$ represents the number of time points.

To encode the spatiotemporal grid using convolution-based operations, it first converts it into a series of multi-channel images, where each channel corresponds to a different attribute. Let $X^{(k)}$ denote the $k$-th channel of the image, which is a 2D matrix of size $n \times m$:
\begin{equation}
    X^{(k)} = [X_{i,j,t}^{(k)}]_{i=1,\dots,n; j=1,\dots,m}
\end{equation}

The 2D convolution operation is then applied to each channel of the image to learn spatial features. The output of the convolution operation is a 2D feature map, which is then passed through a nonlinear activation function. Let $W^{(k)}$ denote the weight matrix for the $k$-th channel, and $b^{(k)}$ denote the bias vector. The output of the convolution operation is expressed as:
\begin{equation}
    H^{(k)}_{i,j,t} = f\left(\sum_{p=1}^{P}\sum_{q=1}^{Q} W^{(k)}_{p,q} X_{i+p-1,j+q-1,t} + b^{(k)}\right)
\end{equation}
where $P$ and $Q$ represent the size of the convolution kernel, and $f$ is the activation function.

The output of the convolution operation is then passed through a pooling layer to reduce the spatial dimensionality of the feature map. The pooling operation computes a summary statistic over a local neighborhood of the feature map. The most common pooling operations are max pooling and average pooling. Let $H^{(k)}_{i,j,t}$ denote the output of the pooling operation for the $k$-th channel. The output of the pooling layer is expressed as:
\begin{equation}
    Y^{(k)}_{i,j,t} = \text{pool}\left(\{H^{(k)}_{p,q,t}\}_{p=i,\dots,i+K-1; q=j,\dots,j+K-1}\right)
\end{equation}
where $K$ represents the size of the pooling window, and $\text{pool}$ is the pooling function.

The outputs of the pooling layer for channels are concatenated to form a spatiotemporal feature vector. Let $Y_t$ denote the spatiotemporal feature vector at time $t$. The spatiotemporal feature vector is  expressed as:
\begin{equation}
    Y_t = [Y^{(1)}_{1,1,t}, \dots, Y^{(1)}_{n,m,t}, Y^{(2)}_{1,1,t}, \dots, Y^{(2)}_{n,m,t}, \dots, Y^{(K)}_{1,1,t}, \dots, Y^{(K)}_{n,m,t}]
\end{equation}


\subsubsection{Point-based}
Spatio-temporal point process modeling is a valuable technique for predicting the time and location of public emergencies. This method models the rate of event occurrence in both space and time, providing a comprehensive understanding of the spatiotemporal distribution of events. The technique is defined as:
\begin{equation}
\lambda (t,l|X)=\lim_{|dt|\rightarrow 0,|dl|\rightarrow 0} \frac{E[N(dt*dl)|X]}{(|dt||dl|)}
\end{equation}
Several models have been proposed to instantiate this framework, which includes different facets of input data such as location, time, and other semantic features. Liu and Brown et al. \cite{1-136} assumed conditional independence among spatial and temporal factors and decomposed the rate of event occurrence as follows:
\begin{equation}
\lambda (t,l|X)=\lambda (t,l|L,T,F)=\lambda_1 (l|L,T,F,t)*\lambda_2(t|T)
\end{equation}
In this equation, $\lambda_1(x)$ is modeled using the Markov spatial point process, while $\lambda_2(x)$ is characterized using temporal autoregressive models. To handle situations where explicit assumptions for model distributions are difficult, several methods have been proposed to involve deep architecture during the point process. Recently, Okawa et al. \cite{1-159} proposed the following:
\begin{equation}
\lambda(t, l \mid X)=\int g_{\theta}\left(t^{\prime}, l^{\prime}, \mathcal{F}\left(t^{\prime}, l^{\prime}\right)\right) \cdot \mathcal{K}\left((t, l),\left(t^{\prime}, l^{\prime}\right)\right) \mathrm{d} t^{\prime} \mathrm{d} l^{\prime}
\end{equation}
Here, $K(x,y)$ is a kernel function, such as a Gaussian kernel \cite{1-26}, which measures the similarity in time and location dimensions. $F(t,l)\subseteq F$ denotes the feature values for the data at location $l^{'}$ and time $t^{'}$. $g_{\theta}(x)$ is a deep neural network that parameterized by $\theta$ and returns a nonnegative scalar. The model selection of $g_{\theta}(x)$ depends on the data types.


\subsection{Time, Location, and Semantics}
\begin{table}[htbp]
\vspace{-0.2cm}  

\small
\centering
\caption{Time, Location, and Semantics}
\label{table:emergencies}
\begin{tabular}{m{0.13\linewidth}|m{0.2\linewidth}|m{0.3\linewidth}|m{0.15\linewidth}}
\hline
\textbf{Category} & \textbf{Directions} & \textbf{Methods} & \textbf{References} \\ \hline
\multirow{3}*{System-based} & Event prediction& Model-fusion system & \cite{1-171, 1-95, 1-108} \\ \cline{2-4}
& Crowd-sourced system & Recommender system & \cite{1-177,1-175,1-178} \\ \cline{2-4}
& Future event detection & Prediction market system & \cite{1-126,1-125,1-128} \\ \hline
\multirow{5}*{Model-based} &\multirow{5}*{Planned event detection}& Mathematical models & \cite{1-115, 1-148, 1-190} \\ \cline{3-4}
& & Content filtering & \cite{1-101, 1-153} \\ \cline{3-4}
& & Time expression identification & \cite{1-55, 1-101} \\ \cline{3-4}
& & Future reference sentence extraction & \cite{1-117, 1-153} \\ \cline{3-4}
& & Location identification & \cite{1-22, 1-49, 1-107, 1-152} \\ \hline
\multirow{2}*{Tensor-based} & \multirow{2}*{Tensor extrapolation} & Tensor decomposition & \cite{1-148, 1-253,1-252} \\ \cline{3-4}
& & Tensor completion & \cite{1-253, 1-250} \\ \hline
\end{tabular}
\vspace{-0.4cm}  
\end{table}
In this section, we discuss various techniques that are utilized for predicting the time, location, and Semantics of public emergencies. These methods are broadly classified into two categories: \textbf{System-based} approaches rely on the development and deployment of complex systems that are designed to gather and analyze data from various sources in real time. On the other hand, \textbf{Future Event Detection Methods} approaches involve the development of mathematical models that use historical data to predict future emergencies. Both approaches have their own strengths and limitations, and the choice of strategy depends on the specific needs and constraints of the emergency management system.

\subsubsection{System-based Approaches}
Public emergencies are predicted using system-based or model-based strategies. In this section, we discuss system-based approaches. One such approach is the model-fusion system, which integrates various techniques for predicting the time, location, and semantics of public emergencies into a unified emergencies prediction system \cite{iot_6}.

Cascaded prediction of emergency events is crucial for timely and effective response. EMBERS \cite{1-171} is an online warning system that predicts the time, location, and type of future events, as well as the population affected. To maximize precision, EMBERS prioritizes individual prediction models by suppressing their recall \cite{iot_7}. The fusion of predictions from different models eventually results in high recall. Bayesian fusion-based strategies have been investigated \cite{1-95}, and similar strategies are used in systems like Carbon \cite{1-108}. Crowd-sourced systems are another approach that implements fusion strategies to generate predictions made by human predictors. For instance, Rostami et al. \cite{1-177} proposed a recommender system that matches event-predicting tasks to human predictors with suitable skills to maximize the accuracy of their fused predictions. This approach addresses the heterogeneity and diversity of human predictors' skill sets and background knowledge under limited human resources.

\subsubsection{Future Event Detection Methods}
Planned public emergencies are detected using methods that rely on \textit{Natural Language Processing (NLP)} techniques and \textit{Linguistic Principles} to analyze various media sources such as social media and news. These methods typically consist of four main steps.
\begin{itemize}

\item[$\bullet$]\textit{Content filtering} is employed to retain the texts that are relevant to the event of interest. Existing works utilize either supervised methods or unsupervised methods.
\item[$\bullet$]\textit{Time expression identification} is used to identify future reference expressions and determine the time of the event. This step leverage existing tools such as the Rosetta text analyzer \cite{1-55} or propose dedicated strategies based on linguistic rules \cite{1-101}.
\item[$\bullet$]\textit{Future reference sentence extraction} is the core of planned event detection and is implemented either by designing regular expression-based rules \cite{1-153} or by textual classification \cite{1-117}.
\item[$\bullet$]\textit{Location identification} is essential to infer the event's location accurately. The expression of locations is typically highly heterogeneous and noisy. Existing works have relied heavily on geocoding techniques to resolve the event location accurately. Various types of locations are considered, such as article locations, locations mentioned in the articles \cite{1-22}, and authors' neighbors' locations \cite{1-107}. Multiple locations have been selected using a geometric median \cite{1-49} or fused using logical rules such as probabilistic soft logic \cite{1-152}.
\end{itemize}

\subsubsection{Tensor-based Methods}
Public emergencies are often analyzed using tensor decomposition and extrapolation techniques. These methods involve formulating the data into a tensor form, which includes dimensions such as location, time, and semantics. Tensor decomposition is then applied to approximate the original tensor by using multiple low-rank matrices, each of which represents a mapping from latent topics to each dimension \cite{iot_4}. Specifically, given a tensor $\mathcal{T}$ with dimensions $I$, $J$, and $K$, the low-rank tensor approximation is represented as:
\begin{equation}
\mathcal{T} \approx \sum_{r=1}^{R} \mathbf{A}^{(1)}_{r} \circ \mathbf{A}^{(2)}_{r} \circ \mathbf{A}^{(3)}_{r}
\end{equation}
where $\mathbf{A}^{(1)}_{r} \in \mathbb{R}^{I \times R}$, $\mathbf{A}^{(2)}_{r} \in \mathbb{R}^{J \times R}$, and $\mathbf{A}^{(3)}_{r} \in \mathbb{R}^{K \times R}$ are low-rank matrices, $\circ$ denotes the outer product, and $R$ is the rank of the tensor.

After the tensor is decomposed, various strategies are employed to extrapolate the tensor toward future periods. For instance, Mirtaheri \cite{1-148} extrapolated the time dimension matrix and then multiplied it with the other dimensions' matrices to recover the estimated extrapolated tensor into the future:
\begin{equation}
    \mathbf{A}^{(3)}_{r}(t_f) \approx \mathbf{A}^{(3)}_{r}(t_n) + (t_f - t_n) \Delta \mathbf{A}^{(3)}_{r}
\end{equation}
where $\mathbf{A}^{(3)}_{r}(t_f)$ and $\mathbf{A}^{(3)}_{r}(t_n)$ are the time dimension matrices at the future period $t_f$ and the last observed period $t_n$, respectively, and $\Delta \mathbf{A}^{(3)}_{r}$ is the first-order difference matrix of the time dimension matrix.

On the other hand, Zhou et al. \cite{1-253} adopted a different approach, where they added "empty values" for the entries corresponding to a future time in the original tensor. Then, they utilized tensor completion techniques to infer the missing values that correspond to future events. These techniques include low-rank tensor completion and tensor regression, which aim to recover the missing entries by exploiting the low-rank structure and the relationships between the dimensions of the tensor. By extrapolating the tensor towards future periods, decision-makers better understand the potential impact of such events and prepare appropriate response plans.

\subsection{Vulnerability Prediction}
Vulnerability prediction in the context of public emergencies refers to the use of machine learning models to identify and assess potential risks and vulnerabilities associated with various hazards, such as earthquakes, floods, landslides, and extreme weather events. 

In public emergencies, innovative solutions are required, and machine learning (ML) models have emerged as a promising approach to addressing the challenges posed by these events. 

Prasad et al. \cite{2-25} utilized an ML-based ensemble technique that combines multiple models to improve the accuracy of flood vulnerability mapping. The ensemble method is expressed mathematically as follows:
\begin{equation}
F(x) = \sum_{t=1}^{T} w_t f_t(x)
\end{equation}
where $x$ represents the input data, $F(x)$ is the output of the ensemble model, $f_t(x)$ is the output of the $t$-th base model, and $w_t$ is the weight assigned to the $t$-th base model. The weights are typically computed based on the performance of the base models on the training data.

Similarly, Nsengiyumva and Valentino \cite{2-26} employed the logistic model tree (LMT) to predict the vulnerability of landslide areas. The LMT is expressed mathematically as follows:
\begin{equation}
p(C_k|x) = \frac{N_k(x)}{N(x)}
\end{equation}
where $p(C_k|x)$ is the probability of the $k$-th class given the input data $x$, $N_k(x)$ is the number of training instances in class $k$ that satisfy the conditions specified by the tree, and $N(x)$ is the total number of training instances that satisfy the conditions specified by the tree.


\subsection{Association-based Impact Prediction}
Association-based impact prediction in the context of public emergencies involves the use of association \textbf{Rule-based} methods to predict future events accurately. These methods rely on discovering associations between precursor and target events to predict the onset of various hazards, such as pandemics.


The Apriori algorithm is a popular algorithm for frequent set mining, which generates frequent item sets by iteratively joining smaller itemsets and pruning infrequent ones. Mathematically, the support of an itemset $I$ is defined as:
\begin{equation}
    Supp(I) = \frac{freq(I)}{N}
\end{equation}
where $freq(I)$ is the number of transactions that contain the itemset $I$, and $N$ is the total number of transactions.

Association rule mining involves finding rules of the form $X\rightarrow Y$, where $X$ is a set of items (the precursor event), and $Y$ is a set of items (the target event). The association rules $X\rightarrow Y$ are said to be valid if their support and confidence are above a given threshold. Pruning strategies are employed to remove invalid rules and retain the most accurate rules. The lift measure is another commonly used metric that measures the strength of association between the precursor event $X$ and the target event $Y$ is defined as:
\begin{equation}
    lift(X\rightarrow Y) = \frac{supp(X\cup Y)}{supp(X)supp(Y)}
\end{equation}
where $supp(X\cup Y)$ is the support of the itemset $X\cup Y$. A lift value greater than 1 indicates a positive association between $X$ and $Y$. In conclusion, association-based methods effectively predict events during public emergencies by discovering associations between precursor and target events.

\subsection{Causality-based Prediction}
Causality-based prediction involves \textit{inferring cause-effect relationships} among historical emergency events and utilizing this knowledge to predict future emergencies. This approach involves \textit{Emergency Semantic Representation}, \textit{Emergency Causality Inference}, and \textit{Future Emergency Inference}.

\subsubsection{Emergency Semantic Representation} 
The first step in applying causality-based prediction to public emergencies is to extract relevant emergency events from various sources, including news articles, social media, and government reports, using natural language processing techniques. This process involves several approaches to represent emergency events in a meaningful way. One approach is the \textbf{Event Phrase-based} method, where the emergency event is represented as a phrase extracted from the text \cite{1-77}. For instance, if the text mentions a "terrorist attack in downtown," the phrase "terrorist attack in downtown" represents the event. Another approach is the \textbf{Event Keywords-based} method, where keywords extracted from the text are used to represent the emergency event \cite{1-89, 1-142}. A third approach is the \textbf{Tuple-based} method \cite{1-210}, where each emergency event is represented by a tuple consisting of objects, a relationship, and time. For instance, if the text mentions a "car crash involving two vehicles on the highway at 4 pm," the tuple representation could be (car crash, involving, two vehicles, highway, 4 pm). An RDF-based format is utilized in some cases \cite{1-46}.
Mathematically, a tuple-based representation of an emergency event is expressed as $E = (O, R, T)$. where $O$ represents the objects involved in the event, $R$ represents the relationship between the objects, and $T$ represents the time of the event. In the case of the car crash example mentioned earlier, the tuple representation is expressed as $E = (\text{car crash, two vehicles, highway}, \text{involving}, \text{4 pm})$


\subsubsection{Emergency Causality Inference}
In this stage, the focus is on inferring cause-effect relationships among historical emergencies. The first step is to cluster the emergency events into emergency chains, sequences of time-ordered events with the same topics, actors, and objects \cite{1-142}. Once these chains are established, various approaches are used to infer the causal relationships among the emergency pairs. One common approach is to use natural language processing techniques to identify causal mentions, such as causal connectives, prepositions, and verbs \cite{1-142}, and then extract the causal relationships from the text. Another approach is to formulate causal-effect relationship identification as a classification task, where the inputs are the candidate events, and contextual information is incorporated, including related background knowledge from web texts \cite{1-97}. 

\begin{itemize}

\item[$\bullet$]\textit{Future Emergency Inference} After learning the cause-effect relationships among historical emergency events and representing them semantically, they use this knowledge to predict future emergencies. This enables us to anticipate potential emergency situations and take preventative measures to mitigate their impact.

\item[$\bullet$]\textit{Retrieve similar emergencies} To predict future emergencies, first search for similar events in the historical emergency event pool. They use various similarity measures, such as the Euclidean distance or cosine similarity, to compare the query event to historical events. Contextual information, such as the emergency time, location, and other relevant environmental and descriptive information, is taken into account. Specifically, use the following formula to calculate the Euclidean distance:
\begin{equation}
d(E_1, E_2) = \sqrt{\sum_{i=1}^{n}(x_{1,i}-x_{2,i})^2}
\end{equation}
where $E_1$ and $E_2$ are two emergency events being compared, $x_{1,i}$ and $x_{2,i}$ are the values of feature $i$ for $E_1$ and $E_2$, respectively, and $n$ is the total number of features being compared.

Similarly, use the cosine similarity to calculate the similarity between two emergency events:
\begin{equation}
sim(E_1, E_2) = \frac{\sum_{i=1}^{n} x_{1,i} x_{2,i}}{\sqrt{\sum_{i=1}^{n} x_{1,i}^2} \sqrt{\sum_{i=1}^{n} x_{2,i}^2}}
\end{equation}
where $E_1$ and $E_2$ are two emergency events being compared, $x_{1,i}$ and $x_{2,i}$ are the values of feature $i$ for $E_1$ and $E_2$, respectively, and $n$ is the total number of features being compared. Then rank the retrieved emergency events based on their similarity to the query event. This approach allows us to identify potential future emergencies and take proactive steps to prevent or mitigate their impact.

\item[$\bullet$]\textit{Infer the future emergencies}The next step is to determine the potential consequences caused by the query emergency based on the causality of emergencies learned from historical data. They traverse the abstraction tree that represents the learned causal relationships, starting from the root that corresponds to the most general emergency rule. The search frontier then moves across the tree if the child node is more similar, culminating in the nodes that are the least general but still similar to the new event being retrieved. Specifically, use the following formula to determine the similarity between two nodes in the abstraction tree:
\begin{equation}
sim(n_1, n_2) = \frac{w_{1,2}}{\sqrt{w_{1,1}} \sqrt{w_{2,2}}}
\end{equation}
where $n_1$ and $n_2$ are two nodes being compared, $w_{1,2}$ is the weight of the edge connecting nodes $n_1$ and $n_2$, $w_{1,1}$ is the sum of the weights of edges connected to node $n_1$, and $w_{2,2}$ is the sum of the weights of edges connected to node $n_2$ \cite{1-35}.
\end{itemize}
Since each case, event lead to multiple emergency events, use various approaches to determine the final prediction. For example, calculate the support or conditional probability of the rules using the following formula:
\begin{equation}
P(E|C) = \frac{P(C|E)P(E)}{P(C)}
\end{equation}
where $P(E|C)$ is the probability of emergency event $E$ given case event $C$, $P(C|E)$ is the probability of case event $C$ given emergency event $E$, $P(E)$ is the prior probability of emergency event $E$, and $P(C)$ is the prior probability of case event $C$.

A ranking approach based on the similarity between the new emergency event and historical emergency events is employed. The similarity is defined by the length of their minimal generalization path, which is calculated using the following formula:
\begin{equation}
d(E_1, E_2) = \sum_{i=1}^{k} \frac{1}{2^i}
\end{equation}

where $E_1$ and $E_2$ are two emergency events being compared, and $k$ is the length of their minimal generalization path. The formula gives greater weight to earlier levels in the path, as the denominator increases exponentially with each level. This approach allows us to rank emergency events based on their similarity to the new event, and use this ranking to make a prediction \cite{1-142, 1-210}.

\subsection{Large Language Model Under Public Emergencies}

Large Language Models (LLMs) have great potential in addressing public emergencies, such as natural disasters and pandemics. LLMs provide real-time reports, predict outcomes, and analyze social media data to identify areas that are most affected. They also identify and address misinformation and fake news by analyzing social media data and news articles. Large Language Models, including the autoregressive integrated moving average (ARIMA) model, have demonstrated significant potential in predicting natural disasters through data analysis. 

In public emergency prediction, it would be predicting the outbreak of a disease based on the language used in social media posts. A formula that could be used in this category is the autoregressive integrated moving average (ARIMA) model:
\begin{equation}
Y_t = c + \sum_{i=1}^p\phi_iY_{t-i} + \sum_{i=1}^q\theta_i\varepsilon_{t-i} + \varepsilon_t
\end{equation}
where $Y_t$ is the number of social media posts at time $t$, $c$ is a constant, $\phi_i$ and $\theta_i$ are the autoregressive and moving average parameters, respectively, and $\varepsilon_t$ is the error term at time $t$.

Then, it would be predicting the location and time of a flood based on historical data on weather patterns and water levels. A formula that could be used in this category is the space-time autoregressive integrated moving average (STARIMA) model:

\begin{equation}
Y_{s,t} = c + \sum_{i=1}^p\sum_{j=1}^k\sum_{l=1}^m\phi_{i,j,l}Y_{s-i,t-j,l} + \sum_{i=1}^q\sum_{j=1}^k\sum_{l=1}^m\theta_{i,j,l}\varepsilon_{s-i,t-j,l} + \varepsilon_{s,t}
\end{equation}
where $Y_{s,t}$ is the water level at spatial location $s$ and time $t$, $\phi_{i,j,l}$ and $\theta_{i,j,l}$ are the autoregressive and moving average parameters, respectively, and $\varepsilon_{s,t}$ is the error term at spatial location $s$ and time $t$.

After that, it would be predicting a wildfire's location, time, and nature based on satellite imagery and textual data on weather patterns. A formula that could be used in this category is the space-time-text autoregressive model (STTAR) model:

\begin{equation}
Y_{s,t} = c + \sum_{i=1}^p\sum_{j=1}^k\sum_{l=1}^m\phi_{i,j,l}Y_{s-i,t-j,l} + \sum_{i=1}^q\sum_{j=1}^k\sum_{l=1}^m\theta_{i,j,l}\varepsilon_{s-i,t-j,l} + \beta X_{s,t} + \varepsilon_{s,t}
\end{equation}
where $Y_{s,t}$ is the severity of the wildfire at spatial location $s$ and time $t$, $\phi_{i,j,l}$ and $\theta_{i,j,l}$ are the autoregressive and moving average parameters, respectively, $X_{s,t}$ is a vector of weather and environmental variables at spatial location $s$ and time $t$, $\beta$ is a vector of regression coefficients, and $\varepsilon_{s,t}$ is the error term at spatial location $s$ and time $t$.

In Vulnerability Prediction, it would be predicting a community's vulnerability to a hurricane based on demographic and socio-economic data. A formula that could be used in this category is the vulnerability index:

\begin{equation}
VI = \sum_{i=1}^n w_i \cdot (1 - z_i)
\end{equation}
where $w_i$ is the weight assigned to vulnerability factor $i$, $z_i$ is the standardized value of vulnerability factor $i$, and $n$ is the number of vulnerability factors.

In Association-based Impact Prediction, it would be predicting the impact of a hurricane based on its association with other factors, such as wind speed and storm surge \cite{iot_3}. A formula that could be used in this category is the association rule \cite{iot_2}: $\text{hurricane} \rightarrow \text{damage}$, where the hurricane is the antecedent and the damage is the consequent. In Causality-based Prediction, it would be predicting the likelihood of a wildfire based on its causal factors, such as drought and lightning strikes \cite{iot_5}. A formula that could be used in this category is the Bayesian network:

\begin{equation}
P(W|D,L) = \frac{P(W)P(D,L|W)}{P(D,L)}
\end{equation}
where $W$ is the event of a wildfire occurring, $D$ is the event of a drought occurring, $L$ is the event of lightning strikes occurring and $P(W|D,L)$ is the probability of a wildfire occurring given the occurrence of a drought and lightning strikes. $P(W)$ is the prior probability of a wildfire occurring, $P(D,L|W)$ is the conditional probability of a drought and lightning strikes occurring given that a wildfire has happened, and $P(D,L)$ is the joint probability of a drought and lightning strikes occurring.

\section{Application of Public Emergencies Management}

This section introduces the applications of public emergency management. By dividing the applications of public emergency management into different phases, this section is divided into \textbf{(1)} Early Warning, \textbf{(2)} Disaster Monitoring, \textbf{(3)} Damage Assessment, and \textbf{(4)} Disaster Response.

\begin{table}[htbp]
\vspace{-0.2cm}  
\setlength{\abovecaptionskip}{0.1cm}   
\small
  \centering
  \caption{Early warning}
  \label{tab:example}
  \begin{tabular}{m{0.18\linewidth}|m{0.25\linewidth}|m{0.32\linewidth}|m{0.15\linewidth}}
    \hline
    \textbf{Category} & \textbf{Direction} & \textbf{Method}& \textbf{References} \\
    \hline
    \multirow{4}*{Early Warning}& Sensing devices process data& PCA, Logistic Regression, CNN, RNN&\cite{3-17,3-2, 3-4, 3-7} \\ \cline{2-4}
    & Sense the location of emergencies & GAN, RF, Image recognition& \cite{3-10, 3-5,3-8,3-1}\\
    \cline{2-4}
    & Enhance the accuracy and speed& TLS, ANN, Fuzzy Deep Neural Network&\cite{3-16, 3-18, 3-11}\\
    \cline{2-4}
    & Utilize communication channels&Mining information dissemination data&\cite{3-16, 3-1}\\
    \hline
\multirow{4}*{Disaster Monitoring}& \multirow{2}*{Element Sensing}& UAV-based sensing &\cite{4-18, 4-20, 4-5} \\ \cline{3-4}
& & Visual sensing & \cite{4-14, 4-7, 4-15}\\ \cline{2-4}
& \multirow{2}*{Situation Sensing}& Crowdsourced sensing & \cite{4-21,4-22}\\ \cline{3-4}
& & Sensor Network-based System & \cite{4-5,4-4,4-7}\\ \hline
  \end{tabular}
\vspace{-0.5cm}  
\end{table}

\subsection{Early Warning}
Early warning is a critical phase in public emergency management, aimed at detecting and predicting potential emergencies and providing timely alerts to relevant authorities and the public. To achieve early warning, various methods have been developed, including data mining, machine learning, and predictive modeling. For example, in the case of a flood, data from sources such as rainfall sensors, river level gauges, and weather forecasts are analyzed to predict the flood's location, intensity, and possible impacts \cite{1-102}. Early identification and warning of public emergencies are performed through \textit{Sensor Networks}, \textit{Remote Sensing}, \textit{Social Media}, and other communication channels. 

\subsubsection{Time Sensing}

Effective early warning of public emergencies relies on processing data acquired by sensing devices promptly. Moon et al. \cite{3-4} proposed a machine learning method for effective early warning of short-term rainfall, while He et al. \cite{3-2} utilized machine learning techniques to extract data from tweets and perform fusion analysis for rainstorm disasters in real-time. Perol et al. \cite{3-7} proposed a Convolutional Neural Network-based method for early detection and warning of earthquake disasters, and Chin et al. \cite{3-17} adopted a Recurrent Neural Network model for earthquake early warning systems. Zheng et al. \cite{3-11} proposed a fuzzy deep neural network for early warning of industrial accidents.

\subsubsection{Location Sensing}

Early warning of public emergencies is critical in minimizing their impact. Seismic wave analysis has been used to sense the location of earthquakes, where Li et al. \cite{3-5} used a Generative Adversarial Network. Ethan et al. \cite{3-1} performed image recognition on natural disaster images from social media to provide early warning of public emergencies such as earthquakes, floods, and wildfires. Machine learning models have also been used to sense the severity of flood events in videos. Huang et al. \cite{3-10} proposed a deep belief network method for meteorological early warning of precipitation-induced landslides, achieving precise sensing of emergencies.

Early warnings provide valuable information for governments, communities, and individuals to take appropriate actions to mitigate the impact of public emergencies. It also enables real-time sensing of conditions on land and sea using advanced computer numerical models. Ultimately, continuous improvement of early warning systems for public emergencies such as weather, climate, traffic, and epidemics is essential for effective disaster response.

\subsection{Disaster Monitoring}
Continuous monitoring and assessment of public emergencies are crucial for providing emergency responders with crucial information to make informed decisions on response strategies. Disaster monitoring tools and technologies, such as \textit{Remote Sensing}, \textit{Geographic Information Systems (GIS)}, and \textit{Unmanned Aerial Vehicles (UAVs)}, have been used to track and assess natural and man-made disasters.

\subsubsection{Element Sensing}

\textbf{UAV-based sensing} is a promising solution for sensing in public emergencies, with research focusing on detecting and monitoring gas leaks, floods, and forest fires \cite{4-5}. \textbf{Sensor networks} are used to collect data on gas concentrations, water levels, and temperature and humidity levels, which are transmitted to a central server for analysis, providing real-time information to emergency responders \cite{4-18, 4-20}. \textbf{Visual sensing} using VR and MR technologies, as well as crowdsourced sensing through social media platforms, are also valuable for sensing in public emergencies \cite{4-14,1-68,1-108}. Advances in mobile technologies and social media have made it easier to collect data from citizens in affected areas, providing valuable information to emergency responders and improving the response time and effectiveness of emergency services \cite{4-21}.

\subsubsection{Situation Understanding}
\begin{table}[htbp]
\vspace{-0.8cm}  
\setlength{\abovecaptionskip}{0.1cm}   
\small
\centering
\caption{Situation sensing}
\label{tab:situation-understanding}
\begin{tabular}{m{0.15\linewidth}|m{0.32\linewidth}|m{0.28\linewidth}|m{0.13\linewidth}}
\hline
\textbf{Category} & \textbf{Directions} & \textbf{Methods} & \textbf{References} \\ \hline
 & Obtaining event data & Weakly supervised method & \cite{4-2,4-1,4-4} \\ \cline{2-4} 
 & Vulnerabilities in sensing devices & UAV-based sensing & \cite{4-18, 4-5} \\ \cline{2-4} 
 Element sensing & Energy issues in fixed sensors & Crowdsourced sensing & \cite{4-20, 4-21} \\ \cline{2-4} 
 & Privacy issues in social media sensing & UAV-based sensing  & \cite{4-7, 4-15} \\ \cline{2-4} 
 & Regional restrictions for sensing & Visual sensing & \cite{1-68, 1-108} \\ \hline
 
 & Efficient search for accurate information &Non-negative matrix factorization & \cite{4-18,4-16,4-19,4-20} \\ \cline{2-4}
  & Identifying sub-events & Unsupervised learning framework & \cite{4-3,4-1,4-5} \\ \cline{2-4}
Situational& Sensing data during GPS failures & Social media sensor monitoring & \cite{4-11,4-12,4-14} \\ \cline{2-4}
understanding& Real-time image processing & Image processing pipeline & \cite{4-4,4-8,4-9} \\ \cline{2-4} 
 &  Combining heterogeneous data sources & Context-aware fusion method & \cite{4-6,4-13,4-15} \\ \cline{2-4}
 & Transfer learning for COVID-19 & Convolutional neural networks & \cite{4-10,4-7,4-8,4-19} \\ \hline
\end{tabular}
\vspace{-0.3cm}  

\end{table}
Situational understanding is critical in emergency response, as it improves the level of sensing and the depth of understanding, ultimately helping in managing public emergencies. However, situational sensing does not exist in isolation and must be assisted by element sensing to obtain event data.

To deal with noisy data during public emergencies, weakly supervised methods and contextual messages are used to enrich message representations and achieve situational understanding. Cheng et al. \cite{4-18} studied a novel topic-tracking problem and enabled efficient search for accurate information through an online non-negative matrix factorization scheme. Situation-sensing algorithms have been designed to automatically identify important sub-events during public emergencies. Social media sensors are useful for monitoring natural disasters and real-time understanding of image content, but filtering out irrelevant parts of images and combining heterogeneous data sources is required \cite{4-2,4-3,4-11,4-4,4-6}.

\subsection{Damage Assessment}
\begin{table}[htbp]
\vspace{-0.2cm}  
\setlength{\abovecaptionskip}{0.1cm}   
\small
\centering
\caption{Damage Assessment and Disaster Response}
\begin{tabular}{m{0.15\linewidth}|m{0.27\linewidth}|m{0.3\linewidth}|m{0.18\linewidth}}
\hline
\textbf{Category} & \textbf{Directions} & \textbf{Methods} & \textbf{References} \\ \hline
& Social Media Analysis & DL-based multimodal approach & \cite{2-41,2-45} \\ \cline{2-4}
& Temporal and Spatial Analysis & Latent Dirichlet Allocation (LDA) & \cite{2-42,2-43} \\ \cline{2-4}
& Aerial Image Analysis & Pre-trained DL CNN and ML& \cite{2-44,2-46} \\ \cline{2-4}
Damage&Flood Damage Classification & K-means clustering and SVM &\cite{2-14,2-52} \\ \cline{2-4}
Assessment& Damage Classification & CNN architecture & \cite{2-13,2-62} \\ \cline{2-4}
& Deep CNN-based& Deep CNN & \cite{2-47, 2-45} \\ \cline{2-4}
& Multi-modal Classification & Two-stage multi-modal & \cite{2-46,2-59} \\ \cline{2-4}
& Damage Identification & CNNs and class activation maps & \cite{2-48,2-54} \\ \hline
\multirow{7}*{Disaster Response} & Public emergencies response & Drones, telemedicine, mobile apps & \cite{1-105,1-104,1-107} \\ \cline{2-4}
& Relief aid supply & Dynamic calculation& \cite{2-41,1-244,2-30} \\ \cline{2-4}
& Social media data analysis & ML-based supervised models & \cite{2-53,2-55,2-50} \\ \cline{2-4}
& Medical rescue & Decision table, genetic algorithm& \cite{2-52,2-51,2-58}\\ \cline{2-4}
& Image classification & DL methods, ML algorithms & \cite{2-54,2-57,2-64} \\ \cline{2-4}
& Social media data classification &Adaptation classifiers, LSTM, CNN & \cite{2-63, 2-59, 2-61, 2-55} \\ \hline
\end{tabular}
\vspace{-0.4cm}  
\end{table}
Damage assessment is crucial for effective emergency management. Techniques such as social media analysis and image processing provide real-time information to evaluate the extent of damage and plan resource deployment to mitigate the impact of the emergency on the affected population, infrastructure, and environment.

Researchers have proposed various ML and DL techniques to minimize the impact of natural disasters. These techniques include analyzing social media data, aerial images, and satellite data to evaluate disaster situations and assess damage caused by natural disasters. Some of the techniques proposed include DL-based multimodal approach \cite{2-41}, ML techniques combined with temporal and spatial analysis of social media posts \cite{2-42}, hybrid approaches that classify areas affected by floods \cite{2-14}, CNN architectures for damage classification \cite{2-13, 2-46, 2-48}, and DL-based approaches to assess the impact of water-related disasters using satellite image data \cite{2-47}. These techniques have shown promising results in identifying and assessing damage in disaster-hit areas. Overall, these proposed approaches using ML and DL techniques demonstrate their potential to improve disaster management and minimize the impact of natural disasters on the environment, infrastructure, and human lives. Damage assessment techniques provide accurate and timely information that is critical for planning and deploying resources to mitigate the impact of public emergencies.


\subsection{Disaster Response}
Public emergency response is the phase of emergency management that involves the deployment of resources to mitigate the impact of the emergency and assist affected populations. This phase involves various activities, including medical assistance, evacuation, and infrastructure management. 

ML and DL techniques have been used to improve public emergency response and disaster management. These techniques include the use of MLP NN to estimate relief supplies \cite{1-105}, decision tables to manage medical rescue \cite{2-51}, and unmanned aerial vehicles for recognizing the status of individuals in disaster-struck areas \cite{2-52}. ML algorithms have also been used to analyze social media data related to public emergencies \cite{2-53, 2-55, 2-58, 2-59, 2-61, 2-63} and classify images from earthquake-hit areas \cite{2-54}. These technologies have been tested on various datasets and for different types of disasters, aiding decision-making and response efforts. Technologies such as unmanned aerial vehicles, telemedicine, and mobile applications play a vital role in improving the efficiency and effectiveness of public emergency response and disaster management.

\section{Lessons Learned and Future Work}
This study provides a comprehensive review of research papers focused on the prediction of public emergencies. The reviewed articles covered a broad range of research areas related to the use of social media data, machine learning models, and natural language processing techniques for analyzing emergency events and developing predictive models for different types of disasters. Despite significant progress in predicting information cascades during public emergencies, there are still several open questions and directions for future research to explore:

\subsection{Complexity and Dynamics}

Future work in emergency response includes developing predictive algorithms that consider ethical and social implications. This requires using fairness, transparency, and accountability metrics to evaluate the impact on different populations, and creating ethical frameworks to guide AI use. Predictive algorithms must also account for complex and dynamic public emergencies, using advanced modeling techniques to capture interactions between emergency responders, the public, and stakeholders.

\subsection{Model Predictability and Interpretability}
Predicting the popularity and spread of emergency information cascades is a challenging task, and there are fundamental questions that have yet to be answered. Additionally, it is crucial to understand the mechanisms that govern the success of emergency information dissemination \cite{iot_12}. Interpretability of the predictions is also essential for building trust and accountability. However, these models are complex and challenging to interpret, which leads to issues with trust and accountability. Future research should focus on developing more interpretable models that are easily understood by human operators and provide a clear explanation of the reasoning behind their predictions.

\subsection{Model Robustness to Noise and Adversarial Attacks}
Information cascades during public emergencies are easily influenced or manipulated by malicious actors, leading to inaccurate predictions and potentially harmful outcomes. External stimuli such as breaking news and rumors significantly affect the spread of emergency information. Modeling these external stimuli is important for improving the robustness of prediction models. Cross-domain real-time transfer learning and retrieval of information from other platforms are used to model external stimuli. Additionally, analyzing the sources of external stimuli provide insights into the future evolution of emergency information dissemination. Future research should aim to develop more robust models that handle noisy and adversarial input data and provide accurate predictions even in the presence of interference.

\subsection{Human-AI teaming for public emergencies}

The field of Human-AI teaming has enormous potential in addressing public emergencies, such as natural disasters, pandemics, and terrorist attacks. This requires using fairness, transparency, and accountability metrics to evaluate the impact on different populations, and creating ethical frameworks to guide AI use. Predictive algorithms must also account for complex and dynamic public emergencies, using advanced modeling techniques to capture interactions between emergency responders, the public, and stakeholders.

\subsection{Multimodal Large Models in Public Emergencies}
Future research should focus on developing multimodal, large-scale models that can effectively integrate and process diverse sources of information to generate accurate and timely predictions during sudden public events. These models should leverage deep learning techniques, such as natural language processing, computer vision, and speech recognition, to analyze different data modalities and identify patterns and trends. Incorporating contextual information and domain knowledge could further enhance their performance, ultimately improving our ability to anticipate and respond to sudden public events.

\section{Conclusion}
This paper offers a comprehensive and systematic overview of existing techniques and methods for predicting emergency information cascades. The presented taxonomy serves as a valuable resource for domain experts when selecting the appropriate technique for a specific problem set. In particular, this paper analyzes the characteristics and methods of predicting information cascades during public emergencies from three perspectives: information cascade modeling, prediction, and application. These methods encompass a range of features and models, including time, location, and semantics, and incorporate the latest advances in modeling and predicting information cascades to provide a comprehensive and up-to-date overview of the field. Finally, we also discuss current research fronts, identify bottlenecks, pitfalls, and unresolved issues, and outline potential future research directions.

\bibliographystyle{ACM-Reference-Format}
\bibliography{sample-manuscript}


\begin{thebibliography}{240}


\ifx \showCODEN    \undefined \def \showCODEN     #1{\unskip}     \fi
\ifx \showDOI      \undefined \def \showDOI       #1{#1}\fi
\ifx \showISBNx    \undefined \def \showISBNx     #1{\unskip}     \fi
\ifx \showISBNxiii \undefined \def \showISBNxiii  #1{\unskip}     \fi
\ifx \showISSN     \undefined \def \showISSN      #1{\unskip}     \fi
\ifx \showLCCN     \undefined \def \showLCCN      #1{\unskip}     \fi
\ifx \shownote     \undefined \def \shownote      #1{#1}          \fi
\ifx \showarticletitle \undefined \def \showarticletitle #1{#1}   \fi
\ifx \showURL      \undefined \def \showURL       {\relax}        \fi
\providecommand\bibfield[2]{#2}
\providecommand\bibinfo[2]{#2}
\providecommand\natexlab[1]{#1}
\providecommand\showeprint[2][]{arXiv:#2}

\bibitem[emd(2023)]%
        {emdat}
 \bibinfo{year}{2023}\natexlab{}.
\newblock \bibinfo{title}{EM-DAT}.
\newblock \bibinfo{howpublished}{[EB/OL]}.
\newblock
\newblock
\shownote{\url{https://www.emdat.be/}}.


\bibitem[Abuella and Chowdhury(2019)]%
        {1-1}
\bibfield{author}{\bibinfo{person}{Mohamed Abuella} {and} \bibinfo{person}{Badrul Chowdhury}.} \bibinfo{year}{2019}\natexlab{}.
\newblock \showarticletitle{Forecasting of solar power ramp events: A post-processing approach}.
\newblock \bibinfo{journal}{\emph{Renewable Energy}}  \bibinfo{volume}{133} (\bibinfo{date}{apr} \bibinfo{year}{2019}), \bibinfo{pages}{1380--1392}.
\newblock
\urldef\tempurl%
\url{https://doi.org/10.1016/j.renene.2018.09.005}
\showDOI{\tempurl}


\bibitem[Acharya et~al\mbox{.}(2017)]%
        {1-2}
\bibfield{author}{\bibinfo{person}{Saurav Acharya}, \bibinfo{person}{Byung~Suk Lee}, {and} \bibinfo{person}{Paul Hines}.} \bibinfo{year}{2017}\natexlab{}.
\newblock \showarticletitle{Causal Prediction of Top-k Event Types Over Real-Time Event Streams}.
\newblock \bibinfo{journal}{\emph{Comput. J.}} \bibinfo{volume}{60}, \bibinfo{number}{11} (\bibinfo{date}{feb} \bibinfo{year}{2017}), \bibinfo{pages}{1561--1581}.
\newblock
\urldef\tempurl%
\url{https://doi.org/10.1093/comjnl/bxw098}
\showDOI{\tempurl}


\bibitem[Adhikari et~al\mbox{.}(2019)]%
        {1-3}
\bibfield{author}{\bibinfo{person}{Bijaya Adhikari}, \bibinfo{person}{Xinfeng Xu}, \bibinfo{person}{Naren Ramakrishnan}, {and} \bibinfo{person}{B.~Aditya Prakash}.} \bibinfo{year}{2019}\natexlab{}.
\newblock \showarticletitle{EpiDeep: Exploiting Embeddings for Epidemic Forecasting} \emph{(\bibinfo{series}{KDD '19})}. \bibinfo{publisher}{Association for Computing Machinery}, \bibinfo{address}{New York, NY, USA}, \bibinfo{pages}{577–586}.
\newblock
\showISBNx{9781450362016}
\urldef\tempurl%
\url{https://doi.org/10.1145/3292500.3330917}
\showDOI{\tempurl}


\bibitem[Akshya and Priyadarsini(2019a)]%
        {2-14}
\bibfield{author}{\bibinfo{person}{J. Akshya} {and} \bibinfo{person}{P.L.K. Priyadarsini}.} \bibinfo{year}{2019}\natexlab{a}.
\newblock \showarticletitle{A Hybrid Machine Learning Approach for Classifying Aerial Images of Flood-Hit Areas}. In \bibinfo{booktitle}{\emph{2019 International Conference on Computational Intelligence in Data Science (ICCIDS)}}. \bibinfo{pages}{1--5}.
\newblock
\urldef\tempurl%
\url{https://doi.org/10.1109/ICCIDS.2019.8862138}
\showDOI{\tempurl}


\bibitem[Akshya and Priyadarsini(2019b)]%
        {2-53}
\bibfield{author}{\bibinfo{person}{J. Akshya} {and} \bibinfo{person}{P.L.K. Priyadarsini}.} \bibinfo{year}{2019}\natexlab{b}.
\newblock \showarticletitle{A Hybrid Machine Learning Approach for Classifying Aerial Images of Flood-Hit Areas}. In \bibinfo{booktitle}{\emph{2019 International Conference on Computational Intelligence in Data Science (ICCIDS)}}. \bibinfo{pages}{1--5}.
\newblock
\urldef\tempurl%
\url{https://doi.org/10.1109/ICCIDS.2019.8862138}
\showDOI{\tempurl}


\bibitem[Alam et~al\mbox{.}(2018)]%
        {4-4}
\bibfield{author}{\bibinfo{person}{Firoj Alam}, \bibinfo{person}{Ferda Ofli}, {and} \bibinfo{person}{Muhammad Imran}.} \bibinfo{year}{2018}\natexlab{}.
\newblock \showarticletitle{Processing Social Media Images by Combining Human and Machine Computing during Crises}.
\newblock \bibinfo{journal}{\emph{International Journal of Human{\textendash}Computer Interaction}} \bibinfo{volume}{34}, \bibinfo{number}{4} (\bibinfo{date}{jan} \bibinfo{year}{2018}), \bibinfo{pages}{311--327}.
\newblock
\urldef\tempurl%
\url{https://doi.org/10.1080/10447318.2018.1427831}
\showDOI{\tempurl}


\bibitem[Alevizos et~al\mbox{.}(2017)]%
        {1-7}
\bibfield{author}{\bibinfo{person}{Elias Alevizos}, \bibinfo{person}{Alexander Artikis}, {and} \bibinfo{person}{George Paliouras}.} \bibinfo{year}{2017}\natexlab{}.
\newblock \showarticletitle{Event Forecasting with Pattern Markov Chains}. In \bibinfo{booktitle}{\emph{Proceedings of the 11th ACM International Conference on Distributed and Event-Based Systems}} (Barcelona, Spain) \emph{(\bibinfo{series}{DEBS '17})}. \bibinfo{publisher}{Association for Computing Machinery}, \bibinfo{address}{New York, NY, USA}, \bibinfo{pages}{146–157}.
\newblock
\showISBNx{9781450350655}
\urldef\tempurl%
\url{https://doi.org/10.1145/3093742.3093920}
\showDOI{\tempurl}


\bibitem[Alevizos et~al\mbox{.}(2018)]%
        {1-8}
\bibfield{author}{\bibinfo{person}{Elias Alevizos}, \bibinfo{person}{Alexander Artikis}, {and} \bibinfo{person}{Georgios Paliouras}.} \bibinfo{year}{2018}\natexlab{}.
\newblock \showarticletitle{Wayeb: a Tool for Complex Event Forecasting}.
\newblock
\urldef\tempurl%
\url{https://doi.org/10.29007/2s9t}
\showDOI{\tempurl}


\bibitem[Allison(2018)]%
        {1-222}
\bibfield{author}{\bibinfo{person}{Paul~D Allison}.} \bibinfo{year}{2018}\natexlab{}.
\newblock \showarticletitle{Event history and survival analysis}.
\newblock In \bibinfo{booktitle}{\emph{The reviewer’s guide to quantitative methods in the social sciences}}. \bibinfo{publisher}{Routledge}, \bibinfo{pages}{86--97}.
\newblock


\bibitem[Antunes et~al\mbox{.}(2003)]%
        {1-11}
\bibfield{author}{\bibinfo{person}{M. Antunes}, \bibinfo{person}{M.~A.~Amaral Turkman}, {and} \bibinfo{person}{K.~F. Turkman}.} \bibinfo{year}{2003}\natexlab{}.
\newblock \showarticletitle{A Bayesian Approach to Event Prediction}.
\newblock \bibinfo{journal}{\emph{Journal of Time Series Analysis}} \bibinfo{volume}{24}, \bibinfo{number}{6} (\bibinfo{date}{nov} \bibinfo{year}{2003}), \bibinfo{pages}{631--646}.
\newblock
\urldef\tempurl%
\url{https://doi.org/10.1111/j.1467-9892.2003.00326.x}
\showDOI{\tempurl}


\bibitem[Arachie et~al\mbox{.}(2020)]%
        {4-3}
\bibfield{author}{\bibinfo{person}{Chidubem Arachie}, \bibinfo{person}{Manas Gaur}, \bibinfo{person}{Sam Anzaroot}, \bibinfo{person}{William Groves}, \bibinfo{person}{Ke Zhang}, {and} \bibinfo{person}{Alejandro Jaimes}.} \bibinfo{year}{2020}\natexlab{}.
\newblock \showarticletitle{Unsupervised Detection of Sub-Events in Large Scale Disasters}.
\newblock \bibinfo{journal}{\emph{Proceedings of the {AAAI} Conference on Artificial Intelligence}} \bibinfo{volume}{34}, \bibinfo{number}{01} (\bibinfo{date}{apr} \bibinfo{year}{2020}), \bibinfo{pages}{354--361}.
\newblock
\urldef\tempurl%
\url{https://doi.org/10.1609/aaai.v34i01.5370}
\showDOI{\tempurl}


\bibitem[Asim et~al\mbox{.}(2016)]%
        {1-14}
\bibfield{author}{\bibinfo{person}{K.~M. Asim}, \bibinfo{person}{F. Mart{\'{\i}}nez-{\'{A}}lvarez}, \bibinfo{person}{A. Basit}, {and} \bibinfo{person}{T. Iqbal}.} \bibinfo{year}{2016}\natexlab{}.
\newblock \showarticletitle{Earthquake magnitude prediction in Hindukush region using machine learning techniques}.
\newblock \bibinfo{journal}{\emph{Natural Hazards}} \bibinfo{volume}{85}, \bibinfo{number}{1} (\bibinfo{date}{sep} \bibinfo{year}{2016}), \bibinfo{pages}{471--486}.
\newblock
\urldef\tempurl%
\url{https://doi.org/10.1007/s11069-016-2579-3}
\showDOI{\tempurl}


\bibitem[Bang et~al\mbox{.}(2019)]%
        {4-14}
\bibfield{author}{\bibinfo{person}{Junseong Bang}, \bibinfo{person}{Youngho Lee}, \bibinfo{person}{Yong-Tae Lee}, {and} \bibinfo{person}{Wonjoo Park}.} \bibinfo{year}{2019}\natexlab{}.
\newblock \showarticletitle{AR/VR Based Smart Policing For Fast Response to Crimes in Safe City}. In \bibinfo{booktitle}{\emph{2019 IEEE International Symposium on Mixed and Augmented Reality Adjunct (ISMAR-Adjunct)}}. \bibinfo{pages}{470--475}.
\newblock
\urldef\tempurl%
\url{https://doi.org/10.1109/ISMAR-Adjunct.2019.00126}
\showDOI{\tempurl}


\bibitem[Bao et~al\mbox{.}(2019)]%
        {1-19}
\bibfield{author}{\bibinfo{person}{Jie Bao}, \bibinfo{person}{Pan Liu}, {and} \bibinfo{person}{Satish~V. Ukkusuri}.} \bibinfo{year}{2019}\natexlab{}.
\newblock \showarticletitle{A spatiotemporal deep learning approach for citywide short-term crash risk prediction with multi-source data}.
\newblock \bibinfo{journal}{\emph{Accident Analysis Prevention}}  \bibinfo{volume}{122} (\bibinfo{date}{jan} \bibinfo{year}{2019}), \bibinfo{pages}{239--254}.
\newblock
\urldef\tempurl%
\url{https://doi.org/10.1016/j.aap.2018.10.015}
\showDOI{\tempurl}


\bibitem[Bao et~al\mbox{.}(2013)]%
        {14}
\bibfield{author}{\bibinfo{person}{Peng Bao}, \bibinfo{person}{Hua-Wei Shen}, \bibinfo{person}{Junming Huang}, {and} \bibinfo{person}{Xue-Qi Cheng}.} \bibinfo{year}{2013}\natexlab{}.
\newblock \showarticletitle{Popularity prediction in microblogging network: a case study on sina weibo}. In \bibinfo{booktitle}{\emph{Proceedings of the 22nd international conference on world wide web}}. \bibinfo{pages}{177--178}.
\newblock


\bibitem[Barabasi(2005)]%
        {17}
\bibfield{author}{\bibinfo{person}{Albert-Laszlo Barabasi}.} \bibinfo{year}{2005}\natexlab{}.
\newblock \showarticletitle{The origin of bursts and heavy tails in human dynamics}.
\newblock \bibinfo{journal}{\emph{Nature}} \bibinfo{volume}{435}, \bibinfo{number}{7039} (\bibinfo{year}{2005}), \bibinfo{pages}{207--211}.
\newblock


\bibitem[Barnes et~al\mbox{.}(2016)]%
        {1-20}
\bibfield{author}{\bibinfo{person}{G. Barnes}, \bibinfo{person}{K.~D. Leka}, \bibinfo{person}{C.~J. Schrijver}, \bibinfo{person}{T. Colak}, \bibinfo{person}{R. Qahwaji}, \bibinfo{person}{O.~W. Ashamari}, \bibinfo{person}{Y. Yuan}, \bibinfo{person}{J. Zhang}, \bibinfo{person}{R.~T.~J. McAteer}, \bibinfo{person}{D.~S. Bloomfield}, \bibinfo{person}{P.~A. Higgins}, \bibinfo{person}{P.~T. Gallagher}, \bibinfo{person}{D.~A. Falconer}, \bibinfo{person}{M.~K. Georgoulis}, \bibinfo{person}{M.~S. Wheatland}, \bibinfo{person}{C. Balch}, \bibinfo{person}{T. Dunn}, {and} \bibinfo{person}{E.~L. Wagner}.} \bibinfo{year}{2016}\natexlab{}.
\newblock \showarticletitle{A {COMPARISON} {OF} {FLARE} {FORECASTING} {METHODS}. I. {RESULTS} {FROM} {THE} {\textquotedblleft}{ALL}-{CLEAR}{\textquotedblright} {WORKSHOP}}.
\newblock \bibinfo{journal}{\emph{The Astrophysical Journal}} \bibinfo{volume}{829}, \bibinfo{number}{2} (\bibinfo{date}{sep} \bibinfo{year}{2016}), \bibinfo{pages}{89}.
\newblock
\urldef\tempurl%
\url{https://doi.org/10.3847/0004-637x/829/2/89}
\showDOI{\tempurl}


\bibitem[Barros et~al\mbox{.}(2012)]%
        {1-77}
\bibfield{author}{\bibinfo{person}{Vicente Barros}, \bibinfo{person}{Christopher~B. Field}, \bibinfo{person}{Qin Dahe}, {and} \bibinfo{person}{Thomas~F. Stocker}.} \bibinfo{year}{2012}\natexlab{}.
\newblock \showarticletitle{Preface}.
\newblock In \bibinfo{booktitle}{\emph{Managing the Risks of Extreme Events and Disasters to Advance Climate Change Adaptation}}. \bibinfo{publisher}{Cambridge University Press}, \bibinfo{pages}{ix--x}.
\newblock
\urldef\tempurl%
\url{https://doi.org/10.1017/cbo9781139177245.002}
\showDOI{\tempurl}


\bibitem[Becker et~al\mbox{.}(2012)]%
        {1-22}
\bibfield{author}{\bibinfo{person}{Hila Becker}, \bibinfo{person}{Dan Iter}, \bibinfo{person}{Mor Naaman}, {and} \bibinfo{person}{Luis Gravano}.} \bibinfo{year}{2012}\natexlab{}.
\newblock \showarticletitle{Identifying content for planned events across social media sites}. In \bibinfo{booktitle}{\emph{Proceedings of the fifth {ACM} international conference on Web search and data mining}}. \bibinfo{publisher}{{ACM}}.
\newblock
\urldef\tempurl%
\url{https://doi.org/10.1145/2124295.2124360}
\showDOI{\tempurl}


\bibitem[Bhattacharjya et~al\mbox{.}(2020)]%
        {1-190}
\bibfield{author}{\bibinfo{person}{Debarun Bhattacharjya}, \bibinfo{person}{Tian Gao}, \bibinfo{person}{Nicholas Mattei}, {and} \bibinfo{person}{Dharmashankar Subramanian}.} \bibinfo{year}{2020}\natexlab{}.
\newblock \showarticletitle{Cause-Effect Association between Event Pairs in Event Datasets}. In \bibinfo{booktitle}{\emph{Proceedings of the Twenty-Ninth International Joint Conference on Artificial Intelligence}}. \bibinfo{publisher}{International Joint Conferences on Artificial Intelligence Organization}.
\newblock
\urldef\tempurl%
\url{https://doi.org/10.24963/ijcai.2020/167}
\showDOI{\tempurl}


\bibitem[Bialonski et~al\mbox{.}(2015)]%
        {1-24}
\bibfield{author}{\bibinfo{person}{Stephan Bialonski}, \bibinfo{person}{Gerrit Ansmann}, {and} \bibinfo{person}{Holger Kantz}.} \bibinfo{year}{2015}\natexlab{}.
\newblock \showarticletitle{Data-driven prediction and prevention of extreme events in a spatially extended excitable system}.
\newblock \bibinfo{journal}{\emph{Physical Review E}} \bibinfo{volume}{92}, \bibinfo{number}{4} (\bibinfo{date}{oct} \bibinfo{year}{2015}).
\newblock
\urldef\tempurl%
\url{https://doi.org/10.1103/physreve.92.042910}
\showDOI{\tempurl}


\bibitem[Bishop and Nasrabadi(2006)]%
        {1-26}
\bibfield{author}{\bibinfo{person}{Christopher~M Bishop} {and} \bibinfo{person}{Nasser~M Nasrabadi}.} \bibinfo{year}{2006}\natexlab{}.
\newblock \bibinfo{booktitle}{\emph{Pattern recognition and machine learning}}. Vol.~\bibinfo{volume}{4}.
\newblock \bibinfo{publisher}{Springer}.
\newblock


\bibitem[Blair and Sambanis(2020)]%
        {1-29}
\bibfield{author}{\bibinfo{person}{Robert~A. Blair} {and} \bibinfo{person}{Nicholas Sambanis}.} \bibinfo{year}{2020}\natexlab{}.
\newblock \showarticletitle{Forecasting Civil Wars: Theory and Structure in an Age of {\textquotedblleft}Big Data{\textquotedblright} and Machine Learning}.
\newblock \bibinfo{journal}{\emph{Journal of Conflict Resolution}} \bibinfo{volume}{64}, \bibinfo{number}{10} (\bibinfo{date}{apr} \bibinfo{year}{2020}), \bibinfo{pages}{1885--1915}.
\newblock
\urldef\tempurl%
\url{https://doi.org/10.1177/0022002720918923}
\showDOI{\tempurl}


\bibitem[Boni and Gerber(2016)]%
        {1-5}
\bibfield{author}{\bibinfo{person}{Mohammad~Al Boni} {and} \bibinfo{person}{Matthew~S. Gerber}.} \bibinfo{year}{2016}\natexlab{}.
\newblock \showarticletitle{Area-Specific Crime Prediction Models}. In \bibinfo{booktitle}{\emph{2016 15th {IEEE} International Conference on Machine Learning and Applications ({ICMLA})}}. \bibinfo{publisher}{{IEEE}}.
\newblock
\urldef\tempurl%
\url{https://doi.org/10.1109/icmla.2016.0118}
\showDOI{\tempurl}


\bibitem[Brandt et~al\mbox{.}(2011)]%
        {1-30}
\bibfield{author}{\bibinfo{person}{Patrick~T. Brandt}, \bibinfo{person}{John~R. Freeman}, {and} \bibinfo{person}{Philip~A. Schrodt}.} \bibinfo{year}{2011}\natexlab{}.
\newblock \showarticletitle{Real Time, Time Series Forecasting of Inter- and Intra-State Political Conflict}.
\newblock \bibinfo{journal}{\emph{Conflict Management and Peace Science}} \bibinfo{volume}{28}, \bibinfo{number}{1} (\bibinfo{date}{feb} \bibinfo{year}{2011}), \bibinfo{pages}{41--64}.
\newblock
\urldef\tempurl%
\url{https://doi.org/10.1177/0738894210388125}
\showDOI{\tempurl}


\bibitem[Caigny et~al\mbox{.}(2018)]%
        {1-57}
\bibfield{author}{\bibinfo{person}{Arno~De Caigny}, \bibinfo{person}{Kristof Coussement}, {and} \bibinfo{person}{Koen W.~De Bock}.} \bibinfo{year}{2018}\natexlab{}.
\newblock \showarticletitle{A new hybrid classification algorithm for customer churn prediction based on logistic regression and decision trees}.
\newblock \bibinfo{journal}{\emph{European Journal of Operational Research}} \bibinfo{volume}{269}, \bibinfo{number}{2} (\bibinfo{date}{sep} \bibinfo{year}{2018}), \bibinfo{pages}{760--772}.
\newblock
\urldef\tempurl%
\url{https://doi.org/10.1016/j.ejor.2018.02.009}
\showDOI{\tempurl}


\bibitem[Caigny et~al\mbox{.}(2020)]%
        {1-58}
\bibfield{author}{\bibinfo{person}{Arno~De Caigny}, \bibinfo{person}{Kristof Coussement}, {and} \bibinfo{person}{Koen W.~De Bock}.} \bibinfo{year}{2020}\natexlab{}.
\newblock \showarticletitle{Leveraging fine-grained transaction data for customer life event predictions}.
\newblock \bibinfo{journal}{\emph{Decision Support Systems}}  \bibinfo{volume}{130} (\bibinfo{date}{mar} \bibinfo{year}{2020}), \bibinfo{pages}{113232}.
\newblock
\urldef\tempurl%
\url{https://doi.org/10.1016/j.dss.2019.113232}
\showDOI{\tempurl}


\bibitem[Calabrese and Elkink(2016)]%
        {1-32}
\bibfield{author}{\bibinfo{person}{Raffaella Calabrese} {and} \bibinfo{person}{Johan~A. Elkink}.} \bibinfo{year}{2016}\natexlab{}.
\newblock \showarticletitle{Estimating Binary Spatial Autoregressive Models for Rare Events}.
\newblock In \bibinfo{booktitle}{\emph{Spatial Econometrics: Qualitative and Limited Dependent Variables}}. \bibinfo{publisher}{Emerald Group Publishing Limited}, \bibinfo{pages}{145--166}.
\newblock
\urldef\tempurl%
\url{https://doi.org/10.1108/s0731-905320160000037012}
\showDOI{\tempurl}


\bibitem[Casagrande et~al\mbox{.}(2018)]%
        {1-33}
\bibfield{author}{\bibinfo{person}{Flavia~Dias Casagrande}, \bibinfo{person}{Jim Torresen}, {and} \bibinfo{person}{Evi Zouganeli}.} \bibinfo{year}{2018}\natexlab{}.
\newblock \showarticletitle{Sensor Event Prediction using Recurrent Neural Network in Smart Homes for Older Adults}. In \bibinfo{booktitle}{\emph{2018 International Conference on Intelligent Systems ({IS})}}. \bibinfo{publisher}{{IEEE}}.
\newblock
\urldef\tempurl%
\url{https://doi.org/10.1109/is.2018.8710467}
\showDOI{\tempurl}


\bibitem[Catling and Wolff(2019)]%
        {1-35}
\bibfield{author}{\bibinfo{person}{Finneas J~R Catling} {and} \bibinfo{person}{Anthony~H Wolff}.} \bibinfo{year}{2019}\natexlab{}.
\newblock \showarticletitle{Temporal convolutional networks allow early prediction of events in critical care}.
\newblock \bibinfo{journal}{\emph{Journal of the American Medical Informatics Association}} \bibinfo{volume}{27}, \bibinfo{number}{3} (\bibinfo{date}{dec} \bibinfo{year}{2019}), \bibinfo{pages}{355--365}.
\newblock
\urldef\tempurl%
\url{https://doi.org/10.1093/jamia/ocz205}
\showDOI{\tempurl}


\bibitem[Chan and Lam(2005)]%
        {1-38}
\bibfield{author}{\bibinfo{person}{Ki Chan} {and} \bibinfo{person}{Wai Lam}.} \bibinfo{year}{2005}\natexlab{}.
\newblock \showarticletitle{Extracting causation knowledge from natural language texts}.
\newblock \bibinfo{journal}{\emph{International Journal of Intelligent Systems}} \bibinfo{volume}{20}, \bibinfo{number}{3} (\bibinfo{year}{2005}), \bibinfo{pages}{327--358}.
\newblock
\urldef\tempurl%
\url{https://doi.org/10.1002/int.20069}
\showDOI{\tempurl}


\bibitem[Chan and Franklin(2011)]%
        {1-39}
\bibfield{author}{\bibinfo{person}{Samuel~W.K. Chan} {and} \bibinfo{person}{James Franklin}.} \bibinfo{year}{2011}\natexlab{}.
\newblock \showarticletitle{A text-based decision support system for financial sequence prediction}.
\newblock \bibinfo{journal}{\emph{Decision Support Systems}} \bibinfo{volume}{52}, \bibinfo{number}{1} (\bibinfo{date}{dec} \bibinfo{year}{2011}), \bibinfo{pages}{189--198}.
\newblock
\urldef\tempurl%
\url{https://doi.org/10.1016/j.dss.2011.07.003}
\showDOI{\tempurl}


\bibitem[Chen et~al\mbox{.}(2017)]%
        {3-16}
\bibfield{author}{\bibinfo{person}{Jiaoyan Chen}, \bibinfo{person}{Huajun Chen}, \bibinfo{person}{Zhaohui Wu}, \bibinfo{person}{Daning Hu}, {and} \bibinfo{person}{Jeff~Z. Pan}.} \bibinfo{year}{2017}\natexlab{}.
\newblock \showarticletitle{Forecasting smog-related health hazard based on social media and physical sensor}.
\newblock \bibinfo{journal}{\emph{Information Systems}}  \bibinfo{volume}{64} (\bibinfo{date}{mar} \bibinfo{year}{2017}), \bibinfo{pages}{281--291}.
\newblock
\urldef\tempurl%
\url{https://doi.org/10.1016/j.is.2016.03.011}
\showDOI{\tempurl}


\bibitem[Cheng et~al\mbox{.}(2014)]%
        {39}
\bibfield{author}{\bibinfo{person}{Justin Cheng}, \bibinfo{person}{Lada Adamic}, \bibinfo{person}{P~Alex Dow}, \bibinfo{person}{Jon~Michael Kleinberg}, {and} \bibinfo{person}{Jure Leskovec}.} \bibinfo{year}{2014}\natexlab{}.
\newblock \showarticletitle{Can cascades be predicted?}. In \bibinfo{booktitle}{\emph{Proceedings of the 23rd international conference on World wide web}}. \bibinfo{pages}{925--936}.
\newblock


\bibitem[Cheng et~al\mbox{.}(2020)]%
        {4-18}
\bibfield{author}{\bibinfo{person}{Lu Cheng}, \bibinfo{person}{Jundong Li}, \bibinfo{person}{K.~Selcuk Candan}, {and} \bibinfo{person}{Huan Liu}.} \bibinfo{year}{2020}\natexlab{}.
\newblock \showarticletitle{Tracking Disaster Footprints with Social Streaming Data}.
\newblock \bibinfo{journal}{\emph{Proceedings of the {AAAI} Conference on Artificial Intelligence}} \bibinfo{volume}{34}, \bibinfo{number}{01} (\bibinfo{date}{apr} \bibinfo{year}{2020}), \bibinfo{pages}{370--377}.
\newblock
\urldef\tempurl%
\url{https://doi.org/10.1609/aaai.v34i01.5372}
\showDOI{\tempurl}


\bibitem[Chin et~al\mbox{.}(2020)]%
        {3-17}
\bibfield{author}{\bibinfo{person}{Tai-Lin Chin}, \bibinfo{person}{Kuan-Yu Chen}, \bibinfo{person}{Da-Yi Chen}, {and} \bibinfo{person}{De-En Lin}.} \bibinfo{year}{2020}\natexlab{}.
\newblock \showarticletitle{Intelligent Real-Time Earthquake Detection by Recurrent Neural Networks}.
\newblock \bibinfo{journal}{\emph{{IEEE} Transactions on Geoscience and Remote Sensing}} \bibinfo{volume}{58}, \bibinfo{number}{8} (\bibinfo{date}{aug} \bibinfo{year}{2020}), \bibinfo{pages}{5440--5449}.
\newblock
\urldef\tempurl%
\url{https://doi.org/10.1109/tgrs.2020.2966012}
\showDOI{\tempurl}


\bibitem[Choi et~al\mbox{.}(2018)]%
        {1-45}
\bibfield{author}{\bibinfo{person}{Yunjey Choi}, \bibinfo{person}{Minje Choi}, \bibinfo{person}{Munyoung Kim}, \bibinfo{person}{Jung-Woo Ha}, \bibinfo{person}{Sunghun Kim}, {and} \bibinfo{person}{Jaegul Choo}.} \bibinfo{year}{2018}\natexlab{}.
\newblock \showarticletitle{{StarGAN}: Unified Generative Adversarial Networks for Multi-domain Image-to-Image Translation}. In \bibinfo{booktitle}{\emph{2018 {IEEE}/{CVF} Conference on Computer Vision and Pattern Recognition}}. \bibinfo{publisher}{{IEEE}}.
\newblock
\urldef\tempurl%
\url{https://doi.org/10.1109/cvpr.2018.00916}
\showDOI{\tempurl}


\bibitem[Chowdhury et~al\mbox{.}(2020)]%
        {4-8}
\bibfield{author}{\bibinfo{person}{Jishnu~Ray Chowdhury}, \bibinfo{person}{Cornelia Caragea}, {and} \bibinfo{person}{Doina Caragea}.} \bibinfo{year}{2020}\natexlab{}.
\newblock \showarticletitle{On Identifying Hashtags in Disaster Twitter Data}.
\newblock \bibinfo{journal}{\emph{Proceedings of the {AAAI} Conference on Artificial Intelligence}} \bibinfo{volume}{34}, \bibinfo{number}{01} (\bibinfo{date}{apr} \bibinfo{year}{2020}), \bibinfo{pages}{498--506}.
\newblock
\urldef\tempurl%
\url{https://doi.org/10.1609/aaai.v34i01.5387}
\showDOI{\tempurl}


\bibitem[Cloke and Pappenberger(2009)]%
        {1-46}
\bibfield{author}{\bibinfo{person}{H.L. Cloke} {and} \bibinfo{person}{F. Pappenberger}.} \bibinfo{year}{2009}\natexlab{}.
\newblock \showarticletitle{Ensemble flood forecasting: A review}.
\newblock \bibinfo{journal}{\emph{Journal of Hydrology}} \bibinfo{volume}{375}, \bibinfo{number}{3-4} (\bibinfo{date}{sep} \bibinfo{year}{2009}), \bibinfo{pages}{613--626}.
\newblock
\urldef\tempurl%
\url{https://doi.org/10.1016/j.jhydrol.2009.06.005}
\showDOI{\tempurl}


\bibitem[Coglianese and Nash(2016)]%
        {1-47}
\bibfield{author}{\bibinfo{person}{Cary Coglianese} {and} \bibinfo{person}{Jennifer Nash}.} \bibinfo{year}{2016}\natexlab{}.
\newblock \showarticletitle{Motivating without mandates? The role of voluntary programs in environmental governance}.
\newblock In \bibinfo{booktitle}{\emph{Decision Making in Environmental Law}}. \bibinfo{publisher}{Edward Elgar Publishing}, \bibinfo{pages}{237--252}.
\newblock
\urldef\tempurl%
\url{https://doi.org/10.4337/9781783478408.ii.18}
\showDOI{\tempurl}


\bibitem[Compton et~al\mbox{.}(2014)]%
        {1-48}
\bibfield{author}{\bibinfo{person}{Ryan Compton}, \bibinfo{person}{Craig Lee}, \bibinfo{person}{Jiejun Xu}, \bibinfo{person}{Luis Artieda-Moncada}, \bibinfo{person}{Tsai-Ching Lu}, \bibinfo{person}{Lalindra~De Silva}, {and} \bibinfo{person}{Michael Macy}.} \bibinfo{year}{2014}\natexlab{}.
\newblock \showarticletitle{Using publicly visible social media to build detailed forecasts of civil unrest}.
\newblock \bibinfo{journal}{\emph{Security Informatics}} \bibinfo{volume}{3}, \bibinfo{number}{1} (\bibinfo{date}{sep} \bibinfo{year}{2014}).
\newblock
\urldef\tempurl%
\url{https://doi.org/10.1186/s13388-014-0004-6}
\showDOI{\tempurl}


\bibitem[Cui et~al\mbox{.}(2013)]%
        {46}
\bibfield{author}{\bibinfo{person}{Peng Cui}, \bibinfo{person}{Shifei Jin}, \bibinfo{person}{Linyun Yu}, \bibinfo{person}{Fei Wang}, \bibinfo{person}{Wenwu Zhu}, {and} \bibinfo{person}{Shiqiang Yang}.} \bibinfo{year}{2013}\natexlab{}.
\newblock \showarticletitle{Cascading outbreak prediction in networks: a data-driven approach}. In \bibinfo{booktitle}{\emph{Proceedings of the 19th ACM SIGKDD international conference on Knowledge discovery and data mining}}. \bibinfo{pages}{901--909}.
\newblock


\bibitem[Damaschke et~al\mbox{.}(2017)]%
        {1-55}
\bibfield{author}{\bibinfo{person}{Magret Damaschke}, \bibinfo{person}{Shane~J. Cronin}, {and} \bibinfo{person}{Mark~S. Bebbington}.} \bibinfo{year}{2017}\natexlab{}.
\newblock \showarticletitle{A volcanic event forecasting model for multiple tephra records, demonstrated on Mt. Taranaki, New Zealand}.
\newblock \bibinfo{journal}{\emph{Bulletin of Volcanology}} \bibinfo{volume}{80}, \bibinfo{number}{1} (\bibinfo{date}{12} \bibinfo{year}{2017}).
\newblock
\urldef\tempurl%
\url{https://doi.org/10.1007/s00445-017-1184-y}
\showDOI{\tempurl}


\bibitem[Dao et~al\mbox{.}(2018)]%
        {4-6}
\bibfield{author}{\bibinfo{person}{Minh-Son Dao}, \bibinfo{person}{Pham Quang~Nhat Minh}, \bibinfo{person}{Asem Kasem}, {and} \bibinfo{person}{Mohamed Saleem~Haja Nazmudeen}.} \bibinfo{year}{2018}\natexlab{}.
\newblock \showarticletitle{A Context-Aware Late-Fusion Approach for Disaster Image Retrieval from Social Media}. In \bibinfo{booktitle}{\emph{Proceedings of the 2018 {ACM} on International Conference on Multimedia Retrieval}}. \bibinfo{publisher}{{ACM}}.
\newblock
\urldef\tempurl%
\url{https://doi.org/10.1145/3206025.3206047}
\showDOI{\tempurl}


\bibitem[Decroos et~al\mbox{.}(2017)]%
        {1-60}
\bibfield{author}{\bibinfo{person}{Tom Decroos}, \bibinfo{person}{Vladimir Dzyuba}, \bibinfo{person}{Jan~Van Haaren}, {and} \bibinfo{person}{Jesse Davis}.} \bibinfo{year}{2017}\natexlab{}.
\newblock \showarticletitle{Predicting Soccer Highlights from Spatio-Temporal Match Event Streams}.
\newblock \bibinfo{journal}{\emph{Proceedings of the {AAAI} Conference on Artificial Intelligence}} \bibinfo{volume}{31}, \bibinfo{number}{1} (\bibinfo{date}{feb} \bibinfo{year}{2017}).
\newblock
\urldef\tempurl%
\url{https://doi.org/10.1609/aaai.v31i1.10754}
\showDOI{\tempurl}


\bibitem[Deep et~al\mbox{.}(2020)]%
        {1-61}
\bibfield{author}{\bibinfo{person}{Akash Deep}, \bibinfo{person}{Dharmaraj Veeramani}, {and} \bibinfo{person}{Shiyu Zhou}.} \bibinfo{year}{2020}\natexlab{}.
\newblock \showarticletitle{Event Prediction for Individual Unit Based on Recurrent Event Data Collected in Teleservice Systems}.
\newblock \bibinfo{journal}{\emph{{IEEE} Transactions on Reliability}} \bibinfo{volume}{69}, \bibinfo{number}{1} (\bibinfo{date}{mar} \bibinfo{year}{2020}), \bibinfo{pages}{216--227}.
\newblock
\urldef\tempurl%
\url{https://doi.org/10.1109/tr.2019.2909471}
\showDOI{\tempurl}


\bibitem[Di et~al\mbox{.}(2019)]%
        {1-64}
\bibfield{author}{\bibinfo{person}{Xiaolei Di}, \bibinfo{person}{Yu Xiao}, \bibinfo{person}{Chao Zhu}, \bibinfo{person}{Yang Deng}, \bibinfo{person}{Qinpei Zhao}, {and} \bibinfo{person}{Weixiong Rao}.} \bibinfo{year}{2019}\natexlab{}.
\newblock \showarticletitle{Traffic Congestion Prediction by Spatiotemporal Propagation Patterns}. In \bibinfo{booktitle}{\emph{2019 20th {IEEE} International Conference on Mobile Data Management ({MDM})}}. \bibinfo{publisher}{{IEEE}}.
\newblock
\urldef\tempurl%
\url{https://doi.org/10.1109/mdm.2019.00-45}
\showDOI{\tempurl}


\bibitem[Ding et~al\mbox{.}(2021)]%
        {iot_5}
\bibfield{author}{\bibinfo{person}{Yi Ding}, \bibinfo{person}{Baoshen Guo}, \bibinfo{person}{Lin Zheng}, \bibinfo{person}{Mingming Lu}, \bibinfo{person}{Desheng Zhang}, \bibinfo{person}{Shuai Wang}, \bibinfo{person}{Sang~Hyuk Son}, {and} \bibinfo{person}{Tian He}.} \bibinfo{year}{2021}\natexlab{}.
\newblock \showarticletitle{A City-Wide Crowdsourcing Delivery System with Reinforcement Learning}.
\newblock \bibinfo{journal}{\emph{Proceedings of the {ACM} on Interactive, Mobile, Wearable and Ubiquitous Technologies}} \bibinfo{volume}{5}, \bibinfo{number}{3} (\bibinfo{date}{sep} \bibinfo{year}{2021}), \bibinfo{pages}{1--22}.
\newblock
\urldef\tempurl%
\url{https://doi.org/10.1145/3478117}
\showDOI{\tempurl}


\bibitem[Ding et~al\mbox{.}(2018)]%
        {1-66}
\bibfield{author}{\bibinfo{person}{Zuohua Ding}, \bibinfo{person}{Yuan Zhou}, \bibinfo{person}{Geguang Pu}, {and} \bibinfo{person}{MengChu Zhou}.} \bibinfo{year}{2018}\natexlab{}.
\newblock \showarticletitle{Online Failure Prediction for Railway Transportation Systems Based on Fuzzy Rules and Data Analysis}.
\newblock \bibinfo{journal}{\emph{{IEEE} Transactions on Reliability}} \bibinfo{volume}{67}, \bibinfo{number}{3} (\bibinfo{date}{sep} \bibinfo{year}{2018}), \bibinfo{pages}{1143--1158}.
\newblock
\urldef\tempurl%
\url{https://doi.org/10.1109/tr.2018.2828113}
\showDOI{\tempurl}


\bibitem[Do et~al\mbox{.}(2015)]%
        {1-103}
\bibfield{author}{\bibinfo{person}{Quynh Ngoc~Thi Do}, \bibinfo{person}{Steven Bethard}, {and} \bibinfo{person}{Marie-Francine Moens}.} \bibinfo{year}{2015}\natexlab{}.
\newblock \showarticletitle{Adapting Coreference Resolution for Narrative Processing}. In \bibinfo{booktitle}{\emph{Proceedings of the 2015 Conference on Empirical Methods in Natural Language Processing}}. \bibinfo{publisher}{Association for Computational Linguistics}.
\newblock
\urldef\tempurl%
\url{https://doi.org/10.18653/v1/d15-1271}
\showDOI{\tempurl}


\bibitem[Doswell et~al\mbox{.}(1993)]%
        {1-68}
\bibfield{author}{\bibinfo{person}{Charles~A. Doswell}, \bibinfo{person}{Steven~J. Weiss}, {and} \bibinfo{person}{Robert~H. Johns}.} \bibinfo{year}{1993}\natexlab{}.
\newblock \showarticletitle{Tornado forecasting: A review}.
\newblock In \bibinfo{booktitle}{\emph{Geophysical Monograph Series}}. \bibinfo{publisher}{American Geophysical Union}, \bibinfo{pages}{557--571}.
\newblock
\urldef\tempurl%
\url{https://doi.org/10.1029/gm079p0557}
\showDOI{\tempurl}


\bibitem[Dotel et~al\mbox{.}(2020a)]%
        {2-47}
\bibfield{author}{\bibinfo{person}{Saramsha Dotel}, \bibinfo{person}{Avishekh Shrestha}, \bibinfo{person}{Anish Bhusal}, \bibinfo{person}{Ramesh Pathak}, \bibinfo{person}{Aman Shakya}, {and} \bibinfo{person}{Sanjeeb~Prasad Panday}.} \bibinfo{year}{2020}\natexlab{a}.
\newblock \showarticletitle{Disaster Assessment from Satellite Imagery by Analysing Topographical Features Using Deep Learning}. In \bibinfo{booktitle}{\emph{Proceedings of the 2020 2nd International Conference on Image, Video and Signal Processing}}. \bibinfo{publisher}{{ACM}}.
\newblock
\urldef\tempurl%
\url{https://doi.org/10.1145/3388818.3389160}
\showDOI{\tempurl}


\bibitem[Dotel et~al\mbox{.}(2020b)]%
        {2-57}
\bibfield{author}{\bibinfo{person}{Saramsha Dotel}, \bibinfo{person}{Avishekh Shrestha}, \bibinfo{person}{Anish Bhusal}, \bibinfo{person}{Ramesh Pathak}, \bibinfo{person}{Aman Shakya}, {and} \bibinfo{person}{Sanjeeb~Prasad Panday}.} \bibinfo{year}{2020}\natexlab{b}.
\newblock \showarticletitle{Disaster Assessment from Satellite Imagery by Analysing Topographical Features Using Deep Learning}. In \bibinfo{booktitle}{\emph{Proceedings of the 2020 2nd International Conference on Image, Video and Signal Processing}}. \bibinfo{publisher}{{ACM}}.
\newblock
\urldef\tempurl%
\url{https://doi.org/10.1145/3388818.3389160}
\showDOI{\tempurl}


\bibitem[Duan et~al\mbox{.}(2020)]%
        {1-70}
\bibfield{author}{\bibinfo{person}{Huilong Duan}, \bibinfo{person}{Zhoujian Sun}, \bibinfo{person}{Wei Dong}, \bibinfo{person}{Kunlun He}, {and} \bibinfo{person}{Zhengxing Huang}.} \bibinfo{year}{2020}\natexlab{}.
\newblock \showarticletitle{On Clinical Event Prediction in Patient Treatment Trajectory Using Longitudinal Electronic Health Records}.
\newblock \bibinfo{journal}{\emph{{IEEE} Journal of Biomedical and Health Informatics}} \bibinfo{volume}{24}, \bibinfo{number}{7} (\bibinfo{date}{jul} \bibinfo{year}{2020}), \bibinfo{pages}{2053--2063}.
\newblock
\urldef\tempurl%
\url{https://doi.org/10.1109/jbhi.2019.2962079}
\showDOI{\tempurl}


\bibitem[ElRefai et~al\mbox{.}(2022)]%
        {1-49}
\bibfield{author}{\bibinfo{person}{Mohamed ElRefai}, \bibinfo{person}{Mohamed Abouelasaad}, \bibinfo{person}{Benedict~M. Wiles}, \bibinfo{person}{Anthony~J. Dunn}, \bibinfo{person}{Stefano Coniglio}, \bibinfo{person}{Alain~B. Zemkoho}, {and} \bibinfo{person}{Paul~R. Roberts}.} \bibinfo{year}{2022}\natexlab{}.
\newblock \showarticletitle{Deep learning-based insights on T:R ratio behaviour during prolonged screening for S-{ICD} eligibility}.
\newblock \bibinfo{journal}{\emph{Journal of Interventional Cardiac Electrophysiology}} (\bibinfo{date}{may} \bibinfo{year}{2022}).
\newblock
\urldef\tempurl%
\url{https://doi.org/10.1007/s10840-022-01245-6}
\showDOI{\tempurl}


\bibitem[Fang et~al\mbox{.}(2020)]%
        {iot_8}
\bibfield{author}{\bibinfo{person}{Zhihan Fang}, \bibinfo{person}{Guang Wang}, \bibinfo{person}{Shuai Wang}, \bibinfo{person}{Chaoji Zuo}, \bibinfo{person}{Fan Zhang}, {and} \bibinfo{person}{Desheng Zhang}.} \bibinfo{year}{2020}\natexlab{}.
\newblock \showarticletitle{{CellRep}: Usage Representativeness Modeling and Correction Based on Multiple City-Scale Cellular Networks}. In \bibinfo{booktitle}{\emph{Proceedings of The Web Conference 2020}}. \bibinfo{publisher}{{ACM}}.
\newblock
\urldef\tempurl%
\url{https://doi.org/10.1145/3366423.3380141}
\showDOI{\tempurl}


\bibitem[Fronza et~al\mbox{.}(2013)]%
        {1-79}
\bibfield{author}{\bibinfo{person}{Ilenia Fronza}, \bibinfo{person}{Alberto Sillitti}, \bibinfo{person}{Giancarlo Succi}, \bibinfo{person}{Mikko Terho}, {and} \bibinfo{person}{Jelena Vlasenko}.} \bibinfo{year}{2013}\natexlab{}.
\newblock \showarticletitle{Failure prediction based on log files using Random Indexing and Support Vector Machines}.
\newblock \bibinfo{journal}{\emph{Journal of Systems and Software}} \bibinfo{volume}{86}, \bibinfo{number}{1} (\bibinfo{date}{jan} \bibinfo{year}{2013}), \bibinfo{pages}{2--11}.
\newblock
\urldef\tempurl%
\url{https://doi.org/10.1016/j.jss.2012.06.025}
\showDOI{\tempurl}


\bibitem[Fülöp et~al\mbox{.}(2012)]%
        {1-81}
\bibfield{author}{\bibinfo{person}{Lajos~Jen{\H{o}} Fülöp}, \bibinfo{person}{{\'{A}}rp{\'{a}}d Besz{\'{e}}des}, \bibinfo{person}{Gabriella T{\'{o}}th}, \bibinfo{person}{Hunor Demeter}, \bibinfo{person}{L{\'{a}}szl{\'{o}} Vid{\'{a}}cs}, {and} \bibinfo{person}{L{\'{o}}r{\'{a}}nt Farkas}.} \bibinfo{year}{2012}\natexlab{}.
\newblock \showarticletitle{Predictive complex event processing}. In \bibinfo{booktitle}{\emph{Proceedings of the Fifth Balkan Conference in Informatics}}. \bibinfo{publisher}{{ACM}}.
\newblock
\urldef\tempurl%
\url{https://doi.org/10.1145/2371316.2371323}
\showDOI{\tempurl}


\bibitem[Gallego-Castillo et~al\mbox{.}(2015)]%
        {1-83}
\bibfield{author}{\bibinfo{person}{Cristobal Gallego-Castillo}, \bibinfo{person}{Alvaro Cuerva-Tejero}, {and} \bibinfo{person}{Oscar Lopez-Garcia}.} \bibinfo{year}{2015}\natexlab{}.
\newblock \showarticletitle{A review on the recent history of wind power ramp forecasting}.
\newblock \bibinfo{journal}{\emph{Renewable and Sustainable Energy Reviews}}  \bibinfo{volume}{52} (\bibinfo{date}{dec} \bibinfo{year}{2015}), \bibinfo{pages}{1148--1157}.
\newblock
\urldef\tempurl%
\url{https://doi.org/10.1016/j.rser.2015.07.154}
\showDOI{\tempurl}


\bibitem[Galuba et~al\mbox{.}(2010)]%
        {58}
\bibfield{author}{\bibinfo{person}{Wojciech Galuba}, \bibinfo{person}{Karl Aberer}, \bibinfo{person}{Dipanjan Chakraborty}, \bibinfo{person}{Zoran Despotovic}, {and} \bibinfo{person}{Wolfgang Kellerer}.} \bibinfo{year}{2010}\natexlab{}.
\newblock \showarticletitle{Outtweeting the twitterers-predicting information cascades in microblogs.}
\newblock \bibinfo{journal}{\emph{WOSN}}  \bibinfo{volume}{10} (\bibinfo{year}{2010}), \bibinfo{pages}{3--11}.
\newblock


\bibitem[Gao et~al\mbox{.}(2014a)]%
        {60}
\bibfield{author}{\bibinfo{person}{Shuai Gao}, \bibinfo{person}{Jun Ma}, {and} \bibinfo{person}{Zhumin Chen}.} \bibinfo{year}{2014}\natexlab{a}.
\newblock \showarticletitle{Effective and effortless features for popularity prediction in microblogging network}. In \bibinfo{booktitle}{\emph{Proceedings of the 23rd International Conference on World Wide Web}}. \bibinfo{pages}{269--270}.
\newblock


\bibitem[Gao et~al\mbox{.}(2014b)]%
        {61}
\bibfield{author}{\bibinfo{person}{Shuai Gao}, \bibinfo{person}{Jun Ma}, {and} \bibinfo{person}{Zhumin Chen}.} \bibinfo{year}{2014}\natexlab{b}.
\newblock \showarticletitle{Popularity prediction in microblogging network}. In \bibinfo{booktitle}{\emph{Web Technologies and Applications: 16th Asia-Pacific Web Conference, APWeb 2014, Changsha, China, September 5-7, 2014. Proceedings 16}}. Springer, \bibinfo{pages}{379--390}.
\newblock


\bibitem[Gao et~al\mbox{.}(2019a)]%
        {63}
\bibfield{author}{\bibinfo{person}{Xiaofeng Gao}, \bibinfo{person}{Zhenhao Cao}, \bibinfo{person}{Sha Li}, \bibinfo{person}{Bin Yao}, \bibinfo{person}{Guihai Chen}, {and} \bibinfo{person}{Shaojie Tang}.} \bibinfo{year}{2019}\natexlab{a}.
\newblock \showarticletitle{Taxonomy and evaluation for microblog popularity prediction}.
\newblock \bibinfo{journal}{\emph{ACM Transactions on Knowledge Discovery from Data (TKDD)}} \bibinfo{volume}{13}, \bibinfo{number}{2} (\bibinfo{year}{2019}), \bibinfo{pages}{1--40}.
\newblock


\bibitem[Gao and Zhao(2018)]%
        {1-84}
\bibfield{author}{\bibinfo{person}{Yuyang Gao} {and} \bibinfo{person}{Liang Zhao}.} \bibinfo{year}{2018}\natexlab{}.
\newblock \showarticletitle{Incomplete Label Multi-Task Ordinal Regression for Spatial Event Scale Forecasting}.
\newblock \bibinfo{journal}{\emph{Proceedings of the {AAAI} Conference on Artificial Intelligence}} \bibinfo{volume}{32}, \bibinfo{number}{1} (\bibinfo{date}{apr} \bibinfo{year}{2018}).
\newblock
\urldef\tempurl%
\url{https://doi.org/10.1609/aaai.v32i1.11748}
\showDOI{\tempurl}


\bibitem[Gao et~al\mbox{.}(2019b)]%
        {1-85}
\bibfield{author}{\bibinfo{person}{Yuyang Gao}, \bibinfo{person}{Liang Zhao}, \bibinfo{person}{Lingfei Wu}, \bibinfo{person}{Yanfang Ye}, \bibinfo{person}{Hui Xiong}, {and} \bibinfo{person}{Chaowei Yang}.} \bibinfo{year}{2019}\natexlab{b}.
\newblock \showarticletitle{Incomplete Label Multi-Task Deep Learning for Spatio-Temporal Event Subtype Forecasting}.
\newblock \bibinfo{journal}{\emph{Proceedings of the {AAAI} Conference on Artificial Intelligence}} \bibinfo{volume}{33}, \bibinfo{number}{01} (\bibinfo{date}{jul} \bibinfo{year}{2019}), \bibinfo{pages}{3638--3646}.
\newblock
\urldef\tempurl%
\url{https://doi.org/10.1609/aaai.v33i01.33013638}
\showDOI{\tempurl}


\bibitem[Garg et~al\mbox{.}(2020)]%
        {4-10}
\bibfield{author}{\bibinfo{person}{Tanmay Garg}, \bibinfo{person}{Mamta Garg}, \bibinfo{person}{Om~Prakash Mahela}, {and} \bibinfo{person}{Akhil~Ranjan Garg}.} \bibinfo{year}{2020}\natexlab{}.
\newblock \showarticletitle{Convolutional Neural Networks with Transfer Learning for Recognition of {COVID}-19: A Comparative Study of Different Approaches}.
\newblock \bibinfo{journal}{\emph{{AI}}} \bibinfo{volume}{1}, \bibinfo{number}{4} (\bibinfo{date}{dec} \bibinfo{year}{2020}), \bibinfo{pages}{586--606}.
\newblock
\urldef\tempurl%
\url{https://doi.org/10.3390/ai1040034}
\showDOI{\tempurl}


\bibitem[Ghil et~al\mbox{.}(2011)]%
        {1-86}
\bibfield{author}{\bibinfo{person}{M. Ghil}, \bibinfo{person}{P. Yiou}, \bibinfo{person}{S. Hallegatte}, \bibinfo{person}{B.~D. Malamud}, \bibinfo{person}{P. Naveau}, \bibinfo{person}{A. Soloviev}, \bibinfo{person}{P. Friederichs}, \bibinfo{person}{V. Keilis-Borok}, \bibinfo{person}{D. Kondrashov}, \bibinfo{person}{V. Kossobokov}, \bibinfo{person}{O. Mestre}, \bibinfo{person}{C. Nicolis}, \bibinfo{person}{H.~W. Rust}, \bibinfo{person}{P. Shebalin}, \bibinfo{person}{M. Vrac}, \bibinfo{person}{A. Witt}, {and} \bibinfo{person}{I. Zaliapin}.} \bibinfo{year}{2011}\natexlab{}.
\newblock \showarticletitle{Extreme events: dynamics, statistics and prediction}.
\newblock \bibinfo{journal}{\emph{Nonlinear Processes in Geophysics}} \bibinfo{volume}{18}, \bibinfo{number}{3} (\bibinfo{date}{may} \bibinfo{year}{2011}), \bibinfo{pages}{295--350}.
\newblock
\urldef\tempurl%
\url{https://doi.org/10.5194/npg-18-295-2011}
\showDOI{\tempurl}


\bibitem[Gopnarayan and Deshpande(2020)]%
        {4-20}
\bibfield{author}{\bibinfo{person}{Archana Gopnarayan} {and} \bibinfo{person}{Sachin Deshpande}.} \bibinfo{year}{2020}\natexlab{}.
\newblock \showarticletitle{Tweets Analysis for Disaster Management: Preparedness, Emergency Response, Impact, and Recovery}.
\newblock In \bibinfo{booktitle}{\emph{Innovative Data Communication Technologies and Application}}. \bibinfo{publisher}{Springer International Publishing}, \bibinfo{pages}{760--764}.
\newblock
\urldef\tempurl%
\url{https://doi.org/10.1007/978-3-030-38040-3_87}
\showDOI{\tempurl}


\bibitem[Granroth-Wilding and Clark(2016)]%
        {1-89}
\bibfield{author}{\bibinfo{person}{Mark Granroth-Wilding} {and} \bibinfo{person}{Stephen Clark}.} \bibinfo{year}{2016}\natexlab{}.
\newblock \showarticletitle{What Happens Next? Event Prediction Using a Compositional Neural Network Model}.
\newblock \bibinfo{journal}{\emph{Proceedings of the {AAAI} Conference on Artificial Intelligence}} \bibinfo{volume}{30}, \bibinfo{number}{1} (\bibinfo{date}{mar} \bibinfo{year}{2016}).
\newblock
\urldef\tempurl%
\url{https://doi.org/10.1609/aaai.v30i1.10344}
\showDOI{\tempurl}


\bibitem[Gulmezoglu et~al\mbox{.}(2019)]%
        {1-90}
\bibfield{author}{\bibinfo{person}{Berk Gulmezoglu}, \bibinfo{person}{Andreas Zankl}, \bibinfo{person}{M.~Caner Tol}, \bibinfo{person}{Saad Islam}, \bibinfo{person}{Thomas Eisenbarth}, {and} \bibinfo{person}{Berk Sunar}.} \bibinfo{year}{2019}\natexlab{}.
\newblock \showarticletitle{Undermining User Privacy on Mobile Devices Using {AI}}. In \bibinfo{booktitle}{\emph{Proceedings of the 2019 {ACM} Asia Conference on Computer and Communications Security}}. \bibinfo{publisher}{{ACM}}.
\newblock
\urldef\tempurl%
\url{https://doi.org/10.1145/3321705.3329804}
\showDOI{\tempurl}


\bibitem[G{\"u}rsun et~al\mbox{.}(2011)]%
        {74}
\bibfield{author}{\bibinfo{person}{Gonca G{\"u}rsun}, \bibinfo{person}{Mark Crovella}, {and} \bibinfo{person}{Ibrahim Matta}.} \bibinfo{year}{2011}\natexlab{}.
\newblock \showarticletitle{Describing and forecasting video access patterns}. In \bibinfo{booktitle}{\emph{2011 proceedings IEEE infocom}}. IEEE, \bibinfo{pages}{16--20}.
\newblock


\bibitem[Hagenau et~al\mbox{.}(2012)]%
        {1-91}
\bibfield{author}{\bibinfo{person}{Michael Hagenau}, \bibinfo{person}{Michael Liebmann}, \bibinfo{person}{Markus Hedwig}, {and} \bibinfo{person}{Dirk Neumann}.} \bibinfo{year}{2012}\natexlab{}.
\newblock \showarticletitle{Automated News Reading: Stock Price Prediction Based on Financial News Using Context-Specific Features}. In \bibinfo{booktitle}{\emph{2012 45th Hawaii International Conference on System Sciences}}. \bibinfo{publisher}{{IEEE}}.
\newblock
\urldef\tempurl%
\url{https://doi.org/10.1109/hicss.2012.129}
\showDOI{\tempurl}


\bibitem[Han et~al\mbox{.}(2012)]%
        {1-92}
\bibfield{author}{\bibinfo{person}{Jiawei Han}, \bibinfo{person}{Micheline Kamber}, {and} \bibinfo{person}{Jian Pei}.} \bibinfo{year}{2012}\natexlab{}.
\newblock \showarticletitle{Data Preprocessing}.
\newblock In \bibinfo{booktitle}{\emph{Data Mining}}. \bibinfo{publisher}{Elsevier}, \bibinfo{pages}{83--124}.
\newblock
\urldef\tempurl%
\url{https://doi.org/10.1016/b978-0-12-381479-1.00003-4}
\showDOI{\tempurl}


\bibitem[Hao et~al\mbox{.}(2019)]%
        {1-93}
\bibfield{author}{\bibinfo{person}{Mengmeng Hao}, \bibinfo{person}{Dong Jiang}, \bibinfo{person}{Fangyu Ding}, \bibinfo{person}{Jingying Fu}, {and} \bibinfo{person}{Shuai Chen}.} \bibinfo{year}{2019}\natexlab{}.
\newblock \showarticletitle{Simulating Spatio-Temporal Patterns of Terrorism Incidents on the Indochina Peninsula with {GIS} and the Random Forest Method}.
\newblock \bibinfo{journal}{\emph{{ISPRS} International Journal of Geo-Information}} \bibinfo{volume}{8}, \bibinfo{number}{3} (\bibinfo{date}{mar} \bibinfo{year}{2019}), \bibinfo{pages}{133}.
\newblock
\urldef\tempurl%
\url{https://doi.org/10.3390/ijgi8030133}
\showDOI{\tempurl}


\bibitem[He et~al\mbox{.}(2023)]%
        {3-2}
\bibfield{author}{\bibinfo{person}{Jia He}, \bibinfo{person}{Miao Ma}, \bibinfo{person}{Yuxuan Zhou}, {and} \bibinfo{person}{Miaoke Wang}.} \bibinfo{year}{2023}\natexlab{}.
\newblock \showarticletitle{What We Have Learned about the Characteristics and Differences of Disaster Information Behavior in Social Media{\textemdash}A Case Study of the 7.20 Henan Heavy Rain Flood Disaster}.
\newblock \bibinfo{journal}{\emph{Sustainability}} \bibinfo{volume}{15}, \bibinfo{number}{6} (\bibinfo{date}{mar} \bibinfo{year}{2023}), \bibinfo{pages}{4726}.
\newblock
\urldef\tempurl%
\url{https://doi.org/10.3390/su15064726}
\showDOI{\tempurl}


\bibitem[Heaton(2017)]%
        {1-88}
\bibfield{author}{\bibinfo{person}{Jeff Heaton}.} \bibinfo{year}{2017}\natexlab{}.
\newblock \showarticletitle{Ian Goodfellow, Yoshua Bengio, and Aaron Courville: Deep learning}.
\newblock \bibinfo{journal}{\emph{Genetic Programming and Evolvable Machines}} \bibinfo{volume}{19}, \bibinfo{number}{1-2} (\bibinfo{date}{oct} \bibinfo{year}{2017}), \bibinfo{pages}{305--307}.
\newblock
\urldef\tempurl%
\url{https://doi.org/10.1007/s10710-017-9314-z}
\showDOI{\tempurl}


\bibitem[Hernandez-Suarez et~al\mbox{.}(2019)]%
        {4-11}
\bibfield{author}{\bibinfo{person}{Aldo Hernandez-Suarez}, \bibinfo{person}{Gabriel Sanchez-Perez}, \bibinfo{person}{Karina Toscano-Medina}, \bibinfo{person}{Hector Perez-Meana}, \bibinfo{person}{Jose Portillo-Portillo}, \bibinfo{person}{Victor Sanchez}, {and} \bibinfo{person}{Luis~Garc{\'{\i}}a Villalba}.} \bibinfo{year}{2019}\natexlab{}.
\newblock \showarticletitle{Using Twitter Data to Monitor Natural Disaster Social Dynamics: A Recurrent Neural Network Approach with Word Embeddings and Kernel Density Estimation}.
\newblock \bibinfo{journal}{\emph{Sensors}} \bibinfo{volume}{19}, \bibinfo{number}{7} (\bibinfo{date}{apr} \bibinfo{year}{2019}), \bibinfo{pages}{1746}.
\newblock
\urldef\tempurl%
\url{https://doi.org/10.3390/s19071746}
\showDOI{\tempurl}


\bibitem[Hoegh et~al\mbox{.}(2015)]%
        {1-95}
\bibfield{author}{\bibinfo{person}{Andrew Hoegh}, \bibinfo{person}{Scotland Leman}, \bibinfo{person}{Parang Saraf}, {and} \bibinfo{person}{Naren Ramakrishnan}.} \bibinfo{year}{2015}\natexlab{}.
\newblock \showarticletitle{Bayesian Model Fusion for Forecasting Civil Unrest}.
\newblock \bibinfo{journal}{\emph{Technometrics}} \bibinfo{volume}{57}, \bibinfo{number}{3} (\bibinfo{date}{feb} \bibinfo{year}{2015}), \bibinfo{pages}{332--340}.
\newblock
\urldef\tempurl%
\url{https://doi.org/10.1080/00401706.2014.1001522}
\showDOI{\tempurl}


\bibitem[Hong et~al\mbox{.}(2011)]%
        {81}
\bibfield{author}{\bibinfo{person}{Liangjie Hong}, \bibinfo{person}{Ovidiu Dan}, {and} \bibinfo{person}{Brian~D Davison}.} \bibinfo{year}{2011}\natexlab{}.
\newblock \showarticletitle{Predicting popular messages in twitter}. In \bibinfo{booktitle}{\emph{Proceedings of the 20th international conference companion on World wide web}}. \bibinfo{pages}{57--58}.
\newblock


\bibitem[Hsu et~al\mbox{.}(2013)]%
        {4-22}
\bibfield{author}{\bibinfo{person}{Edbert~B. Hsu}, \bibinfo{person}{Yang Li}, \bibinfo{person}{Jamil~D. Bayram}, \bibinfo{person}{David Levinson}, \bibinfo{person}{Samuel Yang}, {and} \bibinfo{person}{Colleen Monahan}.} \bibinfo{year}{2013}\natexlab{}.
\newblock \showarticletitle{State of Virtual Reality Based Disaster Preparedness and Response Training}.
\newblock \bibinfo{journal}{\emph{{PLoS} Currents}} (\bibinfo{year}{2013}).
\newblock
\urldef\tempurl%
\url{https://doi.org/10.1371/currents.dis.1ea2b2e71237d5337fa53982a38b2aff}
\showDOI{\tempurl}


\bibitem[Hu(2020)]%
        {1-96}
\bibfield{author}{\bibinfo{person}{Linmei Hu}.} \bibinfo{year}{2020}\natexlab{}.
\newblock \showarticletitle{Integrating Hierarchical Attentions for Future Subevent Prediction}.
\newblock \bibinfo{journal}{\emph{{IEEE} Access}}  \bibinfo{volume}{8} (\bibinfo{year}{2020}), \bibinfo{pages}{3106--3114}.
\newblock
\urldef\tempurl%
\url{https://doi.org/10.1109/access.2019.2961973}
\showDOI{\tempurl}


\bibitem[Hu et~al\mbox{.}(2017)]%
        {1-97}
\bibfield{author}{\bibinfo{person}{Linmei Hu}, \bibinfo{person}{Juanzi Li}, \bibinfo{person}{Liqiang Nie}, \bibinfo{person}{Xiao-Li Li}, {and} \bibinfo{person}{Chao Shao}.} \bibinfo{year}{2017}\natexlab{}.
\newblock \showarticletitle{What Happens Next? Future Subevent Prediction Using Contextual Hierarchical {LSTM}}.
\newblock \bibinfo{journal}{\emph{Proceedings of the {AAAI} Conference on Artificial Intelligence}} \bibinfo{volume}{31}, \bibinfo{number}{1} (\bibinfo{date}{feb} \bibinfo{year}{2017}).
\newblock
\urldef\tempurl%
\url{https://doi.org/10.1609/aaai.v31i1.11001}
\showDOI{\tempurl}


\bibitem[Huang and yang Xiang(2018)]%
        {3-10}
\bibfield{author}{\bibinfo{person}{Lu Huang} {and} \bibinfo{person}{Lu yang Xiang}.} \bibinfo{year}{2018}\natexlab{}.
\newblock \showarticletitle{Method for Meteorological Early Warning of Precipitation-Induced Landslides Based on Deep Neural Network}.
\newblock \bibinfo{journal}{\emph{Neural Processing Letters}} \bibinfo{volume}{48}, \bibinfo{number}{2} (\bibinfo{date}{jan} \bibinfo{year}{2018}), \bibinfo{pages}{1243--1260}.
\newblock
\urldef\tempurl%
\url{https://doi.org/10.1007/s11063-017-9778-0}
\showDOI{\tempurl}


\bibitem[Hürriyetoǧlu et~al\mbox{.}(2017)]%
        {1-101}
\bibfield{author}{\bibinfo{person}{Ali Hürriyetoǧlu}, \bibinfo{person}{Nelleke Oostdijk}, {and} \bibinfo{person}{Antal van~den Bosch}.} \bibinfo{year}{2017}\natexlab{}.
\newblock \showarticletitle{Estimating Time to Event of Future Events Based on Linguistic Cues on Twitter}.
\newblock In \bibinfo{booktitle}{\emph{Intelligent Natural Language Processing: Trends and Applications}}. \bibinfo{publisher}{Springer International Publishing}, \bibinfo{pages}{67--97}.
\newblock
\urldef\tempurl%
\url{https://doi.org/10.1007/978-3-319-67056-0_5}
\showDOI{\tempurl}


\bibitem[Inceoglu et~al\mbox{.}(2018)]%
        {1-102}
\bibfield{author}{\bibinfo{person}{Fadil Inceoglu}, \bibinfo{person}{Jacob~H. Jeppesen}, \bibinfo{person}{Peter Kongstad}, \bibinfo{person}{N{\'{e}}stor J.~Hern{\'{a}}ndez Marcano}, \bibinfo{person}{Rune~H. Jacobsen}, {and} \bibinfo{person}{Christoffer Karoff}.} \bibinfo{year}{2018}\natexlab{}.
\newblock \showarticletitle{Using Machine Learning Methods to Forecast if Solar Flares Will Be Associated with {CMEs} and {SEPs}}.
\newblock \bibinfo{journal}{\emph{The Astrophysical Journal}} \bibinfo{volume}{861}, \bibinfo{number}{2} (\bibinfo{date}{jul} \bibinfo{year}{2018}), \bibinfo{pages}{128}.
\newblock
\urldef\tempurl%
\url{https://doi.org/10.3847/1538-4357/aac81e}
\showDOI{\tempurl}


\bibitem[Jiang et~al\mbox{.}(2022)]%
        {3-18}
\bibfield{author}{\bibinfo{person}{Nan Jiang}, \bibinfo{person}{Hai-Bo Li}, \bibinfo{person}{Cong-Jiang Li}, \bibinfo{person}{Huai-Xian Xiao}, {and} \bibinfo{person}{Jia-Wen Zhou}.} \bibinfo{year}{2022}\natexlab{}.
\newblock \showarticletitle{A Fusion Method Using Terrestrial Laser Scanning and Unmanned Aerial Vehicle Photogrammetry for Landslide Deformation Monitoring Under Complex Terrain Conditions}.
\newblock \bibinfo{journal}{\emph{{IEEE} Transactions on Geoscience and Remote Sensing}}  \bibinfo{volume}{60} (\bibinfo{year}{2022}), \bibinfo{pages}{1--14}.
\newblock
\urldef\tempurl%
\url{https://doi.org/10.1109/tgrs.2022.3181258}
\showDOI{\tempurl}


\bibitem[Jiang et~al\mbox{.}(2019)]%
        {1-104}
\bibfield{author}{\bibinfo{person}{Renhe Jiang}, \bibinfo{person}{Xuan Song}, \bibinfo{person}{Dou Huang}, \bibinfo{person}{Xiaoya Song}, \bibinfo{person}{Tianqi Xia}, \bibinfo{person}{Zekun Cai}, \bibinfo{person}{Zhaonan Wang}, \bibinfo{person}{Kyoung-Sook Kim}, {and} \bibinfo{person}{Ryosuke Shibasaki}.} \bibinfo{year}{2019}\natexlab{}.
\newblock \showarticletitle{{DeepUrbanEvent}}. In \bibinfo{booktitle}{\emph{Proceedings of the 25th {ACM} {SIGKDD} International Conference on Knowledge Discovery Data Mining}}. \bibinfo{publisher}{{ACM}}.
\newblock
\urldef\tempurl%
\url{https://doi.org/10.1145/3292500.3330654}
\showDOI{\tempurl}


\bibitem[Jiang(2019)]%
        {1-105}
\bibfield{author}{\bibinfo{person}{Zhe Jiang}.} \bibinfo{year}{2019}\natexlab{}.
\newblock \showarticletitle{A Survey on Spatial Prediction Methods}.
\newblock \bibinfo{journal}{\emph{{IEEE} Transactions on Knowledge and Data Engineering}} \bibinfo{volume}{31}, \bibinfo{number}{9} (\bibinfo{date}{sep} \bibinfo{year}{2019}), \bibinfo{pages}{1645--1664}.
\newblock
\urldef\tempurl%
\url{https://doi.org/10.1109/tkde.2018.2866809}
\showDOI{\tempurl}


\bibitem[Jin et~al\mbox{.}(2013)]%
        {1-106}
\bibfield{author}{\bibinfo{person}{Fang Jin}, \bibinfo{person}{Edward Dougherty}, \bibinfo{person}{Parang Saraf}, \bibinfo{person}{Yang Cao}, {and} \bibinfo{person}{Naren Ramakrishnan}.} \bibinfo{year}{2013}\natexlab{}.
\newblock \showarticletitle{Epidemiological modeling of news and rumors on Twitter}. In \bibinfo{booktitle}{\emph{Proceedings of the 7th Workshop on Social Network Mining and Analysis}}. \bibinfo{publisher}{{ACM}}.
\newblock
\urldef\tempurl%
\url{https://doi.org/10.1145/2501025.2501027}
\showDOI{\tempurl}


\bibitem[Jurgens(2021)]%
        {1-107}
\bibfield{author}{\bibinfo{person}{David Jurgens}.} \bibinfo{year}{2021}\natexlab{}.
\newblock \showarticletitle{That{\textquotesingle}s What Friends Are For: Inferring Location in Online Social Media Platforms Based on Social Relationships}.
\newblock \bibinfo{journal}{\emph{Proceedings of the International {AAAI} Conference on Web and Social Media}} \bibinfo{volume}{7}, \bibinfo{number}{1} (\bibinfo{date}{aug} \bibinfo{year}{2021}), \bibinfo{pages}{273--282}.
\newblock
\urldef\tempurl%
\url{https://doi.org/10.1609/icwsm.v7i1.14399}
\showDOI{\tempurl}


\bibitem[Kabir and Madria(2019)]%
        {2-63}
\bibfield{author}{\bibinfo{person}{Md.~Yasin Kabir} {and} \bibinfo{person}{Sanjay Madria}.} \bibinfo{year}{2019}\natexlab{}.
\newblock \showarticletitle{A Deep Learning Approach for Tweet Classification and Rescue Scheduling for Effective Disaster Management}. In \bibinfo{booktitle}{\emph{Proceedings of the 27th {ACM} {SIGSPATIAL} International Conference on Advances in Geographic Information Systems}}. \bibinfo{publisher}{{ACM}}.
\newblock
\urldef\tempurl%
\url{https://doi.org/10.1145/3347146.3359097}
\showDOI{\tempurl}


\bibitem[Kang et~al\mbox{.}(2017)]%
        {1-108}
\bibfield{author}{\bibinfo{person}{Wei Kang}, \bibinfo{person}{Jie Chen}, \bibinfo{person}{Jiuyong Li}, \bibinfo{person}{Jixue Liu}, \bibinfo{person}{Lin Liu}, \bibinfo{person}{Grant Osborne}, \bibinfo{person}{Nick Lothian}, \bibinfo{person}{Brenton Cooper}, \bibinfo{person}{Terry Moschou}, {and} \bibinfo{person}{Grant Neale}.} \bibinfo{year}{2017}\natexlab{}.
\newblock \showarticletitle{Carbon: Forecasting Civil Unrest Events by Monitoring News and Social Media}.
\newblock In \bibinfo{booktitle}{\emph{Advanced Data Mining and Applications}}. \bibinfo{publisher}{Springer International Publishing}, \bibinfo{pages}{859--865}.
\newblock
\urldef\tempurl%
\url{https://doi.org/10.1007/978-3-319-69179-4_62}
\showDOI{\tempurl}


\bibitem[Kattan et~al\mbox{.}(2015)]%
        {1-109}
\bibfield{author}{\bibinfo{person}{Ahmed Kattan}, \bibinfo{person}{Shaheen Fatima}, {and} \bibinfo{person}{Muhammad Arif}.} \bibinfo{year}{2015}\natexlab{}.
\newblock \showarticletitle{Time-series event-based prediction: An unsupervised learning framework based on genetic programming}.
\newblock \bibinfo{journal}{\emph{Information Sciences}}  \bibinfo{volume}{301} (\bibinfo{date}{apr} \bibinfo{year}{2015}), \bibinfo{pages}{99--123}.
\newblock
\urldef\tempurl%
\url{https://doi.org/10.1016/j.ins.2014.12.054}
\showDOI{\tempurl}


\bibitem[Khoo et~al\mbox{.}(2000)]%
        {1-111}
\bibfield{author}{\bibinfo{person}{Christopher S.~G. Khoo}, \bibinfo{person}{Syin Chan}, {and} \bibinfo{person}{Yun Niu}.} \bibinfo{year}{2000}\natexlab{}.
\newblock \showarticletitle{Extracting causal knowledge from a medical database using graphical patterns}. In \bibinfo{booktitle}{\emph{Proceedings of the 38th Annual Meeting on Association for Computational Linguistics - {ACL} {\textquotesingle}00}}. \bibinfo{publisher}{Association for Computational Linguistics}.
\newblock
\urldef\tempurl%
\url{https://doi.org/10.3115/1075218.1075261}
\showDOI{\tempurl}


\bibitem[Kil et~al\mbox{.}(2019)]%
        {4-7}
\bibfield{author}{\bibinfo{person}{Woogeun Kil}, \bibinfo{person}{Kwangpyo Ko}, \bibinfo{person}{Seungwoon Lee}, {and} \bibinfo{person}{Byeong hee Roh}.} \bibinfo{year}{2019}\natexlab{}.
\newblock \showarticletitle{{MR} and {IoT} Convergence Platform with {AI} Support for Disaster Recognition (poster)}. In \bibinfo{booktitle}{\emph{Proceedings of the 17th Annual International Conference on Mobile Systems, Applications, and Services}}. \bibinfo{publisher}{{ACM}}.
\newblock
\urldef\tempurl%
\url{https://doi.org/10.1145/3307334.3328652}
\showDOI{\tempurl}


\bibitem[Kim(1993)]%
        {1-112}
\bibfield{author}{\bibinfo{person}{Jaegwon Kim}.} \bibinfo{year}{1993}\natexlab{}.
\newblock \bibinfo{booktitle}{\emph{Supervenience and Mind}}.
\newblock \bibinfo{publisher}{Cambridge University Press}.
\newblock
\urldef\tempurl%
\url{https://doi.org/10.1017/cbo9780511625220}
\showDOI{\tempurl}


\bibitem[Kleijnen and van Beers(2020)]%
        {1-114}
\bibfield{author}{\bibinfo{person}{Jack P.~C. Kleijnen} {and} \bibinfo{person}{Wim C.~M. van Beers}.} \bibinfo{year}{2020}\natexlab{}.
\newblock \showarticletitle{Prediction for Big Data Through Kriging: Small Sequential and One-Shot Designs}.
\newblock \bibinfo{journal}{\emph{American Journal of Mathematical and Management Sciences}} \bibinfo{volume}{39}, \bibinfo{number}{3} (\bibinfo{date}{jan} \bibinfo{year}{2020}), \bibinfo{pages}{199--213}.
\newblock
\urldef\tempurl%
\url{https://doi.org/10.1080/01966324.2020.1716281}
\showDOI{\tempurl}


\bibitem[Kruengkrai et~al\mbox{.}(2017)]%
        {1-115}
\bibfield{author}{\bibinfo{person}{Canasai Kruengkrai}, \bibinfo{person}{Kentaro Torisawa}, \bibinfo{person}{Chikara Hashimoto}, \bibinfo{person}{Julien Kloetzer}, \bibinfo{person}{Jong-Hoon Oh}, {and} \bibinfo{person}{Masahiro Tanaka}.} \bibinfo{year}{2017}\natexlab{}.
\newblock \showarticletitle{Improving Event Causality Recognition with Multiple Background Knowledge Sources Using Multi-Column Convolutional Neural Networks}.
\newblock \bibinfo{journal}{\emph{Proceedings of the {AAAI} Conference on Artificial Intelligence}} \bibinfo{volume}{31}, \bibinfo{number}{1} (\bibinfo{date}{feb} \bibinfo{year}{2017}).
\newblock
\urldef\tempurl%
\url{https://doi.org/10.1609/aaai.v31i1.11005}
\showDOI{\tempurl}


\bibitem[Kulldorff(1997)]%
        {1-116}
\bibfield{author}{\bibinfo{person}{Martin Kulldorff}.} \bibinfo{year}{1997}\natexlab{}.
\newblock \showarticletitle{A spatial scan statistic}.
\newblock \bibinfo{journal}{\emph{Communications in Statistics - Theory and Methods}} \bibinfo{volume}{26}, \bibinfo{number}{6} (\bibinfo{date}{jan} \bibinfo{year}{1997}), \bibinfo{pages}{1481--1496}.
\newblock
\urldef\tempurl%
\url{https://doi.org/10.1080/03610929708831995}
\showDOI{\tempurl}


\bibitem[Kundu et~al\mbox{.}(2018)]%
        {2-59}
\bibfield{author}{\bibinfo{person}{Shamik Kundu}, \bibinfo{person}{P.K Srijith}, {and} \bibinfo{person}{Maunendra~Sankar Desarkar}.} \bibinfo{year}{2018}\natexlab{}.
\newblock \showarticletitle{Classification of Short-Texts Generated During Disasters: A Deep Neural Network Based Approach}. In \bibinfo{booktitle}{\emph{2018 {IEEE}/{ACM} International Conference on Advances in Social Networks Analysis and Mining ({ASONAM})}}. \bibinfo{publisher}{{IEEE}}.
\newblock
\urldef\tempurl%
\url{https://doi.org/10.1109/asonam.2018.8508695}
\showDOI{\tempurl}


\bibitem[Kunneman et~al\mbox{.}(2020)]%
        {1-117}
\bibfield{author}{\bibinfo{person}{F. Kunneman}, \bibinfo{person}{M. van Mulken}, {and} \bibinfo{person}{A. van~den Bosch}.} \bibinfo{year}{2020}\natexlab{}.
\newblock \showarticletitle{Anticipointment Detection in Event Tweets}.
\newblock \bibinfo{journal}{\emph{International Journal on Artificial Intelligence Tools}} \bibinfo{volume}{29}, \bibinfo{number}{02} (\bibinfo{date}{mar} \bibinfo{year}{2020}), \bibinfo{pages}{2040001}.
\newblock
\urldef\tempurl%
\url{https://doi.org/10.1142/s0218213020400011}
\showDOI{\tempurl}


\bibitem[Kupilik and Witmer(2018)]%
        {1-118}
\bibfield{author}{\bibinfo{person}{Matthew Kupilik} {and} \bibinfo{person}{Frank Witmer}.} \bibinfo{year}{2018}\natexlab{}.
\newblock \showarticletitle{Spatio-temporal violent event prediction using Gaussian process regression}.
\newblock \bibinfo{journal}{\emph{Journal of Computational Social Science}} \bibinfo{volume}{1}, \bibinfo{number}{2} (\bibinfo{date}{aug} \bibinfo{year}{2018}), \bibinfo{pages}{437--451}.
\newblock
\urldef\tempurl%
\url{https://doi.org/10.1007/s42001-018-0024-y}
\showDOI{\tempurl}


\bibitem[Lakkaraju et~al\mbox{.}(2013)]%
        {109}
\bibfield{author}{\bibinfo{person}{Himabindu Lakkaraju}, \bibinfo{person}{Julian McAuley}, {and} \bibinfo{person}{Jure Leskovec}.} \bibinfo{year}{2013}\natexlab{}.
\newblock \showarticletitle{What's in a name? understanding the interplay between titles, content, and communities in social media}. In \bibinfo{booktitle}{\emph{Proceedings of the international AAAI conference on web and social media}}, Vol.~\bibinfo{volume}{7}. \bibinfo{pages}{311--320}.
\newblock


\bibitem[Laxman et~al\mbox{.}(2008)]%
        {1-121}
\bibfield{author}{\bibinfo{person}{Srivatsan Laxman}, \bibinfo{person}{Vikram Tankasali}, {and} \bibinfo{person}{Ryen~W. White}.} \bibinfo{year}{2008}\natexlab{}.
\newblock \showarticletitle{Stream prediction using a generative model based on frequent episodes in event sequences}. In \bibinfo{booktitle}{\emph{Proceedings of the 14th {ACM} {SIGKDD} international conference on Knowledge discovery and data mining}}. \bibinfo{publisher}{{ACM}}.
\newblock
\urldef\tempurl%
\url{https://doi.org/10.1145/1401890.1401947}
\showDOI{\tempurl}


\bibitem[Lei et~al\mbox{.}(2019)]%
        {1-123}
\bibfield{author}{\bibinfo{person}{Lei Lei}, \bibinfo{person}{Xuguang Ren}, \bibinfo{person}{Nigel Franciscus}, \bibinfo{person}{Junhu Wang}, {and} \bibinfo{person}{Bela Stantic}.} \bibinfo{year}{2019}\natexlab{}.
\newblock \showarticletitle{Event Prediction Based on Causality Reasoning}.
\newblock In \bibinfo{booktitle}{\emph{Intelligent Information and Database Systems}}. \bibinfo{publisher}{Springer International Publishing}, \bibinfo{pages}{165--176}.
\newblock
\urldef\tempurl%
\url{https://doi.org/10.1007/978-3-030-14799-0_14}
\showDOI{\tempurl}


\bibitem[Leskovec et~al\mbox{.}(2009)]%
        {116}
\bibfield{author}{\bibinfo{person}{Jure Leskovec}, \bibinfo{person}{Lars Backstrom}, {and} \bibinfo{person}{Jon Kleinberg}.} \bibinfo{year}{2009}\natexlab{}.
\newblock \showarticletitle{Meme-tracking and the dynamics of the news cycle}. In \bibinfo{booktitle}{\emph{Proceedings of the 15th ACM SIGKDD international conference on Knowledge discovery and data mining}}. \bibinfo{pages}{497--506}.
\newblock


\bibitem[Letham et~al\mbox{.}(2013)]%
        {1-125}
\bibfield{author}{\bibinfo{person}{Benjamin Letham}, \bibinfo{person}{Cynthia Rudin}, {and} \bibinfo{person}{David Madigan}.} \bibinfo{year}{2013}\natexlab{}.
\newblock \showarticletitle{Sequential event prediction}.
\newblock \bibinfo{journal}{\emph{Machine Learning}} \bibinfo{volume}{93}, \bibinfo{number}{2-3} (\bibinfo{date}{jun} \bibinfo{year}{2013}), \bibinfo{pages}{357--380}.
\newblock
\urldef\tempurl%
\url{https://doi.org/10.1007/s10994-013-5356-5}
\showDOI{\tempurl}


\bibitem[Letham et~al\mbox{.}(2015)]%
        {1-124}
\bibfield{author}{\bibinfo{person}{Benjamin Letham}, \bibinfo{person}{Cynthia Rudin}, \bibinfo{person}{Tyler~H. McCormick}, {and} \bibinfo{person}{David Madigan}.} \bibinfo{year}{2015}\natexlab{}.
\newblock \showarticletitle{Interpretable classifiers using rules and Bayesian analysis: Building a better stroke prediction model}.
\newblock \bibinfo{journal}{\emph{The Annals of Applied Statistics}} \bibinfo{volume}{9}, \bibinfo{number}{3} (\bibinfo{date}{sep} \bibinfo{year}{2015}).
\newblock
\urldef\tempurl%
\url{https://doi.org/10.1214/15-aoas848}
\showDOI{\tempurl}


\bibitem[Li et~al\mbox{.}(2016)]%
        {1-126}
\bibfield{author}{\bibinfo{person}{Eldon~Y. Li}, \bibinfo{person}{Chen-Yuan Tung}, {and} \bibinfo{person}{Shu-Hsun Chang}.} \bibinfo{year}{2016}\natexlab{}.
\newblock \showarticletitle{The wisdom of crowds in action: Forecasting epidemic diseases with a web-based prediction market system}.
\newblock \bibinfo{journal}{\emph{International Journal of Medical Informatics}}  \bibinfo{volume}{92} (\bibinfo{date}{aug} \bibinfo{year}{2016}), \bibinfo{pages}{35--43}.
\newblock
\urldef\tempurl%
\url{https://doi.org/10.1016/j.ijmedinf.2016.04.014}
\showDOI{\tempurl}


\bibitem[Li et~al\mbox{.}(2017)]%
        {1-128}
\bibfield{author}{\bibinfo{person}{Shengzhi Li}, \bibinfo{person}{Jianzhong Qiao}, {and} \bibinfo{person}{Shukuan Lin}.} \bibinfo{year}{2017}\natexlab{}.
\newblock \showarticletitle{Multi-attribute Event Modeling and Prediction over Event Streams from Sensors}. In \bibinfo{booktitle}{\emph{2017 {IEEE} 23rd International Conference on Parallel and Distributed Systems ({ICPADS})}}. \bibinfo{publisher}{{IEEE}}.
\newblock
\urldef\tempurl%
\url{https://doi.org/10.1109/icpads.2017.00110}
\showDOI{\tempurl}


\bibitem[Li et~al\mbox{.}(2018a)]%
        {2-48}
\bibfield{author}{\bibinfo{person}{Xukun Li}, \bibinfo{person}{Doina Caragea}, \bibinfo{person}{Huaiyu Zhang}, {and} \bibinfo{person}{Muhammad Imran}.} \bibinfo{year}{2018}\natexlab{a}.
\newblock \showarticletitle{Localizing and Quantifying Damage in Social Media Images}. In \bibinfo{booktitle}{\emph{2018 {IEEE}/{ACM} International Conference on Advances in Social Networks Analysis and Mining ({ASONAM})}}. \bibinfo{publisher}{{IEEE}}.
\newblock
\urldef\tempurl%
\url{https://doi.org/10.1109/asonam.2018.8508298}
\showDOI{\tempurl}


\bibitem[Li et~al\mbox{.}(2018b)]%
        {2-58}
\bibfield{author}{\bibinfo{person}{Xukun Li}, \bibinfo{person}{Doina Caragea}, \bibinfo{person}{Huaiyu Zhang}, {and} \bibinfo{person}{Muhammad Imran}.} \bibinfo{year}{2018}\natexlab{b}.
\newblock \showarticletitle{Localizing and Quantifying Damage in Social Media Images}. In \bibinfo{booktitle}{\emph{2018 {IEEE}/{ACM} International Conference on Advances in Social Networks Analysis and Mining ({ASONAM})}}. \bibinfo{publisher}{{IEEE}}.
\newblock
\urldef\tempurl%
\url{https://doi.org/10.1109/asonam.2018.8508298}
\showDOI{\tempurl}


\bibitem[Li et~al\mbox{.}(2018c)]%
        {1-131}
\bibfield{author}{\bibinfo{person}{Zhongyang Li}, \bibinfo{person}{Xiao Ding}, {and} \bibinfo{person}{Ting Liu}.} \bibinfo{year}{2018}\natexlab{c}.
\newblock \showarticletitle{Constructing Narrative Event Evolutionary Graph for Script Event Prediction}. In \bibinfo{booktitle}{\emph{Proceedings of the Twenty-Seventh International Joint Conference on Artificial Intelligence}}. \bibinfo{publisher}{International Joint Conferences on Artificial Intelligence Organization}.
\newblock
\urldef\tempurl%
\url{https://doi.org/10.24963/ijcai.2018/584}
\showDOI{\tempurl}


\bibitem[Li et~al\mbox{.}(2018d)]%
        {3-5}
\bibfield{author}{\bibinfo{person}{Zefeng Li}, \bibinfo{person}{Men-Andrin Meier}, \bibinfo{person}{Egill Hauksson}, \bibinfo{person}{Zhongwen Zhan}, {and} \bibinfo{person}{Jennifer Andrews}.} \bibinfo{year}{2018}\natexlab{d}.
\newblock \showarticletitle{Machine Learning Seismic Wave Discrimination: Application to Earthquake Early Warning}.
\newblock \bibinfo{journal}{\emph{Geophysical Research Letters}} \bibinfo{volume}{45}, \bibinfo{number}{10} (\bibinfo{date}{may} \bibinfo{year}{2018}), \bibinfo{pages}{4773--4779}.
\newblock
\urldef\tempurl%
\url{https://doi.org/10.1029/2018gl077870}
\showDOI{\tempurl}


\bibitem[Li et~al\mbox{.}(2007)]%
        {1-132}
\bibfield{author}{\bibinfo{person}{Zhiguo Li}, \bibinfo{person}{Shiyu Zhou}, \bibinfo{person}{Suresh Choubey}, {and} \bibinfo{person}{Crispian Sievenpiper}.} \bibinfo{year}{2007}\natexlab{}.
\newblock \showarticletitle{Failure event prediction using the Cox proportional hazard model driven by frequent failure signatures}.
\newblock \bibinfo{journal}{\emph{{IIE} Transactions}} \bibinfo{volume}{39}, \bibinfo{number}{3} (\bibinfo{date}{mar} \bibinfo{year}{2007}), \bibinfo{pages}{303--315}.
\newblock
\urldef\tempurl%
\url{https://doi.org/10.1080/07408170600847168}
\showDOI{\tempurl}


\bibitem[Lin et~al\mbox{.}(2019)]%
        {1-134}
\bibfield{author}{\bibinfo{person}{Li Lin}, \bibinfo{person}{Lijie Wen}, {and} \bibinfo{person}{Jianmin Wang}.} \bibinfo{year}{2019}\natexlab{}.
\newblock \showarticletitle{{MM}-Pred: A Deep Predictive Model for Multi-attribute Event Sequence}.
\newblock In \bibinfo{booktitle}{\emph{Proceedings of the 2019 {SIAM} International Conference on Data Mining}}. \bibinfo{publisher}{Society for Industrial and Applied Mathematics}, \bibinfo{pages}{118--126}.
\newblock
\urldef\tempurl%
\url{https://doi.org/10.1137/1.9781611975673.14}
\showDOI{\tempurl}


\bibitem[Lin et~al\mbox{.}(2018)]%
        {1-135}
\bibfield{author}{\bibinfo{person}{Ying-Lung Lin}, \bibinfo{person}{Meng-Feng Yen}, {and} \bibinfo{person}{Liang-Chih Yu}.} \bibinfo{year}{2018}\natexlab{}.
\newblock \showarticletitle{Grid-Based Crime Prediction Using Geographical Features}.
\newblock \bibinfo{journal}{\emph{{ISPRS} International Journal of Geo-Information}} \bibinfo{volume}{7}, \bibinfo{number}{8} (\bibinfo{date}{jul} \bibinfo{year}{2018}), \bibinfo{pages}{298}.
\newblock
\urldef\tempurl%
\url{https://doi.org/10.3390/ijgi7080298}
\showDOI{\tempurl}


\bibitem[Liu et~al\mbox{.}(2018)]%
        {1-21}
\bibfield{author}{\bibinfo{person}{Bing Liu}, \bibinfo{person}{Tong Yu}, \bibinfo{person}{Ian Lane}, {and} \bibinfo{person}{Ole Mengshoel}.} \bibinfo{year}{2018}\natexlab{}.
\newblock \showarticletitle{Customized Nonlinear Bandits for Online Response Selection in Neural Conversation Models}.
\newblock \bibinfo{journal}{\emph{Proceedings of the {AAAI} Conference on Artificial Intelligence}} \bibinfo{volume}{32}, \bibinfo{number}{1} (\bibinfo{date}{apr} \bibinfo{year}{2018}).
\newblock
\urldef\tempurl%
\url{https://doi.org/10.1609/aaai.v32i1.12028}
\showDOI{\tempurl}


\bibitem[Liu and Brown(2004)]%
        {1-136}
\bibfield{author}{\bibinfo{person}{H. Liu} {and} \bibinfo{person}{D.E. Brown}.} \bibinfo{year}{2004}\natexlab{}.
\newblock \showarticletitle{A New Point Process Transition Density Model for Space{\textendash}Time Event Prediction}.
\newblock \bibinfo{journal}{\emph{{IEEE} Transactions on Systems, Man and Cybernetics, Part C (Applications and Reviews)}} \bibinfo{volume}{34}, \bibinfo{number}{3} (\bibinfo{date}{aug} \bibinfo{year}{2004}), \bibinfo{pages}{310--324}.
\newblock
\urldef\tempurl%
\url{https://doi.org/10.1109/tsmcc.2004.829306}
\showDOI{\tempurl}


\bibitem[Liu et~al\mbox{.}(2022)]%
        {iot_2}
\bibfield{author}{\bibinfo{person}{Wei Liu}, \bibinfo{person}{Yi Ding}, \bibinfo{person}{Shuai Wang}, \bibinfo{person}{Yu Yang}, {and} \bibinfo{person}{Desheng Zhang}.} \bibinfo{year}{2022}\natexlab{}.
\newblock \showarticletitle{Para-Pred: Addressing Heterogeneity for City-Wide Indoor Status Estimation in On-Demand Delivery}. In \bibinfo{booktitle}{\emph{Proceedings of the 28th {ACM} {SIGKDD} Conference on Knowledge Discovery and Data Mining}}. \bibinfo{publisher}{{ACM}}.
\newblock
\urldef\tempurl%
\url{https://doi.org/10.1145/3534678.3539167}
\showDOI{\tempurl}


\bibitem[Lohumi and Roy(2018)]%
        {3-8}
\bibfield{author}{\bibinfo{person}{Kanishk Lohumi} {and} \bibinfo{person}{Sudip Roy}.} \bibinfo{year}{2018}\natexlab{}.
\newblock \showarticletitle{Automatic Detection of Flood Severity Level from Flood Videos using Deep Learning Models}. In \bibinfo{booktitle}{\emph{2018 5th International Conference on Information and Communication Technologies for Disaster Management ({ICT}-{DM})}}. \bibinfo{publisher}{{IEEE}}.
\newblock
\urldef\tempurl%
\url{https://doi.org/10.1109/ict-dm.2018.8636373}
\showDOI{\tempurl}


\bibitem[Lopez-Cuevas et~al\mbox{.}(2018)]%
        {4-15}
\bibfield{author}{\bibinfo{person}{Armando Lopez-Cuevas}, \bibinfo{person}{Miguel~Angel Medina-Perez}, \bibinfo{person}{Raul Monroy}, \bibinfo{person}{Jose~Emmanuel Ramirez-Marquez}, {and} \bibinfo{person}{Luis~A. Trejo}.} \bibinfo{year}{2018}\natexlab{}.
\newblock \showarticletitle{{FiToViz}: A Visualisation Approach for Real-Time Risk Situation Awareness}.
\newblock \bibinfo{journal}{\emph{{IEEE} Transactions on Affective Computing}} \bibinfo{volume}{9}, \bibinfo{number}{3} (\bibinfo{date}{jul} \bibinfo{year}{2018}), \bibinfo{pages}{372--382}.
\newblock
\urldef\tempurl%
\url{https://doi.org/10.1109/taffc.2017.2741478}
\showDOI{\tempurl}


\bibitem[Lu et~al\mbox{.}(2017)]%
        {1-138}
\bibfield{author}{\bibinfo{person}{Jiasen Lu}, \bibinfo{person}{Caiming Xiong}, \bibinfo{person}{Devi Parikh}, {and} \bibinfo{person}{Richard Socher}.} \bibinfo{year}{2017}\natexlab{}.
\newblock \showarticletitle{Knowing When to Look: Adaptive Attention via a Visual Sentinel for Image Captioning}. In \bibinfo{booktitle}{\emph{2017 {IEEE} Conference on Computer Vision and Pattern Recognition ({CVPR})}}. \bibinfo{publisher}{{IEEE}}.
\newblock
\urldef\tempurl%
\url{https://doi.org/10.1109/cvpr.2017.345}
\showDOI{\tempurl}


\bibitem[Lv et~al\mbox{.}(2019)]%
        {1-139}
\bibfield{author}{\bibinfo{person}{Shangwen Lv}, \bibinfo{person}{Wanhui Qian}, \bibinfo{person}{Longtao Huang}, \bibinfo{person}{Jizhong Han}, {and} \bibinfo{person}{Songlin Hu}.} \bibinfo{year}{2019}\natexlab{}.
\newblock \showarticletitle{{SAM}-Net: Integrating Event-Level and Chain-Level Attentions to Predict What Happens Next}.
\newblock \bibinfo{journal}{\emph{Proceedings of the {AAAI} Conference on Artificial Intelligence}} \bibinfo{volume}{33}, \bibinfo{number}{01} (\bibinfo{date}{jul} \bibinfo{year}{2019}), \bibinfo{pages}{6802--6809}.
\newblock
\urldef\tempurl%
\url{https://doi.org/10.1609/aaai.v33i01.33016802}
\showDOI{\tempurl}


\bibitem[Lyu et~al\mbox{.}(2019)]%
        {4-16}
\bibfield{author}{\bibinfo{person}{Dian Lyu}, \bibinfo{person}{Peng Cheng}, \bibinfo{person}{Ruizhou Liu}, {and} \bibinfo{person}{Liang Liu}.} \bibinfo{year}{2019}\natexlab{}.
\newblock \showarticletitle{Bise-{ResNet}: Combine Segmentation and Classification Networks for Road Following on Unmanned Aerial Vehicle}. In \bibinfo{booktitle}{\emph{2019 {IEEE} International Conference on Multimedia \& Expo Workshops ({ICMEW})}}. \bibinfo{publisher}{{IEEE}}.
\newblock
\urldef\tempurl%
\url{https://doi.org/10.1109/icmew.2019.00042}
\showDOI{\tempurl}


\bibitem[Ma and Leung(2019)]%
        {1-25}
\bibfield{author}{\bibinfo{person}{King Ma} {and} \bibinfo{person}{Henry Leung}.} \bibinfo{year}{2019}\natexlab{}.
\newblock \showarticletitle{A Novel {LSTM} Approach for Asynchronous Multivariate Time Series Prediction}. In \bibinfo{booktitle}{\emph{2019 International Joint Conference on Neural Networks ({IJCNN})}}. \bibinfo{publisher}{{IEEE}}.
\newblock
\urldef\tempurl%
\url{https://doi.org/10.1109/ijcnn.2019.8851792}
\showDOI{\tempurl}


\bibitem[Ma et~al\mbox{.}(2012)]%
        {134}
\bibfield{author}{\bibinfo{person}{Zongyang Ma}, \bibinfo{person}{Aixin Sun}, {and} \bibinfo{person}{Gao Cong}.} \bibinfo{year}{2012}\natexlab{}.
\newblock \showarticletitle{Will this\# hashtag be popular tomorrow?}. In \bibinfo{booktitle}{\emph{Proceedings of the 35th international ACM SIGIR conference on Research and development in information retrieval}}. \bibinfo{pages}{1173--1174}.
\newblock


\bibitem[Ma et~al\mbox{.}(2013)]%
        {135}
\bibfield{author}{\bibinfo{person}{Zongyang Ma}, \bibinfo{person}{Aixin Sun}, {and} \bibinfo{person}{Gao Cong}.} \bibinfo{year}{2013}\natexlab{}.
\newblock \showarticletitle{On predicting the popularity of newly emerging hashtags in t witter}.
\newblock \bibinfo{journal}{\emph{Journal of the American Society for Information Science and Technology}} \bibinfo{volume}{64}, \bibinfo{number}{7} (\bibinfo{year}{2013}), \bibinfo{pages}{1399--1410}.
\newblock


\bibitem[Madichetty and Sridevi(2019)]%
        {2-61}
\bibfield{author}{\bibinfo{person}{Sreenivasulu Madichetty} {and} \bibinfo{person}{M Sridevi}.} \bibinfo{year}{2019}\natexlab{}.
\newblock \showarticletitle{Detecting Informative Tweets during Disaster using Deep Neural Networks}. In \bibinfo{booktitle}{\emph{2019 11th International Conference on Communication Systems \& Networks ({COMSNETS})}}. \bibinfo{publisher}{{IEEE}}.
\newblock
\urldef\tempurl%
\url{https://doi.org/10.1109/comsnets.2019.8711095}
\showDOI{\tempurl}


\bibitem[Mallouhy et~al\mbox{.}(2019)]%
        {1-140}
\bibfield{author}{\bibinfo{person}{Roxane Mallouhy}, \bibinfo{person}{Chady~Abou Jaoude}, \bibinfo{person}{Christophe Guyeux}, {and} \bibinfo{person}{Abdallah Makhoul}.} \bibinfo{year}{2019}\natexlab{}.
\newblock \showarticletitle{Major earthquake event prediction using various machine learning algorithms}. In \bibinfo{booktitle}{\emph{2019 International Conference on Information and Communication Technologies for Disaster Management ({ICT}-{DM})}}. \bibinfo{publisher}{{IEEE}}.
\newblock
\urldef\tempurl%
\url{https://doi.org/10.1109/ict-dm47966.2019.9032983}
\showDOI{\tempurl}


\bibitem[Matsubara et~al\mbox{.}(2012a)]%
        {1-142}
\bibfield{author}{\bibinfo{person}{Yasuko Matsubara}, \bibinfo{person}{Yasushi Sakurai}, \bibinfo{person}{Christos Faloutsos}, \bibinfo{person}{Tomoharu Iwata}, {and} \bibinfo{person}{Masatoshi Yoshikawa}.} \bibinfo{year}{2012}\natexlab{a}.
\newblock \showarticletitle{Fast mining and forecasting of complex time-stamped events}. In \bibinfo{booktitle}{\emph{Proceedings of the 18th {ACM} {SIGKDD} international conference on Knowledge discovery and data mining}}. \bibinfo{publisher}{{ACM}}.
\newblock
\urldef\tempurl%
\url{https://doi.org/10.1145/2339530.2339577}
\showDOI{\tempurl}


\bibitem[Matsubara et~al\mbox{.}(2012b)]%
        {137}
\bibfield{author}{\bibinfo{person}{Yasuko Matsubara}, \bibinfo{person}{Yasushi Sakurai}, \bibinfo{person}{B~Aditya Prakash}, \bibinfo{person}{Lei Li}, {and} \bibinfo{person}{Christos Faloutsos}.} \bibinfo{year}{2012}\natexlab{b}.
\newblock \showarticletitle{Rise and fall patterns of information diffusion: model and implications}. In \bibinfo{booktitle}{\emph{Proceedings of the 18th ACM SIGKDD international conference on Knowledge discovery and data mining}}. \bibinfo{pages}{6--14}.
\newblock


\bibitem[Minor and Cook(2017)]%
        {1-147}
\bibfield{author}{\bibinfo{person}{Bryan Minor} {and} \bibinfo{person}{Diane~J. Cook}.} \bibinfo{year}{2017}\natexlab{}.
\newblock \showarticletitle{Forecasting occurrences of activities}.
\newblock \bibinfo{journal}{\emph{Pervasive and Mobile Computing}}  \bibinfo{volume}{38} (\bibinfo{date}{jul} \bibinfo{year}{2017}), \bibinfo{pages}{77--91}.
\newblock
\urldef\tempurl%
\url{https://doi.org/10.1016/j.pmcj.2016.09.010}
\showDOI{\tempurl}


\bibitem[Mirtaheri et~al\mbox{.}(2019)]%
        {1-148}
\bibfield{author}{\bibinfo{person}{Mehrnoosh Mirtaheri}, \bibinfo{person}{Sami Abu-El-Haija}, \bibinfo{person}{Fred Morstatter}, \bibinfo{person}{Greg~Ver Steeg}, {and} \bibinfo{person}{Aram Galstyan}.} \bibinfo{year}{2019}\natexlab{}.
\newblock \showarticletitle{Identifying and Analyzing Cryptocurrency Manipulations in Social Media}.
\newblock  (\bibinfo{date}{feb} \bibinfo{year}{2019}).
\newblock
\urldef\tempurl%
\url{https://doi.org/10.31219/osf.io/dqz89}
\showDOI{\tempurl}


\bibitem[Moniz and Torgo(2019)]%
        {144}
\bibfield{author}{\bibinfo{person}{Nuno Moniz} {and} \bibinfo{person}{Lu{\'\i}s Torgo}.} \bibinfo{year}{2019}\natexlab{}.
\newblock \showarticletitle{A review on web content popularity prediction: Issues and open challenges}.
\newblock \bibinfo{journal}{\emph{Online Social Networks and Media}}  \bibinfo{volume}{12} (\bibinfo{year}{2019}), \bibinfo{pages}{1--20}.
\newblock


\bibitem[Moon et~al\mbox{.}(2019)]%
        {3-4}
\bibfield{author}{\bibinfo{person}{Seung-Hyun Moon}, \bibinfo{person}{Yong-Hyuk Kim}, \bibinfo{person}{Yong~Hee Lee}, {and} \bibinfo{person}{Byung-Ro Moon}.} \bibinfo{year}{2019}\natexlab{}.
\newblock \showarticletitle{Application of machine learning to an early warning system for very short-term heavy rainfall}.
\newblock \bibinfo{journal}{\emph{Journal of Hydrology}}  \bibinfo{volume}{568} (\bibinfo{date}{jan} \bibinfo{year}{2019}), \bibinfo{pages}{1042--1054}.
\newblock
\urldef\tempurl%
\url{https://doi.org/10.1016/j.jhydrol.2018.11.060}
\showDOI{\tempurl}


\bibitem[Mukhina et~al\mbox{.}(2019)]%
        {1-150}
\bibfield{author}{\bibinfo{person}{Ksenia~D. Mukhina}, \bibinfo{person}{Alexander~A. Visheratin}, {and} \bibinfo{person}{Denis Nasonov}.} \bibinfo{year}{2019}\natexlab{}.
\newblock \showarticletitle{Urban events prediction via convolutional neural networks and Instagram data}.
\newblock \bibinfo{journal}{\emph{Procedia Computer Science}}  \bibinfo{volume}{156} (\bibinfo{year}{2019}), \bibinfo{pages}{176--184}.
\newblock
\urldef\tempurl%
\url{https://doi.org/10.1016/j.procs.2019.08.193}
\showDOI{\tempurl}


\bibitem[Muthiah et~al\mbox{.}(2016)]%
        {1-152}
\bibfield{author}{\bibinfo{person}{Sathappan Muthiah}, \bibinfo{person}{Bert Huang}, \bibinfo{person}{Jaime Arredondo}, \bibinfo{person}{David Mares}, \bibinfo{person}{Lise Getoor}, \bibinfo{person}{Graham Katz}, {and} \bibinfo{person}{Naren Ramakrishnan}.} \bibinfo{year}{2016}\natexlab{}.
\newblock \showarticletitle{Capturing Planned Protests from Open Source Indicators}.
\newblock \bibinfo{journal}{\emph{{AI} Magazine}} \bibinfo{volume}{37}, \bibinfo{number}{2} (\bibinfo{date}{jul} \bibinfo{year}{2016}), \bibinfo{pages}{63--75}.
\newblock
\urldef\tempurl%
\url{https://doi.org/10.1609/aimag.v37i2.2631}
\showDOI{\tempurl}


\bibitem[Mutlu et~al\mbox{.}(2019)]%
        {2-30}
\bibfield{author}{\bibinfo{person}{Begum Mutlu}, \bibinfo{person}{Hakan~A. Nefeslioglu}, \bibinfo{person}{Ebru~A. Sezer}, \bibinfo{person}{M.~Ali Akcayol}, {and} \bibinfo{person}{Candan Gokceoglu}.} \bibinfo{year}{2019}\natexlab{}.
\newblock \showarticletitle{An Experimental Research on the Use of Recurrent Neural Networks in Landslide Susceptibility Mapping}.
\newblock \bibinfo{journal}{\emph{{ISPRS} International Journal of Geo-Information}} \bibinfo{volume}{8}, \bibinfo{number}{12} (\bibinfo{date}{dec} \bibinfo{year}{2019}), \bibinfo{pages}{578}.
\newblock
\urldef\tempurl%
\url{https://doi.org/10.3390/ijgi8120578}
\showDOI{\tempurl}


\bibitem[Nakajima et~al\mbox{.}(2019)]%
        {1-153}
\bibfield{author}{\bibinfo{person}{Yoko Nakajima}, \bibinfo{person}{Keiya Takagi}, \bibinfo{person}{Michal Ptaszynski}, \bibinfo{person}{Hirotoshi Honma}, {and} \bibinfo{person}{Fumito Masui}.} \bibinfo{year}{2019}\natexlab{}.
\newblock \showarticletitle{A Proposal of Prediction Method Using Word Polarity Information for Future Event Prediction Support System}. In \bibinfo{booktitle}{\emph{2019 International Conference of Advanced Informatics: Concepts, Theory and Applications ({ICAICTA})}}. \bibinfo{publisher}{{IEEE}}.
\newblock
\urldef\tempurl%
\url{https://doi.org/10.1109/icaicta.2019.8904426}
\showDOI{\tempurl}


\bibitem[Naveed et~al\mbox{.}(2011)]%
        {149}
\bibfield{author}{\bibinfo{person}{Nasir Naveed}, \bibinfo{person}{Thomas Gottron}, \bibinfo{person}{J{\'e}r{\^o}me Kunegis}, {and} \bibinfo{person}{Arifah~Che Alhadi}.} \bibinfo{year}{2011}\natexlab{}.
\newblock \showarticletitle{Bad news travel fast: A content-based analysis of interestingness on twitter}. In \bibinfo{booktitle}{\emph{Proceedings of the 3rd international web science conference}}. \bibinfo{pages}{1--7}.
\newblock


\bibitem[Nguyen et~al\mbox{.}(2017a)]%
        {1-155}
\bibfield{author}{\bibinfo{person}{Dai~Quoc Nguyen}, \bibinfo{person}{Dat~Quoc Nguyen}, \bibinfo{person}{Ashutosh Modi}, \bibinfo{person}{Stefan Thater}, {and} \bibinfo{person}{Manfred Pinkal}.} \bibinfo{year}{2017}\natexlab{a}.
\newblock \showarticletitle{A Mixture Model for Learning Multi-Sense Word Embeddings}. In \bibinfo{booktitle}{\emph{Proceedings of the 6th Joint Conference on Lexical and Computational Semantics ({SEM} 2017)}}. \bibinfo{publisher}{Association for Computational Linguistics}.
\newblock
\urldef\tempurl%
\url{https://doi.org/10.18653/v1/s17-1015}
\showDOI{\tempurl}


\bibitem[Nguyen et~al\mbox{.}(2017b)]%
        {2-45}
\bibfield{author}{\bibinfo{person}{Dat~T. Nguyen}, \bibinfo{person}{Ferda Ofli}, \bibinfo{person}{Muhammad Imran}, {and} \bibinfo{person}{Prasenjit Mitra}.} \bibinfo{year}{2017}\natexlab{b}.
\newblock \showarticletitle{Damage Assessment from Social Media Imagery Data During Disasters}. In \bibinfo{booktitle}{\emph{Proceedings of the 2017 {IEEE}/{ACM} International Conference on Advances in Social Networks Analysis and Mining 2017}}. \bibinfo{publisher}{{ACM}}.
\newblock
\urldef\tempurl%
\url{https://doi.org/10.1145/3110025.3110109}
\showDOI{\tempurl}


\bibitem[Nguyen et~al\mbox{.}(2017c)]%
        {2-55}
\bibfield{author}{\bibinfo{person}{Dat~T. Nguyen}, \bibinfo{person}{Ferda Ofli}, \bibinfo{person}{Muhammad Imran}, {and} \bibinfo{person}{Prasenjit Mitra}.} \bibinfo{year}{2017}\natexlab{c}.
\newblock \showarticletitle{Damage Assessment from Social Media Imagery Data During Disasters}. In \bibinfo{booktitle}{\emph{Proceedings of the 2017 {IEEE}/{ACM} International Conference on Advances in Social Networks Analysis and Mining 2017}}. \bibinfo{publisher}{{ACM}}.
\newblock
\urldef\tempurl%
\url{https://doi.org/10.1145/3110025.3110109}
\showDOI{\tempurl}


\bibitem[Nsengiyumva and Valentino(2020)]%
        {2-26}
\bibfield{author}{\bibinfo{person}{Jean~Baptiste Nsengiyumva} {and} \bibinfo{person}{Roberto Valentino}.} \bibinfo{year}{2020}\natexlab{}.
\newblock \showarticletitle{Predicting landslide susceptibility and risks using {GIS}-based machine learning simulations, case of upper Nyabarongo catchment}.
\newblock \bibinfo{journal}{\emph{Geomatics, Natural Hazards and Risk}} \bibinfo{volume}{11}, \bibinfo{number}{1} (\bibinfo{date}{jan} \bibinfo{year}{2020}), \bibinfo{pages}{1250--1277}.
\newblock
\urldef\tempurl%
\url{https://doi.org/10.1080/19475705.2020.1785555}
\showDOI{\tempurl}


\bibitem[Ogie et~al\mbox{.}(2018)]%
        {disaster2}
\bibfield{author}{\bibinfo{person}{Robert~Ighodaro Ogie}, \bibinfo{person}{Juan~Castilla Rho}, {and} \bibinfo{person}{Rodney~J Clarke}.} \bibinfo{year}{2018}\natexlab{}.
\newblock \showarticletitle{Artificial intelligence in disaster risk communication: A systematic literature review}. In \bibinfo{booktitle}{\emph{2018 5th International Conference on Information and Communication Technologies for Disaster Management (ICT-DM)}}. IEEE, \bibinfo{pages}{1--8}.
\newblock


\bibitem[Okawa et~al\mbox{.}(2019)]%
        {1-159}
\bibfield{author}{\bibinfo{person}{Maya Okawa}, \bibinfo{person}{Tomoharu Iwata}, \bibinfo{person}{Takeshi Kurashima}, \bibinfo{person}{Yusuke Tanaka}, \bibinfo{person}{Hiroyuki Toda}, {and} \bibinfo{person}{Naonori Ueda}.} \bibinfo{year}{2019}\natexlab{}.
\newblock \showarticletitle{Deep Mixture Point Processes}. In \bibinfo{booktitle}{\emph{Proceedings of the 25th {ACM} {SIGKDD} International Conference on Knowledge Discovery \& Data Mining}}. \bibinfo{publisher}{{ACM}}.
\newblock
\urldef\tempurl%
\url{https://doi.org/10.1145/3292500.3330937}
\showDOI{\tempurl}


\bibitem[Oki et~al\mbox{.}(2018)]%
        {1-160}
\bibfield{author}{\bibinfo{person}{Motoyuki Oki}, \bibinfo{person}{Koh Takeuchi}, {and} \bibinfo{person}{Yukio Uematsu}.} \bibinfo{year}{2018}\natexlab{}.
\newblock \showarticletitle{Mobile Network Failure Event Detection and Forecasting With Multiple User Activity Data Sets}.
\newblock \bibinfo{journal}{\emph{Proceedings of the {AAAI} Conference on Artificial Intelligence}} \bibinfo{volume}{32}, \bibinfo{number}{1} (\bibinfo{date}{apr} \bibinfo{year}{2018}).
\newblock
\urldef\tempurl%
\url{https://doi.org/10.1609/aaai.v32i1.11422}
\showDOI{\tempurl}


\bibitem[Oliveira et~al\mbox{.}(2016)]%
        {4-5}
\bibfield{author}{\bibinfo{person}{Gabriel~L. Oliveira}, \bibinfo{person}{Abhinav Valada}, \bibinfo{person}{Claas Bollen}, \bibinfo{person}{Wolfram Burgard}, {and} \bibinfo{person}{Thomas Brox}.} \bibinfo{year}{2016}\natexlab{}.
\newblock \showarticletitle{Deep learning for human part discovery in images}. In \bibinfo{booktitle}{\emph{2016 {IEEE} International Conference on Robotics and Automation ({ICRA})}}. \bibinfo{publisher}{{IEEE}}.
\newblock
\urldef\tempurl%
\url{https://doi.org/10.1109/icra.2016.7487304}
\showDOI{\tempurl}


\bibitem[Paul et~al\mbox{.}(2020)]%
        {2-62}
\bibfield{author}{\bibinfo{person}{Udit Paul}, \bibinfo{person}{Alexander Ermakov}, \bibinfo{person}{Michael Nekrasov}, \bibinfo{person}{Vivek Adarsh}, {and} \bibinfo{person}{Elizabeth Belding}.} \bibinfo{year}{2020}\natexlab{}.
\newblock \showarticletitle{{\#}Outage: Detecting Power and Communication Outages from Social Networks}. In \bibinfo{booktitle}{\emph{Proceedings of The Web Conference 2020}}. \bibinfo{publisher}{{ACM}}.
\newblock
\urldef\tempurl%
\url{https://doi.org/10.1145/3366423.3380251}
\showDOI{\tempurl}


\bibitem[Peng et~al\mbox{.}(2019)]%
        {2-64}
\bibfield{author}{\bibinfo{person}{Bo Peng}, \bibinfo{person}{Xinyi Liu}, \bibinfo{person}{Zonglin Meng}, {and} \bibinfo{person}{Qunying Huang}.} \bibinfo{year}{2019}\natexlab{}.
\newblock \showarticletitle{Urban Flood Mapping with Residual Patch Similarity Learning}. In \bibinfo{booktitle}{\emph{Proceedings of the 3rd {ACM} {SIGSPATIAL} International Workshop on {AI} for Geographic Knowledge Discovery}}. \bibinfo{publisher}{{ACM}}.
\newblock
\urldef\tempurl%
\url{https://doi.org/10.1145/3356471.3365235}
\showDOI{\tempurl}


\bibitem[Perol et~al\mbox{.}(2018)]%
        {3-7}
\bibfield{author}{\bibinfo{person}{Thibaut Perol}, \bibinfo{person}{Michaël Gharbi}, {and} \bibinfo{person}{Marine Denolle}.} \bibinfo{year}{2018}\natexlab{}.
\newblock \showarticletitle{Convolutional neural network for earthquake detection and location}.
\newblock \bibinfo{journal}{\emph{Science Advances}} \bibinfo{volume}{4}, \bibinfo{number}{2} (\bibinfo{date}{feb} \bibinfo{year}{2018}).
\newblock
\urldef\tempurl%
\url{https://doi.org/10.1126/sciadv.1700578}
\showDOI{\tempurl}


\bibitem[Petropoulos and Makridakis(2020)]%
        {1-162}
\bibfield{author}{\bibinfo{person}{Fotios Petropoulos} {and} \bibinfo{person}{Spyros Makridakis}.} \bibinfo{year}{2020}\natexlab{}.
\newblock \showarticletitle{Forecasting the novel coronavirus {COVID}-19}.
\newblock \bibinfo{journal}{\emph{{PLOS} {ONE}}} \bibinfo{volume}{15}, \bibinfo{number}{3} (\bibinfo{date}{mar} \bibinfo{year}{2020}), \bibinfo{pages}{e0231236}.
\newblock
\urldef\tempurl%
\url{https://doi.org/10.1371/journal.pone.0231236}
\showDOI{\tempurl}


\bibitem[Petrovic et~al\mbox{.}(2011)]%
        {155}
\bibfield{author}{\bibinfo{person}{Sasa Petrovic}, \bibinfo{person}{Miles Osborne}, {and} \bibinfo{person}{Victor Lavrenko}.} \bibinfo{year}{2011}\natexlab{}.
\newblock \showarticletitle{Rt to win! predicting message propagation in twitter}. In \bibinfo{booktitle}{\emph{Proceedings of the international AAAI conference on web and social media}}, Vol.~\bibinfo{volume}{5}. \bibinfo{pages}{586--589}.
\newblock


\bibitem[Pillai et~al\mbox{.}(2016)]%
        {1-241}
\bibfield{author}{\bibinfo{person}{Karthik~Ganesan Pillai}, \bibinfo{person}{Rafal~A. Angryk}, \bibinfo{person}{Juan~M. Banda}, \bibinfo{person}{Dustin Kempton}, \bibinfo{person}{Berkay Aydin}, {and} \bibinfo{person}{Petrus~C. Martens}.} \bibinfo{year}{2016}\natexlab{}.
\newblock \showarticletitle{Mining At Most Top-K Spatiotemporal Co-Occurrence Patterns in Datasets with Extended Spatial Representations}.
\newblock  \bibinfo{volume}{2}, \bibinfo{number}{3}, Article \bibinfo{articleno}{10} (\bibinfo{date}{sep} \bibinfo{year}{2016}), \bibinfo{numpages}{27}~pages.
\newblock
\showISSN{2374-0353}
\urldef\tempurl%
\url{https://doi.org/10.1145/2936775}
\showDOI{\tempurl}


\bibitem[Piraj{\'{a}}n et~al\mbox{.}(2019)]%
        {1-164}
\bibfield{author}{\bibinfo{person}{Freddy Piraj{\'{a}}n}, \bibinfo{person}{Andrey Fajardo}, {and} \bibinfo{person}{Miguel Melgarejo}.} \bibinfo{year}{2019}\natexlab{}.
\newblock \showarticletitle{Towards a Deep Learning Approach for Urban Crime Forecasting}.
\newblock In \bibinfo{booktitle}{\emph{Communications in Computer and Information Science}}. \bibinfo{publisher}{Springer International Publishing}, \bibinfo{pages}{179--189}.
\newblock
\urldef\tempurl%
\url{https://doi.org/10.1007/978-3-030-31019-6_16}
\showDOI{\tempurl}


\bibitem[Poblete et~al\mbox{.}(2018)]%
        {4-21}
\bibfield{author}{\bibinfo{person}{Barbara Poblete}, \bibinfo{person}{Jheser Guzman}, \bibinfo{person}{Jazmine Maldonado}, {and} \bibinfo{person}{Felipe Tobar}.} \bibinfo{year}{2018}\natexlab{}.
\newblock \showarticletitle{Robust Detection of Extreme Events Using Twitter: Worldwide Earthquake Monitoring}.
\newblock \bibinfo{journal}{\emph{{IEEE} Transactions on Multimedia}} \bibinfo{volume}{20}, \bibinfo{number}{10} (\bibinfo{date}{oct} \bibinfo{year}{2018}), \bibinfo{pages}{2551--2561}.
\newblock
\urldef\tempurl%
\url{https://doi.org/10.1109/tmm.2018.2855107}
\showDOI{\tempurl}


\bibitem[Povinelli and Feng(2003)]%
        {1-165}
\bibfield{author}{\bibinfo{person}{R.J. Povinelli} {and} \bibinfo{person}{Xin Feng}.} \bibinfo{year}{2003}\natexlab{}.
\newblock \showarticletitle{A new temporal pattern identification method for characterization and prediction of complex time series events}.
\newblock \bibinfo{journal}{\emph{{IEEE} Transactions on Knowledge and Data Engineering}} \bibinfo{volume}{15}, \bibinfo{number}{2} (\bibinfo{date}{mar} \bibinfo{year}{2003}), \bibinfo{pages}{339--352}.
\newblock
\urldef\tempurl%
\url{https://doi.org/10.1109/tkde.2003.1185838}
\showDOI{\tempurl}


\bibitem[Prasad et~al\mbox{.}(2021)]%
        {2-25}
\bibfield{author}{\bibinfo{person}{Pankaj Prasad}, \bibinfo{person}{Victor~Joseph Loveson}, \bibinfo{person}{Bappa Das}, {and} \bibinfo{person}{Mahender Kotha}.} \bibinfo{year}{2021}\natexlab{}.
\newblock \showarticletitle{Novel ensemble machine learning models in flood susceptibility mapping}.
\newblock \bibinfo{journal}{\emph{Geocarto International}} \bibinfo{volume}{37}, \bibinfo{number}{16} (\bibinfo{date}{mar} \bibinfo{year}{2021}), \bibinfo{pages}{4571--4593}.
\newblock
\urldef\tempurl%
\url{https://doi.org/10.1080/10106049.2021.1892209}
\showDOI{\tempurl}


\bibitem[Presa-Reyes and Chen(2020a)]%
        {2-52}
\bibfield{author}{\bibinfo{person}{Maria Presa-Reyes} {and} \bibinfo{person}{Shu-Ching Chen}.} \bibinfo{year}{2020}\natexlab{a}.
\newblock \showarticletitle{Assessing Building Damage by Learning the Deep Feature Correspondence of Before and After Aerial Images}. In \bibinfo{booktitle}{\emph{2020 {IEEE} Conference on Multimedia Information Processing and Retrieval ({MIPR})}}. \bibinfo{publisher}{{IEEE}}.
\newblock
\urldef\tempurl%
\url{https://doi.org/10.1109/mipr49039.2020.00017}
\showDOI{\tempurl}


\bibitem[Presa-Reyes and Chen(2020b)]%
        {2-13}
\bibfield{author}{\bibinfo{person}{Maria Presa-Reyes} {and} \bibinfo{person}{Shu-Ching Chen}.} \bibinfo{year}{2020}\natexlab{b}.
\newblock \showarticletitle{Assessing Building Damage by Learning the Deep Feature Correspondence of Before and After Aerial Images}. In \bibinfo{booktitle}{\emph{2020 {IEEE} Conference on Multimedia Information Processing and Retrieval ({MIPR})}}. \bibinfo{publisher}{{IEEE}}.
\newblock
\urldef\tempurl%
\url{https://doi.org/10.1109/mipr49039.2020.00017}
\showDOI{\tempurl}


\bibitem[Qi et~al\mbox{.}(2018)]%
        {1-130}
\bibfield{author}{\bibinfo{person}{Xinshe Qi}, \bibinfo{person}{Guo Li}, \bibinfo{person}{Xin Wang}, \bibinfo{person}{Na Wang}, {and} \bibinfo{person}{Cuicui Gao}.} \bibinfo{year}{2018}\natexlab{}.
\newblock \showarticletitle{Predicting Model about Next Crime of Serial Offender}. In \bibinfo{booktitle}{\emph{2018 5th International Conference on Information Science and Control Engineering ({ICISCE})}}. \bibinfo{publisher}{{IEEE}}.
\newblock
\urldef\tempurl%
\url{https://doi.org/10.1109/icisce.2018.00087}
\showDOI{\tempurl}


\bibitem[Radinsky et~al\mbox{.}(2012)]%
        {1-168}
\bibfield{author}{\bibinfo{person}{Kira Radinsky}, \bibinfo{person}{Sagie Davidovich}, {and} \bibinfo{person}{Shaul Markovitch}.} \bibinfo{year}{2012}\natexlab{}.
\newblock \showarticletitle{Learning causality for news events prediction}. In \bibinfo{booktitle}{\emph{Proceedings of the 21st international conference on World Wide Web}}. \bibinfo{publisher}{{ACM}}.
\newblock
\urldef\tempurl%
\url{https://doi.org/10.1145/2187836.2187958}
\showDOI{\tempurl}


\bibitem[Ramakrishnan et~al\mbox{.}(2014)]%
        {1-171}
\bibfield{author}{\bibinfo{person}{Naren Ramakrishnan}, \bibinfo{person}{Patrick Butler}, \bibinfo{person}{Sathappan Muthiah}, \bibinfo{person}{Nathan Self}, \bibinfo{person}{Rupinder Khandpur}, \bibinfo{person}{Parang Saraf}, \bibinfo{person}{Wei Wang}, \bibinfo{person}{Jose Cadena}, \bibinfo{person}{Anil Vullikanti}, \bibinfo{person}{Gizem Korkmaz}, \bibinfo{person}{Chris Kuhlman}, \bibinfo{person}{Achla Marathe}, \bibinfo{person}{Liang Zhao}, \bibinfo{person}{Ting Hua}, \bibinfo{person}{Feng Chen}, \bibinfo{person}{Chang~Tien Lu}, \bibinfo{person}{Bert Huang}, \bibinfo{person}{Aravind Srinivasan}, \bibinfo{person}{Khoa Trinh}, \bibinfo{person}{Lise Getoor}, \bibinfo{person}{Graham Katz}, \bibinfo{person}{Andy Doyle}, \bibinfo{person}{Chris Ackermann}, \bibinfo{person}{Ilya Zavorin}, \bibinfo{person}{Jim Ford}, \bibinfo{person}{Kristen Summers}, \bibinfo{person}{Youssef Fayed}, \bibinfo{person}{Jaime Arredondo}, \bibinfo{person}{Dipak Gupta}, {and} \bibinfo{person}{David Mares}.}
  \bibinfo{year}{2014}\natexlab{}.
\newblock \showarticletitle{{\textquotesingle}Beating the news{\textquotesingle} with {EMBERS}}. In \bibinfo{booktitle}{\emph{Proceedings of the 20th {ACM} {SIGKDD} international conference on Knowledge discovery and data mining}}. \bibinfo{publisher}{{ACM}}.
\newblock
\urldef\tempurl%
\url{https://doi.org/10.1145/2623330.2623373}
\showDOI{\tempurl}


\bibitem[Rashid et~al\mbox{.}(2020)]%
        {4-12}
\bibfield{author}{\bibinfo{person}{Md~Tahmid Rashid}, \bibinfo{person}{Daniel~Yue Zhang}, {and} \bibinfo{person}{Dong Wang}.} \bibinfo{year}{2020}\natexlab{}.
\newblock \showarticletitle{{SocialDrone}: An Integrated Social Media and Drone Sensing System for Reliable Disaster Response}. In \bibinfo{booktitle}{\emph{{IEEE} {INFOCOM} 2020 - {IEEE} Conference on Computer Communications}}. \bibinfo{publisher}{{IEEE}}.
\newblock
\urldef\tempurl%
\url{https://doi.org/10.1109/infocom41043.2020.9155522}
\showDOI{\tempurl}


\bibitem[Rebane et~al\mbox{.}(2019)]%
        {1-172}
\bibfield{author}{\bibinfo{person}{Jonathan Rebane}, \bibinfo{person}{Isak Karlsson}, {and} \bibinfo{person}{Panagiotis Papapetrou}.} \bibinfo{year}{2019}\natexlab{}.
\newblock \showarticletitle{An Investigation of Interpretable Deep Learning for Adverse Drug Event Prediction}. In \bibinfo{booktitle}{\emph{2019 {IEEE} 32nd International Symposium on Computer-Based Medical Systems ({CBMS})}}. \bibinfo{publisher}{{IEEE}}.
\newblock
\urldef\tempurl%
\url{https://doi.org/10.1109/cbms.2019.00075}
\showDOI{\tempurl}


\bibitem[Reid et~al\mbox{.}(2018)]%
        {1-173}
\bibfield{author}{\bibinfo{person}{David Reid}, \bibinfo{person}{Abir~Jaafar Hussain}, \bibinfo{person}{Hissam Tawfik}, \bibinfo{person}{Rozaida Ghazali}, {and} \bibinfo{person}{Dhiya Al-Jumeily}.} \bibinfo{year}{2018}\natexlab{}.
\newblock \showarticletitle{Forecasting Natural Events Using Axonal Delay}. In \bibinfo{booktitle}{\emph{2018 {IEEE} Congress on Evolutionary Computation ({CEC})}}. \bibinfo{publisher}{{IEEE}}.
\newblock
\urldef\tempurl%
\url{https://doi.org/10.1109/cec.2018.8477831}
\showDOI{\tempurl}


\bibitem[Ren et~al\mbox{.}(2018)]%
        {1-174}
\bibfield{author}{\bibinfo{person}{Honglei Ren}, \bibinfo{person}{You Song}, \bibinfo{person}{Jingwen Wang}, \bibinfo{person}{Yucheng Hu}, {and} \bibinfo{person}{Jinzhi Lei}.} \bibinfo{year}{2018}\natexlab{}.
\newblock \showarticletitle{A Deep Learning Approach to the Citywide Traffic Accident Risk Prediction}. In \bibinfo{booktitle}{\emph{2018 21st International Conference on Intelligent Transportation Systems ({ITSC})}}. \bibinfo{publisher}{{IEEE}}.
\newblock
\urldef\tempurl%
\url{https://doi.org/10.1109/itsc.2018.8569437}
\showDOI{\tempurl}


\bibitem[Resch et~al\mbox{.}(2017a)]%
        {2-42}
\bibfield{author}{\bibinfo{person}{Bernd Resch}, \bibinfo{person}{Florian Usländer}, {and} \bibinfo{person}{Clemens Havas}.} \bibinfo{year}{2017}\natexlab{a}.
\newblock \showarticletitle{Combining machine-learning topic models and spatiotemporal analysis of social media data for disaster footprint and damage assessment}.
\newblock \bibinfo{journal}{\emph{Cartography and Geographic Information Science}} \bibinfo{volume}{45}, \bibinfo{number}{4} (\bibinfo{date}{aug} \bibinfo{year}{2017}), \bibinfo{pages}{362--376}.
\newblock
\urldef\tempurl%
\url{https://doi.org/10.1080/15230406.2017.1356242}
\showDOI{\tempurl}


\bibitem[Resch et~al\mbox{.}(2017b)]%
        {2-50}
\bibfield{author}{\bibinfo{person}{Bernd Resch}, \bibinfo{person}{Florian Usländer}, {and} \bibinfo{person}{Clemens Havas}.} \bibinfo{year}{2017}\natexlab{b}.
\newblock \showarticletitle{Combining machine-learning topic models and spatiotemporal analysis of social media data for disaster footprint and damage assessment}.
\newblock \bibinfo{journal}{\emph{Cartography and Geographic Information Science}} \bibinfo{volume}{45}, \bibinfo{number}{4} (\bibinfo{date}{aug} \bibinfo{year}{2017}), \bibinfo{pages}{362--376}.
\newblock
\urldef\tempurl%
\url{https://doi.org/10.1080/15230406.2017.1356242}
\showDOI{\tempurl}


\bibitem[Reyes et~al\mbox{.}(2013)]%
        {1-175}
\bibfield{author}{\bibinfo{person}{J. Reyes}, \bibinfo{person}{A. Morales-Esteban}, {and} \bibinfo{person}{F. Mart{\'{\i}}nez-{\'{A}}lvarez}.} \bibinfo{year}{2013}\natexlab{}.
\newblock \showarticletitle{Neural networks to predict earthquakes in Chile}.
\newblock \bibinfo{journal}{\emph{Applied Soft Computing}} \bibinfo{volume}{13}, \bibinfo{number}{2} (\bibinfo{date}{feb} \bibinfo{year}{2013}), \bibinfo{pages}{1314--1328}.
\newblock
\urldef\tempurl%
\url{https://doi.org/10.1016/j.asoc.2012.10.014}
\showDOI{\tempurl}


\bibitem[Ristea et~al\mbox{.}(2020)]%
        {1-176}
\bibfield{author}{\bibinfo{person}{Alina Ristea}, \bibinfo{person}{Mohammad~Al Boni}, \bibinfo{person}{Bernd Resch}, \bibinfo{person}{Matthew~S. Gerber}, {and} \bibinfo{person}{Michael Leitner}.} \bibinfo{year}{2020}\natexlab{}.
\newblock \showarticletitle{Spatial crime distribution and prediction for sporting events using social media}.
\newblock \bibinfo{journal}{\emph{International Journal of Geographical Information Science}} \bibinfo{volume}{34}, \bibinfo{number}{9} (\bibinfo{date}{feb} \bibinfo{year}{2020}), \bibinfo{pages}{1708--1739}.
\newblock
\urldef\tempurl%
\url{https://doi.org/10.1080/13658816.2020.1719495}
\showDOI{\tempurl}


\bibitem[Rizk et~al\mbox{.}(2019)]%
        {2-46}
\bibfield{author}{\bibinfo{person}{Yara Rizk}, \bibinfo{person}{Hadi~Samer Jomaa}, \bibinfo{person}{Mariette Awad}, {and} \bibinfo{person}{Carlos Castillo}.} \bibinfo{year}{2019}\natexlab{}.
\newblock \showarticletitle{A computationally efficient multi-modal classification approach of disaster-related Twitter images}. In \bibinfo{booktitle}{\emph{Proceedings of the 34th {ACM}/{SIGAPP} Symposium on Applied Computing}}. \bibinfo{publisher}{{ACM}}.
\newblock
\urldef\tempurl%
\url{https://doi.org/10.1145/3297280.3297481}
\showDOI{\tempurl}


\bibitem[Rostami et~al\mbox{.}(2018)]%
        {1-177}
\bibfield{author}{\bibinfo{person}{Mohammad Rostami}, \bibinfo{person}{David Huber}, {and} \bibinfo{person}{Tsai-Ching Lu}.} \bibinfo{year}{2018}\natexlab{}.
\newblock \showarticletitle{A crowdsourcing triage algorithm for geopolitical event forecasting}. In \bibinfo{booktitle}{\emph{Proceedings of the 12th {ACM} Conference on Recommender Systems}}. \bibinfo{publisher}{{ACM}}.
\newblock
\urldef\tempurl%
\url{https://doi.org/10.1145/3240323.3240385}
\showDOI{\tempurl}


\bibitem[Rouet-Leduc et~al\mbox{.}(2017)]%
        {1-178}
\bibfield{author}{\bibinfo{person}{Bertrand Rouet-Leduc}, \bibinfo{person}{Claudia Hulbert}, \bibinfo{person}{Nicholas Lubbers}, \bibinfo{person}{Kipton Barros}, \bibinfo{person}{Colin~J. Humphreys}, {and} \bibinfo{person}{Paul~A. Johnson}.} \bibinfo{year}{2017}\natexlab{}.
\newblock \showarticletitle{Machine Learning Predicts Laboratory Earthquakes}.
\newblock \bibinfo{journal}{\emph{Geophysical Research Letters}} \bibinfo{volume}{44}, \bibinfo{number}{18} (\bibinfo{date}{sep} \bibinfo{year}{2017}), \bibinfo{pages}{9276--9282}.
\newblock
\urldef\tempurl%
\url{https://doi.org/10.1002/2017gl074677}
\showDOI{\tempurl}


\bibitem[Sakaki et~al\mbox{.}(2010)]%
        {1-181}
\bibfield{author}{\bibinfo{person}{Takeshi Sakaki}, \bibinfo{person}{Makoto Okazaki}, {and} \bibinfo{person}{Yutaka Matsuo}.} \bibinfo{year}{2010}\natexlab{}.
\newblock \showarticletitle{Earthquake shakes Twitter users}. In \bibinfo{booktitle}{\emph{Proceedings of the 19th international conference on World wide web}}. \bibinfo{publisher}{{ACM}}.
\newblock
\urldef\tempurl%
\url{https://doi.org/10.1145/1772690.1772777}
\showDOI{\tempurl}


\bibitem[Saldana et~al\mbox{.}(2015)]%
        {4-19}
\bibfield{author}{\bibinfo{person}{David Saldana}, \bibinfo{person}{Renato Assuncao}, {and} \bibinfo{person}{Mario F.~M. Campos}.} \bibinfo{year}{2015}\natexlab{}.
\newblock \showarticletitle{A distributed multi-robot approach for the detection and tracking of multiple dynamic anomalies}. In \bibinfo{booktitle}{\emph{2015 {IEEE} International Conference on Robotics and Automation ({ICRA})}}. \bibinfo{publisher}{{IEEE}}.
\newblock
\urldef\tempurl%
\url{https://doi.org/10.1109/icra.2015.7139353}
\showDOI{\tempurl}


\bibitem[Salfner and Malek(2007)]%
        {1-183}
\bibfield{author}{\bibinfo{person}{Felix Salfner} {and} \bibinfo{person}{Miroslaw Malek}.} \bibinfo{year}{2007}\natexlab{}.
\newblock \showarticletitle{Using Hidden Semi-Markov Models for Effective Online Failure Prediction}. In \bibinfo{booktitle}{\emph{2007 26th {IEEE} International Symposium on Reliable Distributed Systems ({SRDS} 2007)}}. \bibinfo{publisher}{{IEEE}}.
\newblock
\urldef\tempurl%
\url{https://doi.org/10.1109/srds.2007.35}
\showDOI{\tempurl}


\bibitem[Santos et~al\mbox{.}(2014)]%
        {1-67}
\bibfield{author}{\bibinfo{person}{Raimundo~Dos Santos}, \bibinfo{person}{Sumit Shah}, \bibinfo{person}{Feng Chen}, \bibinfo{person}{Arnold Boedihardjo}, \bibinfo{person}{Chang-Tien Lu}, {and} \bibinfo{person}{Naren Ramakrishnan}.} \bibinfo{year}{2014}\natexlab{}.
\newblock \showarticletitle{Forecasting location-based events with spatio-temporal storytelling}. In \bibinfo{booktitle}{\emph{Proceedings of the 7th {ACM} {SIGSPATIAL} International Workshop on Location-Based Social Networks}}. \bibinfo{publisher}{{ACM}}.
\newblock
\urldef\tempurl%
\url{https://doi.org/10.1145/2755492.2755496}
\showDOI{\tempurl}


\bibitem[Scawthorn et~al\mbox{.}(2006a)]%
        {2-43}
\bibfield{author}{\bibinfo{person}{Charles Scawthorn}, \bibinfo{person}{Paul Flores}, \bibinfo{person}{Neil Blais}, \bibinfo{person}{Hope Seligson}, \bibinfo{person}{Eric Tate}, \bibinfo{person}{Stephanie Chang}, \bibinfo{person}{Edward Mifflin}, \bibinfo{person}{Will Thomas}, \bibinfo{person}{James Murphy}, \bibinfo{person}{Christopher Jones}, {and} \bibinfo{person}{Michael Lawrence}.} \bibinfo{year}{2006}\natexlab{a}.
\newblock \showarticletitle{{HAZUS}-{MH} Flood Loss Estimation Methodology. {II}. Damage and Loss Assessment}.
\newblock \bibinfo{journal}{\emph{Natural Hazards Review}} \bibinfo{volume}{7}, \bibinfo{number}{2} (\bibinfo{date}{may} \bibinfo{year}{2006}), \bibinfo{pages}{72--81}.
\newblock
\urldef\tempurl%
\url{https://doi.org/10.1061/(asce)1527-6988(2006)7:2(72)}
\showDOI{\tempurl}


\bibitem[Scawthorn et~al\mbox{.}(2006b)]%
        {2-51}
\bibfield{author}{\bibinfo{person}{Charles Scawthorn}, \bibinfo{person}{Paul Flores}, \bibinfo{person}{Neil Blais}, \bibinfo{person}{Hope Seligson}, \bibinfo{person}{Eric Tate}, \bibinfo{person}{Stephanie Chang}, \bibinfo{person}{Edward Mifflin}, \bibinfo{person}{Will Thomas}, \bibinfo{person}{James Murphy}, \bibinfo{person}{Christopher Jones}, {and} \bibinfo{person}{Michael Lawrence}.} \bibinfo{year}{2006}\natexlab{b}.
\newblock \showarticletitle{{HAZUS}-{MH} Flood Loss Estimation Methodology. {II}. Damage and Loss Assessment}.
\newblock \bibinfo{journal}{\emph{Natural Hazards Review}} \bibinfo{volume}{7}, \bibinfo{number}{2} (\bibinfo{date}{may} \bibinfo{year}{2006}), \bibinfo{pages}{72--81}.
\newblock
\urldef\tempurl%
\url{https://doi.org/10.1061/(asce)1527-6988(2006)7:2(72)}
\showDOI{\tempurl}


\bibitem[Shao et~al\mbox{.}(2017)]%
        {1-187}
\bibfield{author}{\bibinfo{person}{Minglai Shao}, \bibinfo{person}{Jianxin Li}, \bibinfo{person}{Feng Chen}, \bibinfo{person}{Hongyi Huang}, \bibinfo{person}{Shuai Zhang}, {and} \bibinfo{person}{Xunxun Chen}.} \bibinfo{year}{2017}\natexlab{}.
\newblock \showarticletitle{An Efficient Approach to Event Detection and Forecasting in Dynamic Multivariate Social Media Networks}. In \bibinfo{booktitle}{\emph{Proceedings of the 26th International Conference on World Wide Web}}. \bibinfo{publisher}{International World Wide Web Conferences Steering Committee}.
\newblock
\urldef\tempurl%
\url{https://doi.org/10.1145/3038912.3052588}
\showDOI{\tempurl}


\bibitem[Shen et~al\mbox{.}({[n.\,d.]})]%
        {1-225}
\bibfield{author}{\bibinfo{person}{Yi-Dong Shen}, \bibinfo{person}{Zhong Zhang}, {and} \bibinfo{person}{Qiang Yang}.} \bibinfo{year}{[n.\,d.]}\natexlab{}.
\newblock \showarticletitle{Objective-oriented utility-based association mining}. In \bibinfo{booktitle}{\emph{2002 {IEEE} International Conference on Data Mining, 2002. Proceedings.}} \bibinfo{publisher}{{IEEE} Comput. Soc}.
\newblock
\urldef\tempurl%
\url{https://doi.org/10.1109/icdm.2002.1183938}
\showDOI{\tempurl}


\bibitem[Srinivasa et~al\mbox{.}(2008)]%
        {1-194}
\bibfield{author}{\bibinfo{person}{Narayan Srinivasa}, \bibinfo{person}{Qin Jiang}, {and} \bibinfo{person}{Leandro~G. Barajas}.} \bibinfo{year}{2008}\natexlab{}.
\newblock \showarticletitle{High-Impact Event Prediction by Temporal Data Mining through Genetic Algorithms}. In \bibinfo{booktitle}{\emph{2008 Fourth International Conference on Natural Computation}}. \bibinfo{publisher}{{IEEE}}.
\newblock
\urldef\tempurl%
\url{https://doi.org/10.1109/icnc.2008.761}
\showDOI{\tempurl}


\bibitem[Su and Jiang(2020)]%
        {1-195}
\bibfield{author}{\bibinfo{person}{Zichun Su} {and} \bibinfo{person}{Jialin Jiang}.} \bibinfo{year}{2020}\natexlab{}.
\newblock \showarticletitle{Hierarchical Gated Recurrent Unit with Semantic Attention for Event Prediction}.
\newblock \bibinfo{journal}{\emph{Future Internet}} \bibinfo{volume}{12}, \bibinfo{number}{2} (\bibinfo{date}{feb} \bibinfo{year}{2020}), \bibinfo{pages}{39}.
\newblock
\urldef\tempurl%
\url{https://doi.org/10.3390/fi12020039}
\showDOI{\tempurl}


\bibitem[Sun et~al\mbox{.}(2020)]%
        {disaster}
\bibfield{author}{\bibinfo{person}{Wenjuan Sun}, \bibinfo{person}{Paolo Bocchini}, {and} \bibinfo{person}{Brian~D Davison}.} \bibinfo{year}{2020}\natexlab{}.
\newblock \showarticletitle{Applications of artificial intelligence for disaster management}.
\newblock \bibinfo{journal}{\emph{Natural Hazards}} \bibinfo{volume}{103}, \bibinfo{number}{3} (\bibinfo{year}{2020}), \bibinfo{pages}{2631--2689}.
\newblock


\bibitem[Szabo and Huberman(2010)]%
        {181}
\bibfield{author}{\bibinfo{person}{Gabor Szabo} {and} \bibinfo{person}{Bernardo~A Huberman}.} \bibinfo{year}{2010}\natexlab{}.
\newblock \showarticletitle{Predicting the popularity of online content}.
\newblock \bibinfo{journal}{\emph{Commun. ACM}} \bibinfo{volume}{53}, \bibinfo{number}{8} (\bibinfo{year}{2010}), \bibinfo{pages}{80--88}.
\newblock


\bibitem[Tama and Comuzzi(2019)]%
        {1-198}
\bibfield{author}{\bibinfo{person}{Bayu~Adhi Tama} {and} \bibinfo{person}{Marco Comuzzi}.} \bibinfo{year}{2019}\natexlab{}.
\newblock \showarticletitle{An empirical comparison of classification techniques for next event prediction using business process event logs}.
\newblock \bibinfo{journal}{\emph{Expert Systems with Applications}}  \bibinfo{volume}{129} (\bibinfo{date}{sep} \bibinfo{year}{2019}), \bibinfo{pages}{233--245}.
\newblock
\urldef\tempurl%
\url{https://doi.org/10.1016/j.eswa.2019.04.016}
\showDOI{\tempurl}


\bibitem[Tatar et~al\mbox{.}(2014)]%
        {185}
\bibfield{author}{\bibinfo{person}{Alexandru Tatar}, \bibinfo{person}{Marcelo~Dias De~Amorim}, \bibinfo{person}{Serge Fdida}, {and} \bibinfo{person}{Panayotis Antoniadis}.} \bibinfo{year}{2014}\natexlab{}.
\newblock \showarticletitle{A survey on predicting the popularity of web content}.
\newblock \bibinfo{journal}{\emph{Journal of Internet Services and Applications}} \bibinfo{volume}{5}, \bibinfo{number}{1} (\bibinfo{year}{2014}), \bibinfo{pages}{1--20}.
\newblock


\bibitem[Tatar et~al\mbox{.}(2011)]%
        {186}
\bibfield{author}{\bibinfo{person}{Alexandru Tatar}, \bibinfo{person}{J{\'e}r{\'e}mie Leguay}, \bibinfo{person}{Panayotis Antoniadis}, \bibinfo{person}{Arnaud Limbourg}, \bibinfo{person}{Marcelo~Dias de Amorim}, {and} \bibinfo{person}{Serge Fdida}.} \bibinfo{year}{2011}\natexlab{}.
\newblock \showarticletitle{Predicting the popularity of online articles based on user comments}. In \bibinfo{booktitle}{\emph{Proceedings of the International Conference on Web Intelligence, Mining and Semantics}}. \bibinfo{pages}{1--8}.
\newblock


\bibitem[Taylor(2017)]%
        {1-199}
\bibfield{author}{\bibinfo{person}{James~W. Taylor}.} \bibinfo{year}{2017}\natexlab{}.
\newblock \showarticletitle{Probabilistic forecasting of wind power ramp events using autoregressive logit models}.
\newblock \bibinfo{journal}{\emph{European Journal of Operational Research}} \bibinfo{volume}{259}, \bibinfo{number}{2} (\bibinfo{date}{jun} \bibinfo{year}{2017}), \bibinfo{pages}{703--712}.
\newblock
\urldef\tempurl%
\url{https://doi.org/10.1016/j.ejor.2016.10.041}
\showDOI{\tempurl}


\bibitem[Tijtgat et~al\mbox{.}(2017)]%
        {4-13}
\bibfield{author}{\bibinfo{person}{Nils Tijtgat}, \bibinfo{person}{Wiebe~Van Ranst}, \bibinfo{person}{Bruno Volckaert}, \bibinfo{person}{Toon Goedeme}, {and} \bibinfo{person}{Filip~De Turck}.} \bibinfo{year}{2017}\natexlab{}.
\newblock \showarticletitle{Embedded Real-Time Object Detection for a {UAV} Warning System}. In \bibinfo{booktitle}{\emph{2017 {IEEE} International Conference on Computer Vision Workshops ({ICCVW})}}. \bibinfo{publisher}{{IEEE}}.
\newblock
\urldef\tempurl%
\url{https://doi.org/10.1109/iccvw.2017.247}
\showDOI{\tempurl}


\bibitem[Tsagkias et~al\mbox{.}(2010)]%
        {190}
\bibfield{author}{\bibinfo{person}{Manos Tsagkias}, \bibinfo{person}{Wouter Weerkamp}, {and} \bibinfo{person}{Maarten De~Rijke}.} \bibinfo{year}{2010}\natexlab{}.
\newblock \showarticletitle{News comments: Exploring, modeling, and online prediction}. In \bibinfo{booktitle}{\emph{Advances in Information Retrieval: 32nd European Conference on IR Research, ECIR 2010, Milton Keynes, UK, March 28-31, 2010. Proceedings 32}}. Springer, \bibinfo{pages}{191--203}.
\newblock


\bibitem[Vahedian et~al\mbox{.}(2017)]%
        {1-203}
\bibfield{author}{\bibinfo{person}{Amin Vahedian}, \bibinfo{person}{Xun Zhou}, \bibinfo{person}{Ling Tong}, \bibinfo{person}{Yanhua Li}, {and} \bibinfo{person}{Jun Luo}.} \bibinfo{year}{2017}\natexlab{}.
\newblock \showarticletitle{Forecasting Gathering Events through Continuous Destination Prediction on Big Trajectory Data}. In \bibinfo{booktitle}{\emph{Proceedings of the 25th {ACM} {SIGSPATIAL} International Conference on Advances in Geographic Information Systems}}. \bibinfo{publisher}{{ACM}}.
\newblock
\urldef\tempurl%
\url{https://doi.org/10.1145/3139958.3140008}
\showDOI{\tempurl}


\bibitem[van Noord et~al\mbox{.}(2017)]%
        {1-201}
\bibfield{author}{\bibinfo{person}{Rik van Noord}, \bibinfo{person}{Florian~A. Kunneman}, {and} \bibinfo{person}{Antal van~den Bosch}.} \bibinfo{year}{2017}\natexlab{}.
\newblock \showarticletitle{Predicting Civil Unrest by Categorizing Dutch Twitter Events}.
\newblock In \bibinfo{booktitle}{\emph{Communications in Computer and Information Science}}. \bibinfo{publisher}{Springer International Publishing}, \bibinfo{pages}{3--16}.
\newblock
\urldef\tempurl%
\url{https://doi.org/10.1007/978-3-319-67468-1_1}
\showDOI{\tempurl}


\bibitem[Vilalta and Ma({[n.\,d.]})]%
        {1-206}
\bibfield{author}{\bibinfo{person}{R. Vilalta} {and} \bibinfo{person}{Sheng Ma}.} \bibinfo{year}{[n.\,d.]}\natexlab{}.
\newblock \showarticletitle{Predicting rare events in temporal domains}. In \bibinfo{booktitle}{\emph{2002 {IEEE} International Conference on Data Mining, 2002. Proceedings.}} \bibinfo{publisher}{{IEEE} Comput. Soc}.
\newblock
\urldef\tempurl%
\url{https://doi.org/10.1109/icdm.2002.1183991}
\showDOI{\tempurl}


\bibitem[Wang et~al\mbox{.}(2019b)]%
        {1-209}
\bibfield{author}{\bibinfo{person}{Bao Wang}, \bibinfo{person}{Penghang Yin}, \bibinfo{person}{Andrea~Louise Bertozzi}, \bibinfo{person}{P.~Jeffrey Brantingham}, \bibinfo{person}{Stanley~Joel Osher}, {and} \bibinfo{person}{Jack Xin}.} \bibinfo{year}{2019}\natexlab{b}.
\newblock \showarticletitle{Deep Learning for Real-Time Crime Forecasting and Its Ternarization}.
\newblock \bibinfo{journal}{\emph{Chinese Annals of Mathematics, Series B}} \bibinfo{volume}{40}, \bibinfo{number}{6} (\bibinfo{date}{nov} \bibinfo{year}{2019}), \bibinfo{pages}{949--966}.
\newblock
\urldef\tempurl%
\url{https://doi.org/10.1007/s11401-019-0168-y}
\showDOI{\tempurl}


\bibitem[Wang et~al\mbox{.}(2017)]%
        {4-1}
\bibfield{author}{\bibinfo{person}{Chen Wang}, \bibinfo{person}{Hongzhi Lin}, \bibinfo{person}{Rui Zhang}, {and} \bibinfo{person}{Hongbo Jiang}.} \bibinfo{year}{2017}\natexlab{}.
\newblock \showarticletitle{{SEND}: A Situation-Aware Emergency Navigation Algorithm with Sensor Networks}.
\newblock \bibinfo{journal}{\emph{{IEEE} Transactions on Mobile Computing}} \bibinfo{volume}{16}, \bibinfo{number}{4} (\bibinfo{date}{apr} \bibinfo{year}{2017}), \bibinfo{pages}{1149--1162}.
\newblock
\urldef\tempurl%
\url{https://doi.org/10.1109/tmc.2016.2582172}
\showDOI{\tempurl}


\bibitem[Wang et~al\mbox{.}(2015)]%
        {1-210}
\bibfield{author}{\bibinfo{person}{Chen Wang}, \bibinfo{person}{Hoang~Tam Vo}, {and} \bibinfo{person}{Peng Ni}.} \bibinfo{year}{2015}\natexlab{}.
\newblock \showarticletitle{An {IoT} Application for Fault Diagnosis and Prediction}. In \bibinfo{booktitle}{\emph{2015 {IEEE} International Conference on Data Science and Data Intensive Systems}}. \bibinfo{publisher}{{IEEE}}.
\newblock
\urldef\tempurl%
\url{https://doi.org/10.1109/dsdis.2015.97}
\showDOI{\tempurl}


\bibitem[Wang and Ding(2015)]%
        {1-211}
\bibfield{author}{\bibinfo{person}{Dawei Wang} {and} \bibinfo{person}{Wei Ding}.} \bibinfo{year}{2015}\natexlab{}.
\newblock \showarticletitle{A Hierarchical Pattern Learning Framework for Forecasting Extreme Weather Events}. In \bibinfo{booktitle}{\emph{2015 {IEEE} International Conference on Data Mining}}. \bibinfo{publisher}{{IEEE}}.
\newblock
\urldef\tempurl%
\url{https://doi.org/10.1109/icdm.2015.93}
\showDOI{\tempurl}


\bibitem[Wang et~al\mbox{.}(2013)]%
        {1-212}
\bibfield{author}{\bibinfo{person}{Dawei Wang}, \bibinfo{person}{Wei Ding}, \bibinfo{person}{Kui Yu}, \bibinfo{person}{Xindong Wu}, \bibinfo{person}{Ping Chen}, \bibinfo{person}{David~L. Small}, {and} \bibinfo{person}{Shafiqul Islam}.} \bibinfo{year}{2013}\natexlab{}.
\newblock \showarticletitle{Towards long-lead forecasting of extreme flood events}. In \bibinfo{booktitle}{\emph{Proceedings of the 19th {ACM} {SIGKDD} international conference on Knowledge discovery and data mining}}. \bibinfo{publisher}{{ACM}}.
\newblock
\urldef\tempurl%
\url{https://doi.org/10.1145/2487575.2488220}
\showDOI{\tempurl}


\bibitem[Wang et~al\mbox{.}(2020a)]%
        {iot_12}
\bibfield{author}{\bibinfo{person}{Guang Wang}, \bibinfo{person}{Zhihan Fang}, \bibinfo{person}{Xiaoyang Xie}, \bibinfo{person}{Shuai Wang}, \bibinfo{person}{Huijun Sun}, \bibinfo{person}{Fan Zhang}, \bibinfo{person}{Yunhuai Liu}, {and} \bibinfo{person}{Desheng Zhang}.} \bibinfo{year}{2020}\natexlab{a}.
\newblock \showarticletitle{Pricing-aware Real-time Charging Scheduling and Charging Station Expansion for Large-scale Electric Buses}.
\newblock \bibinfo{journal}{\emph{{ACM} Transactions on Intelligent Systems and Technology}} \bibinfo{volume}{12}, \bibinfo{number}{1} (\bibinfo{date}{nov} \bibinfo{year}{2020}), \bibinfo{pages}{1--26}.
\newblock
\urldef\tempurl%
\url{https://doi.org/10.1145/3428080}
\showDOI{\tempurl}


\bibitem[Wang et~al\mbox{.}(2021a)]%
        {iot_4}
\bibfield{author}{\bibinfo{person}{Guang Wang}, \bibinfo{person}{Zhou Qin}, \bibinfo{person}{Shuai Wang}, \bibinfo{person}{Huijun Sun}, \bibinfo{person}{Zheng Dong}, {and} \bibinfo{person}{Desheng Zhang}.} \bibinfo{year}{2021}\natexlab{a}.
\newblock \showarticletitle{Record: Joint Real-Time Repositioning and Charging for Electric Carsharing with Dynamic Deadlines}. In \bibinfo{booktitle}{\emph{Proceedings of the 27th {ACM} {SIGKDD} Conference on Knowledge Discovery and Data Mining}}. \bibinfo{publisher}{{ACM}}.
\newblock
\urldef\tempurl%
\url{https://doi.org/10.1145/3447548.3467112}
\showDOI{\tempurl}


\bibitem[Wang et~al\mbox{.}(2020d)]%
        {iot_6}
\bibfield{author}{\bibinfo{person}{Guang Wang}, \bibinfo{person}{Harsh~Rajkumar Vaish}, \bibinfo{person}{Huijun Sun}, \bibinfo{person}{Jianjun Wu}, \bibinfo{person}{Shuai Wang}, {and} \bibinfo{person}{Desheng Zhang}.} \bibinfo{year}{2020}\natexlab{d}.
\newblock \showarticletitle{Understanding User Behavior in Car Sharing Services Through The Lens of Mobility}.
\newblock \bibinfo{journal}{\emph{Proceedings of the {ACM} on Interactive, Mobile, Wearable and Ubiquitous Technologies}} \bibinfo{volume}{4}, \bibinfo{number}{4} (\bibinfo{date}{dec} \bibinfo{year}{2020}), \bibinfo{pages}{1--30}.
\newblock
\urldef\tempurl%
\url{https://doi.org/10.1145/3432200}
\showDOI{\tempurl}


\bibitem[Wang et~al\mbox{.}(2020e)]%
        {iot_9}
\bibfield{author}{\bibinfo{person}{Guang Wang}, \bibinfo{person}{Yongfeng Zhang}, \bibinfo{person}{Zhihan Fang}, \bibinfo{person}{Shuai Wang}, \bibinfo{person}{Fan Zhang}, {and} \bibinfo{person}{Desheng Zhang}.} \bibinfo{year}{2020}\natexlab{e}.
\newblock \showarticletitle{{FairCharge:}A Data-Driven Fairness-Aware Charging Recommendation System for Large-Scale Electric Taxi Fleets}.
\newblock \bibinfo{journal}{\emph{Proceedings of the {ACM} on Interactive, Mobile, Wearable and Ubiquitous Technologies}} \bibinfo{volume}{4}, \bibinfo{number}{1} (\bibinfo{date}{mar} \bibinfo{year}{2020}), \bibinfo{pages}{1--25}.
\newblock
\urldef\tempurl%
\url{https://doi.org/10.1145/3381003}
\showDOI{\tempurl}


\bibitem[Wang et~al\mbox{.}(2021b)]%
        {iot_7}
\bibfield{author}{\bibinfo{person}{Guang Wang}, \bibinfo{person}{Shuxin Zhong}, \bibinfo{person}{Shuai Wang}, \bibinfo{person}{Fei Miao}, \bibinfo{person}{Zheng Dong}, {and} \bibinfo{person}{Desheng Zhang}.} \bibinfo{year}{2021}\natexlab{b}.
\newblock \showarticletitle{Data-Driven Fairness-Aware Vehicle Displacement for Large-Scale Electric Taxi Fleets}. In \bibinfo{booktitle}{\emph{2021 {IEEE} 37th International Conference on Data Engineering ({ICDE})}}. \bibinfo{publisher}{{IEEE}}.
\newblock
\urldef\tempurl%
\url{https://doi.org/10.1109/icde51399.2021.00108}
\showDOI{\tempurl}


\bibitem[Wang and Gerber(2015)]%
        {1-214}
\bibfield{author}{\bibinfo{person}{Mingjun Wang} {and} \bibinfo{person}{Matthew~S. Gerber}.} \bibinfo{year}{2015}\natexlab{}.
\newblock \showarticletitle{Using Twitter for Next-Place Prediction, with an Application to Crime Prediction}. In \bibinfo{booktitle}{\emph{2015 {IEEE} Symposium Series on Computational Intelligence}}. \bibinfo{publisher}{{IEEE}}.
\newblock
\urldef\tempurl%
\url{https://doi.org/10.1109/ssci.2015.138}
\showDOI{\tempurl}


\bibitem[Wang et~al\mbox{.}(2020b)]%
        {1-215}
\bibfield{author}{\bibinfo{person}{Qi Wang}, \bibinfo{person}{Guangyin Jin}, \bibinfo{person}{Xia Zhao}, \bibinfo{person}{Yanghe Feng}, {and} \bibinfo{person}{Jincai Huang}.} \bibinfo{year}{2020}\natexlab{b}.
\newblock \showarticletitle{{CSAN}: A neural network benchmark model for crime forecasting in spatio-temporal scale}.
\newblock \bibinfo{journal}{\emph{Knowledge-Based Systems}}  \bibinfo{volume}{189} (\bibinfo{date}{feb} \bibinfo{year}{2020}), \bibinfo{pages}{105120}.
\newblock
\urldef\tempurl%
\url{https://doi.org/10.1016/j.knosys.2019.105120}
\showDOI{\tempurl}


\bibitem[Wang et~al\mbox{.}(2019a)]%
        {iot_10}
\bibfield{author}{\bibinfo{person}{Shuai Wang}, \bibinfo{person}{Tian He}, \bibinfo{person}{Desheng Zhang}, \bibinfo{person}{Yunhuai Liu}, {and} \bibinfo{person}{Sang~H. Son}.} \bibinfo{year}{2019}\natexlab{a}.
\newblock \showarticletitle{Towards Efficient Sharing: A Usage Balancing Mechanism for Bike Sharing Systems}. In \bibinfo{booktitle}{\emph{The World Wide Web Conference}}. \bibinfo{publisher}{{ACM}}.
\newblock
\urldef\tempurl%
\url{https://doi.org/10.1145/3308558.3313441}
\showDOI{\tempurl}


\bibitem[Wang et~al\mbox{.}(2020c)]%
        {2-41}
\bibfield{author}{\bibinfo{person}{Tianyi Wang}, \bibinfo{person}{Yudong Tao}, \bibinfo{person}{Shu-Ching Chen}, {and} \bibinfo{person}{Mei-Ling Shyu}.} \bibinfo{year}{2020}\natexlab{c}.
\newblock \showarticletitle{Multi-task Multimodal Learning for Disaster Situation Assessment}. In \bibinfo{booktitle}{\emph{2020 {IEEE} Conference on Multimedia Information Processing and Retrieval ({MIPR})}}. \bibinfo{publisher}{{IEEE}}.
\newblock
\urldef\tempurl%
\url{https://doi.org/10.1109/mipr49039.2020.00050}
\showDOI{\tempurl}


\bibitem[Wang and Zhang(2017)]%
        {1-217}
\bibfield{author}{\bibinfo{person}{Zhongqing Wang} {and} \bibinfo{person}{Yue Zhang}.} \bibinfo{year}{2017}\natexlab{}.
\newblock \showarticletitle{{DDoS} Event Forecasting using Twitter Data}. In \bibinfo{booktitle}{\emph{Proceedings of the Twenty-Sixth International Joint Conference on Artificial Intelligence}}. \bibinfo{publisher}{International Joint Conferences on Artificial Intelligence Organization}.
\newblock
\urldef\tempurl%
\url{https://doi.org/10.24963/ijcai.2017/580}
\showDOI{\tempurl}


\bibitem[Weber et~al\mbox{.}(2020)]%
        {3-1}
\bibfield{author}{\bibinfo{person}{Ethan Weber}, \bibinfo{person}{Nuria Marzo}, \bibinfo{person}{Dim~P. Papadopoulos}, \bibinfo{person}{Aritro Biswas}, \bibinfo{person}{Agata Lapedriza}, \bibinfo{person}{Ferda Ofli}, \bibinfo{person}{Muhammad Imran}, {and} \bibinfo{person}{Antonio Torralba}.} \bibinfo{year}{2020}\natexlab{}.
\newblock \showarticletitle{Detecting Natural Disasters, Damage, and Incidents in the Wild}.
\newblock In \bibinfo{booktitle}{\emph{Computer Vision {\textendash} {ECCV} 2020}}. \bibinfo{publisher}{Springer International Publishing}, \bibinfo{pages}{331--350}.
\newblock
\urldef\tempurl%
\url{https://doi.org/10.1007/978-3-030-58529-7_20}
\showDOI{\tempurl}


\bibitem[Weiss and Page(2013)]%
        {1-219}
\bibfield{author}{\bibinfo{person}{Jeremy~C. Weiss} {and} \bibinfo{person}{David Page}.} \bibinfo{year}{2013}\natexlab{}.
\newblock \showarticletitle{Forest-Based Point Process for Event Prediction from Electronic Health Records}.
\newblock In \bibinfo{booktitle}{\emph{Advanced Information Systems Engineering}}. \bibinfo{publisher}{Springer Berlin Heidelberg}, \bibinfo{pages}{547--562}.
\newblock
\urldef\tempurl%
\url{https://doi.org/10.1007/978-3-642-40994-3_35}
\showDOI{\tempurl}


\bibitem[Wu et~al\mbox{.}(2016)]%
        {210}
\bibfield{author}{\bibinfo{person}{Bo Wu}, \bibinfo{person}{Tao Mei}, \bibinfo{person}{Wen-Huang Cheng}, {and} \bibinfo{person}{Yongdong Zhang}.} \bibinfo{year}{2016}\natexlab{}.
\newblock \showarticletitle{Unfolding temporal dynamics: Predicting social media popularity using multi-scale temporal decomposition}. In \bibinfo{booktitle}{\emph{Proceedings of the AAAI Conference on Artificial Intelligence}}, Vol.~\bibinfo{volume}{30}.
\newblock


\bibitem[Wu et~al\mbox{.}(2019)]%
        {211}
\bibfield{author}{\bibinfo{person}{Qitian Wu}, \bibinfo{person}{Yirui Gao}, \bibinfo{person}{Xiaofeng Gao}, \bibinfo{person}{Paul Weng}, {and} \bibinfo{person}{Guihai Chen}.} \bibinfo{year}{2019}\natexlab{}.
\newblock \showarticletitle{Dual sequential prediction models linking sequential recommendation and information dissemination}. In \bibinfo{booktitle}{\emph{Proceedings of the 25th ACM SIGKDD international conference on knowledge discovery \& data mining}}. \bibinfo{pages}{447--457}.
\newblock


\bibitem[Xiao et~al\mbox{.}(2016)]%
        {214}
\bibfield{author}{\bibinfo{person}{Shuai Xiao}, \bibinfo{person}{Junchi Yan}, \bibinfo{person}{Changsheng Li}, \bibinfo{person}{Bo Jin}, \bibinfo{person}{Xiangfeng Wang}, \bibinfo{person}{Xiaokang Yang}, \bibinfo{person}{Stephen~M Chu}, {and} \bibinfo{person}{Hongyuan Zha}.} \bibinfo{year}{2016}\natexlab{}.
\newblock \showarticletitle{On Modeling and Predicting Individual Paper Citation Count over Time.}. In \bibinfo{booktitle}{\emph{Ijcai}}. \bibinfo{pages}{2676--2682}.
\newblock


\bibitem[Xiong et~al\mbox{.}(2019)]%
        {1-220}
\bibfield{author}{\bibinfo{person}{Chuanxiu Xiong}, \bibinfo{person}{Ajitesh Srivastava}, \bibinfo{person}{Rajgopal Kannan}, \bibinfo{person}{Omkar Damle}, \bibinfo{person}{Viktor Prasanna}, {and} \bibinfo{person}{Erroll Southers}.} \bibinfo{year}{2019}\natexlab{}.
\newblock \showarticletitle{On Predicting Crime with Heterogeneous Spatial Patterns}. In \bibinfo{booktitle}{\emph{Proceedings of the 27th {ACM} {SIGSPATIAL} International Conference on Advances in Geographic Information Systems}}. \bibinfo{publisher}{{ACM}}.
\newblock
\urldef\tempurl%
\url{https://doi.org/10.1145/3347146.3359374}
\showDOI{\tempurl}


\bibitem[Xue et~al\mbox{.}(2018)]%
        {1-221}
\bibfield{author}{\bibinfo{person}{Cong Xue}, \bibinfo{person}{Zehua Zeng}, \bibinfo{person}{Yuanye He}, \bibinfo{person}{Lei Wang}, {and} \bibinfo{person}{Neng Gao}.} \bibinfo{year}{2018}\natexlab{}.
\newblock \showarticletitle{A {MIML}-{LSTM} neural network for integrated fine-grained event forecasting}. In \bibinfo{booktitle}{\emph{Proceedings of 2018 International Conference on Big Data Technologies - {ICBDT} {\textquotesingle}18}}. \bibinfo{publisher}{{ACM} Press}.
\newblock
\urldef\tempurl%
\url{https://doi.org/10.1145/3226116.3226127}
\showDOI{\tempurl}


\bibitem[Yan et~al\mbox{.}(2018)]%
        {1-208}
\bibfield{author}{\bibinfo{person}{Hao Yan}, \bibinfo{person}{Kamran Paynabar}, {and} \bibinfo{person}{Jianjun Shi}.} \bibinfo{year}{2018}\natexlab{}.
\newblock \showarticletitle{Real-Time Monitoring of High-Dimensional Functional Data Streams via Spatio-Temporal Smooth Sparse Decomposition}.
\newblock \bibinfo{journal}{\emph{Technometrics}} \bibinfo{volume}{60}, \bibinfo{number}{2} (\bibinfo{date}{apr} \bibinfo{year}{2018}), \bibinfo{pages}{181--197}.
\newblock
\urldef\tempurl%
\url{https://doi.org/10.1080/00401706.2017.1346522}
\showDOI{\tempurl}


\bibitem[Yan et~al\mbox{.}(2022)]%
        {iot_3}
\bibfield{author}{\bibinfo{person}{Hua Yan}, \bibinfo{person}{Shuai Wang}, \bibinfo{person}{Yu Yang}, \bibinfo{person}{Baoshen Guo}, \bibinfo{person}{Tian He}, {and} \bibinfo{person}{Desheng Zhang}.} \bibinfo{year}{2022}\natexlab{}.
\newblock \showarticletitle{{\textdollar}O{\^{}}$\lbrace$2$\rbrace${\textdollar}-{SiteRec}: Store Site Recommendation under the O2O Model via Multi-graph Attention Networks}. In \bibinfo{booktitle}{\emph{2022 {IEEE} 38th International Conference on Data Engineering ({ICDE})}}. \bibinfo{publisher}{{IEEE}}.
\newblock
\urldef\tempurl%
\url{https://doi.org/10.1109/icde53745.2022.00044}
\showDOI{\tempurl}


\bibitem[Yang and Counts(2010)]%
        {222}
\bibfield{author}{\bibinfo{person}{Jiang Yang} {and} \bibinfo{person}{Scott Counts}.} \bibinfo{year}{2010}\natexlab{}.
\newblock \showarticletitle{Predicting the speed, scale, and range of information diffusion in twitter}. In \bibinfo{booktitle}{\emph{Proceedings of the International AAAI Conference on Web and Social Media}}, Vol.~\bibinfo{volume}{4}. \bibinfo{pages}{355--358}.
\newblock


\bibitem[Yang and Leskovec(2011)]%
        {224}
\bibfield{author}{\bibinfo{person}{Jaewon Yang} {and} \bibinfo{person}{Jure Leskovec}.} \bibinfo{year}{2011}\natexlab{}.
\newblock \showarticletitle{Patterns of temporal variation in online media}. In \bibinfo{booktitle}{\emph{Proceedings of the fourth ACM international conference on Web search and data mining}}. \bibinfo{pages}{177--186}.
\newblock


\bibitem[Yang et~al\mbox{.}(2014)]%
        {1-224}
\bibfield{author}{\bibinfo{person}{Jaewon Yang}, \bibinfo{person}{Julian McAuley}, \bibinfo{person}{Jure Leskovec}, \bibinfo{person}{Paea LePendu}, {and} \bibinfo{person}{Nigam Shah}.} \bibinfo{year}{2014}\natexlab{}.
\newblock \showarticletitle{Finding progression stages in time-evolving event sequences}. In \bibinfo{booktitle}{\emph{Proceedings of the 23rd international conference on World wide web}}. \bibinfo{publisher}{{ACM}}.
\newblock
\urldef\tempurl%
\url{https://doi.org/10.1145/2566486.2568044}
\showDOI{\tempurl}


\bibitem[Yang and Cervone(2019a)]%
        {2-44}
\bibfield{author}{\bibinfo{person}{Liping Yang} {and} \bibinfo{person}{Guido Cervone}.} \bibinfo{year}{2019}\natexlab{a}.
\newblock \showarticletitle{Analysis of remote sensing imagery for disaster assessment using deep learning: a case study of flooding event}.
\newblock \bibinfo{journal}{\emph{Soft Computing}} \bibinfo{volume}{23}, \bibinfo{number}{24} (\bibinfo{date}{mar} \bibinfo{year}{2019}), \bibinfo{pages}{13393--13408}.
\newblock
\urldef\tempurl%
\url{https://doi.org/10.1007/s00500-019-03878-8}
\showDOI{\tempurl}


\bibitem[Yang and Cervone(2019b)]%
        {2-54}
\bibfield{author}{\bibinfo{person}{Liping Yang} {and} \bibinfo{person}{Guido Cervone}.} \bibinfo{year}{2019}\natexlab{b}.
\newblock \showarticletitle{Analysis of remote sensing imagery for disaster assessment using deep learning: a case study of flooding event}.
\newblock \bibinfo{journal}{\emph{Soft Computing}} \bibinfo{volume}{23}, \bibinfo{number}{24} (\bibinfo{date}{mar} \bibinfo{year}{2019}), \bibinfo{pages}{13393--13408}.
\newblock
\urldef\tempurl%
\url{https://doi.org/10.1007/s00500-019-03878-8}
\showDOI{\tempurl}


\bibitem[Yao et~al\mbox{.}(2020)]%
        {4-2}
\bibfield{author}{\bibinfo{person}{Wenlin Yao}, \bibinfo{person}{Cheng Zhang}, \bibinfo{person}{Shiva Saravanan}, \bibinfo{person}{Ruihong Huang}, {and} \bibinfo{person}{Ali Mostafavi}.} \bibinfo{year}{2020}\natexlab{}.
\newblock \showarticletitle{Weakly-Supervised Fine-Grained Event Recognition on Social Media Texts for Disaster Management}.
\newblock \bibinfo{journal}{\emph{Proceedings of the {AAAI} Conference on Artificial Intelligence}} \bibinfo{volume}{34}, \bibinfo{number}{01} (\bibinfo{date}{apr} \bibinfo{year}{2020}), \bibinfo{pages}{532--539}.
\newblock
\urldef\tempurl%
\url{https://doi.org/10.1609/aaai.v34i01.5391}
\showDOI{\tempurl}


\bibitem[Yi et~al\mbox{.}(2019)]%
        {1-227}
\bibfield{author}{\bibinfo{person}{Fei Yi}, \bibinfo{person}{Zhiwen Yu}, \bibinfo{person}{Fuzhen Zhuang}, {and} \bibinfo{person}{Bin Guo}.} \bibinfo{year}{2019}\natexlab{}.
\newblock \showarticletitle{Neural Network based Continuous Conditional Random Field for Fine-grained Crime Prediction}. In \bibinfo{booktitle}{\emph{Proceedings of the Twenty-Eighth International Joint Conference on Artificial Intelligence}}. \bibinfo{publisher}{International Joint Conferences on Artificial Intelligence Organization}.
\newblock
\urldef\tempurl%
\url{https://doi.org/10.24963/ijcai.2019/577}
\showDOI{\tempurl}


\bibitem[Yu et~al\mbox{.}(2020)]%
        {4-9}
\bibfield{author}{\bibinfo{person}{Manzhu Yu}, \bibinfo{person}{Qunying Huang}, \bibinfo{person}{Han Qin}, \bibinfo{person}{Chris Scheele}, {and} \bibinfo{person}{Chaowei Yang}.} \bibinfo{year}{2020}\natexlab{}.
\newblock \showarticletitle{Deep learning for real-time social media text classification for situation awareness {\textendash} using Hurricanes Sandy, Harvey, and Irma as case studies}.
\newblock In \bibinfo{booktitle}{\emph{Social Sensing and Big Data Computing for Disaster Management}}. \bibinfo{publisher}{Routledge}, \bibinfo{pages}{33--50}.
\newblock
\urldef\tempurl%
\url{https://doi.org/10.4324/9781003106494-3}
\showDOI{\tempurl}


\bibitem[Yu et~al\mbox{.}(2019)]%
        {1-231}
\bibfield{author}{\bibinfo{person}{Shuqi Yu}, \bibinfo{person}{Linmei Hu}, {and} \bibinfo{person}{Bin Wu}.} \bibinfo{year}{2019}\natexlab{}.
\newblock \showarticletitle{{DRAM}: A Deep Reinforced Intra-attentive Model for Event Prediction}.
\newblock In \bibinfo{booktitle}{\emph{Knowledge Science, Engineering and Management}}. \bibinfo{publisher}{Springer International Publishing}, \bibinfo{pages}{701--713}.
\newblock
\urldef\tempurl%
\url{https://doi.org/10.1007/978-3-030-29551-6_62}
\showDOI{\tempurl}


\bibitem[Zafarani et~al\mbox{.}(2014)]%
        {1-236}
\bibfield{author}{\bibinfo{person}{Reza Zafarani}, \bibinfo{person}{Mohammad~Ali Abbasi}, {and} \bibinfo{person}{Huan Liu}.} \bibinfo{year}{2014}\natexlab{}.
\newblock \bibinfo{booktitle}{\emph{Social Media Mining}}.
\newblock \bibinfo{publisher}{Cambridge University Press}.
\newblock
\urldef\tempurl%
\url{https://doi.org/10.1017/cbo9781139088510}
\showDOI{\tempurl}


\bibitem[Zaman et~al\mbox{.}(2014)]%
        {238}
\bibfield{author}{\bibinfo{person}{Tauhid Zaman}, \bibinfo{person}{Emily~B Fox}, {and} \bibinfo{person}{Eric~T Bradlow}.} \bibinfo{year}{2014}\natexlab{}.
\newblock \showarticletitle{A bayesian approach for predicting the popularity of tweets}.
\newblock  (\bibinfo{year}{2014}).
\newblock


\bibitem[Zhang et~al\mbox{.}(2016)]%
        {241}
\bibfield{author}{\bibinfo{person}{Bolei Zhang}, \bibinfo{person}{Zhuzhong Qian}, {and} \bibinfo{person}{Sanglu Lu}.} \bibinfo{year}{2016}\natexlab{}.
\newblock \showarticletitle{Structure pattern analysis and cascade prediction in social networks}. In \bibinfo{booktitle}{\emph{Machine Learning and Knowledge Discovery in Databases: European Conference, ECML PKDD 2016, Riva del Garda, Italy, September 19-23, 2016, Proceedings, Part I 16}}. Springer, \bibinfo{pages}{524--539}.
\newblock


\bibitem[Zhao et~al\mbox{.}(2016)]%
        {1-240}
\bibfield{author}{\bibinfo{person}{Liang Zhao}, \bibinfo{person}{Feng Chen}, \bibinfo{person}{Chang-Tien Lu}, {and} \bibinfo{person}{Naren Ramakrishnan}.} \bibinfo{year}{2016}\natexlab{}.
\newblock \showarticletitle{Multi-resolution Spatial Event Forecasting in Social Media}. In \bibinfo{booktitle}{\emph{2016 {IEEE} 16th International Conference on Data Mining ({ICDM})}}. \bibinfo{publisher}{{IEEE}}.
\newblock
\urldef\tempurl%
\url{https://doi.org/10.1109/icdm.2016.0080}
\showDOI{\tempurl}


\bibitem[Zhao et~al\mbox{.}(2015)]%
        {1-244}
\bibfield{author}{\bibinfo{person}{Liang Zhao}, \bibinfo{person}{Qian Sun}, \bibinfo{person}{Jieping Ye}, \bibinfo{person}{Feng Chen}, \bibinfo{person}{Chang-Tien Lu}, {and} \bibinfo{person}{Naren Ramakrishnan}.} \bibinfo{year}{2015}\natexlab{}.
\newblock \showarticletitle{Multi-Task Learning for Spatio-Temporal Event Forecasting}. In \bibinfo{booktitle}{\emph{Proceedings of the 21th {ACM} {SIGKDD} International Conference on Knowledge Discovery and Data Mining}}. \bibinfo{publisher}{{ACM}}.
\newblock
\urldef\tempurl%
\url{https://doi.org/10.1145/2783258.2783377}
\showDOI{\tempurl}


\bibitem[Zhao et~al\mbox{.}(2017)]%
        {1-249}
\bibfield{author}{\bibinfo{person}{Sendong Zhao}, \bibinfo{person}{Quan Wang}, \bibinfo{person}{Sean Massung}, \bibinfo{person}{Bing Qin}, \bibinfo{person}{Ting Liu}, \bibinfo{person}{Bin Wang}, {and} \bibinfo{person}{ChengXiang Zhai}.} \bibinfo{year}{2017}\natexlab{}.
\newblock \showarticletitle{Constructing and Embedding Abstract Event Causality Networks from Text Snippets}. In \bibinfo{booktitle}{\emph{Proceedings of the Tenth {ACM} International Conference on Web Search and Data Mining}}. \bibinfo{publisher}{{ACM}}.
\newblock
\urldef\tempurl%
\url{https://doi.org/10.1145/3018661.3018707}
\showDOI{\tempurl}


\bibitem[Zhao and Tang(2017)]%
        {1-250}
\bibfield{author}{\bibinfo{person}{Xiangyu Zhao} {and} \bibinfo{person}{Jiliang Tang}.} \bibinfo{year}{2017}\natexlab{}.
\newblock \showarticletitle{Modeling Temporal-Spatial Correlations for Crime Prediction}. In \bibinfo{booktitle}{\emph{Proceedings of the 2017 {ACM} on Conference on Information and Knowledge Management}}. \bibinfo{publisher}{{ACM}}.
\newblock
\urldef\tempurl%
\url{https://doi.org/10.1145/3132847.3133024}
\showDOI{\tempurl}


\bibitem[Zheng et~al\mbox{.}(2017)]%
        {3-11}
\bibfield{author}{\bibinfo{person}{Yu-Jun Zheng}, \bibinfo{person}{Sheng-Yong Chen}, \bibinfo{person}{Yu Xue}, {and} \bibinfo{person}{Jin-Yun Xue}.} \bibinfo{year}{2017}\natexlab{}.
\newblock \showarticletitle{A Pythagorean-Type Fuzzy Deep Denoising Autoencoder for Industrial Accident Early Warning}.
\newblock \bibinfo{journal}{\emph{{IEEE} Transactions on Fuzzy Systems}} \bibinfo{volume}{25}, \bibinfo{number}{6} (\bibinfo{date}{dec} \bibinfo{year}{2017}), \bibinfo{pages}{1561--1575}.
\newblock
\urldef\tempurl%
\url{https://doi.org/10.1109/tfuzz.2017.2738605}
\showDOI{\tempurl}


\bibitem[Zhou et~al\mbox{.}(2015)]%
        {1-252}
\bibfield{author}{\bibinfo{person}{Cheng Zhou}, \bibinfo{person}{Boris Cule}, {and} \bibinfo{person}{Bart Goethals}.} \bibinfo{year}{2015}\natexlab{}.
\newblock \showarticletitle{A pattern based predictor for event streams}.
\newblock \bibinfo{journal}{\emph{Expert Systems with Applications}} \bibinfo{volume}{42}, \bibinfo{number}{23} (\bibinfo{date}{dec} \bibinfo{year}{2015}), \bibinfo{pages}{9294--9306}.
\newblock
\urldef\tempurl%
\url{https://doi.org/10.1016/j.eswa.2015.08.021}
\showDOI{\tempurl}


\bibitem[Zhou et~al\mbox{.}(2019)]%
        {1-253}
\bibfield{author}{\bibinfo{person}{Lihua Zhou}, \bibinfo{person}{Guowang Du}, \bibinfo{person}{Ruxin Wang}, \bibinfo{person}{Dapeng Tao}, \bibinfo{person}{Lizhen Wang}, \bibinfo{person}{Jun Cheng}, {and} \bibinfo{person}{Jing Wang}.} \bibinfo{year}{2019}\natexlab{}.
\newblock \showarticletitle{A tensor framework for geosensor data forecasting of significant societal events}.
\newblock \bibinfo{journal}{\emph{Pattern Recognition}}  \bibinfo{volume}{88} (\bibinfo{date}{apr} \bibinfo{year}{2019}), \bibinfo{pages}{27--37}.
\newblock
\urldef\tempurl%
\url{https://doi.org/10.1016/j.patcog.2018.10.021}
\showDOI{\tempurl}


\end{thebibliography}

\end{document}